%% file: ANA-SUSY-2018-05-PAPER.tex
\documentclass[cernpreprint, atlasdraft=false, texlive=2023, UKenglish, texmf, orcidlogo]{atlasdoc}
 
\usepackage[subfig]{atlaspackage}
\usepackage{bm}
\usepackage{adjustbox}

\usepackage{dcolumn} 
\newcolumntype{d}[1]{D{.}{.}{#1}}
\newcolumntype{P}[1]{D{,}{\,\pm\,}{#1}}
 
\usepackage[acronym,shortcuts]{glossaries}
\glsdisablehyper
 
\usepackage{atlasbiblatex}
 
\usepackage{atlasphysics}
 
\graphicspath{{logos/}{figures/}}
 
\usepackage{ANA-SUSY-2018-05-PAPER-defs}

\AtlasTitle{Searches for new phenomena in events with two leptons, jets, and missing transverse momentum in 139~fb$^{-1}$ of $\boldmath\sqrt{s}=13$~TeV $\boldmath pp$ collisions with the ATLAS detector}
 
\AtlasAbstract{
Searches for new phenomena inspired by supersymmetry in final states containing an $e^+e^-$ or $\mu^+\mu^-$ pair, jets, and missing transverse momentum are presented.
These searches make use of proton--proton collision data with an integrated luminosity of $139~\text{fb}^{-1}$, collected during 2015--2018 at a centre-of-mass
energy $\sqrt{s}=13~$TeV by the ATLAS detector at the Large Hadron Collider.
Two searches target the pair production of charginos and neutralinos.
One uses the recursive-jigsaw reconstruction technique to follow up on excesses observed in $36.1~\text{fb}^{-1}$ of data, and the other uses conventional event variables.
The third search targets pair production of coloured supersymmetric particles (squarks or gluinos) decaying through the next-to-lightest neutralino $(\tilde\chi_2^0)$ via a slepton $(\tilde\ell)$ or $Z$ boson into $\ell^+\ell^-\tilde\chi_1^0$, resulting in a kinematic endpoint or peak in the dilepton invariant mass spectrum.
The data are found to be consistent with the Standard Model expectations.
Results are interpreted using simplified models and exclude masses up to 900~GeV for electroweakinos,
1550~GeV for squarks, and 2250~GeV for gluinos.}
 
\author{The ATLAS Collaboration}
 
\AtlasRefCode{SUSY-2018-05}
 
\PreprintIdNumber{CERN-EP-2022-014}

\AtlasJournal{EPJC}
 
\AtlasJournalRef{Eur. Phys. J. C 83 (2023) 515}
 
\AtlasDOI{10.1140/epjc/s10052-023-11434-w}

\hypersetup{pdftitle={ATLAS document},pdfauthor={The ATLAS Collaboration}}
 
\begin{document}
 
\maketitle
 
\tableofcontents
 
\clearpage
 
\section{Introduction}
\label{sec:intro}


Supersymmetric (SUSY) \cite{Golfand:1971iw,Volkov:1973ix,Wess:1974tw,Wess:1974jb,Ferrara:1974pu,Salam:1974ig} extensions to the \ac{SM} have the potential to resolve the SM gauge hierarchy problem, explain the origin
of dark matter \cite{Goldberg:1983nd,Ellis:1983ew}, and lead to a grand unified theory of nature \cite{Sakai:1981gr,Dimopoulos:1981yj,Ibanez:1981yh,Dimopoulos:1981zb}.
Partner particles, called sparticles, differ from their \ac{SM} counterparts by half a unit of spin.
The scalar partners of quarks and leptons are squarks (\squark) and sleptons (\slepton) respectively.
Gluinos (\gluino) are the fermionic partners of gluons.
Charginos ($\tilde\chi_i^\pm$) and neutralinos ($\tilde\chi_i^0$) are the mass eigenstates of the mixing between the SUSY partners of the Higgs (higgsinos) and electroweak bosons, where $i$ runs from the lightest to heaviest mass.
If $R$-parity is conserved \cite{Farrar:1978xj}, then the \ac{LSP} is stable and a good candidate for dark matter.
 
This paper reports on searches for electroweak and strong production of sparticles in events with exactly two \ac{SF} \ac{OS} electrons or muons, jets, and missing transverse momentum (\ptmiss, with magnitude \MET).
The full Run~2 dataset of 13~\TeV\ proton--proton collisions collected by the ATLAS detector~\cite{PERF-2007-01} at the \ac{LHC} \cite{Evans:2008zzb} is used, corresponding to an integrated luminosity of 139~\ifb.
The same-flavour lepton final state is used in order to make use of the dilepton system's invariant mass (\mll) as a discriminant to search for events in models with leptonic $Z$ boson decays or models where the \mll\ distribution has a kinematic endpoint.
Two searches targeting electroweak production of sparticles and one targeting strong production are considered.
The first consists of two \acp{SR}, using \ac{RJR} variables~\cite{RJR} targeting electroweak production, which check whether previously observed excesses of 2.0$\sigma$ and 1.4$\sigma$ above the \ac{SM} expectations in the 36~\ifb\ 13~\TeV\ dataset collected during 2015--2016 \cite{SUSY-2017-03} persist with more data.
This is referred to as the RJR search, and only model-independent upper limits are presented for it.
The same 36~\ifb\ 13~\TeV\ analysis also included three-lepton regions which had 3.0$\sigma$ and 2.1$\sigma$ excesses above the Standard Model expectations, which were not observed with more data \cite{SUSY-2019-09}.
This previous search, and this update, used RJR variables designed to target electroweak production of SUSY particles.
The second targets electroweak (EWK) production of chargino--neutralino pairs decaying to $W$ and $Z$ bosons along with two {\chionezero} neutralinos, and also includes a new search inspired by \ac{GMSB} \cite{Dine:1981gu,AlvarezGaume:1981wy,Nappi:1982hm} targeting the pair production of higgsino \acp{NLSP} decaying into a $ZZ$ or $Zh$ pair and gravitino \acp{LSP}.
This is referred to as the EWK search, and it follows a methodology similar to that used in the two-lepton channel in a previous search \cite{SUSY-2016-24} using the 36~\ifb\ 13~\TeV\ dataset, but with optimizations for the full Run~2 dataset and a new region targeting off-shell $Z$ boson decays.
The third, the Strong search, targets the production of gluino or squark pairs that produce lepton pairs from $\tilde{\chi}^0_2 \rightarrow \ell^+\ell^- \tilde{\chi}^0_1$ decays and follows a methodology similar to that in a previous search~\cite{SUSY-2016-33} based on 36~\ifb\ of 13~\TeV\ data, also with optimizations for the full Run~2 dataset.
These updated EWK and Strong searches benefit from a larger dataset and an optimization of the analysis, generally resulting in tighter selection requirements for signal-like events.
The EWK search now includes additional binning of the \acp{SR}, further improving sensitivity to the considered signal models.
 
The EWK and Strong searches interpret the results with simplified models, described in the next section, by performing separate model-dependent profile likelihood fits \cite{Cowan:2010js} in their respective regions.
All three searches also report upper limits on possible \ac{BSM} event yields from model-independent fits to single-bin regions, where the \ac{BSM} signal is assumed to only populate the \ac{SR}.
 
The EWK search presented here extends the sensitivity to \ac{GMSB} models with \chionezero\ masses in the 400--500~\GeV\ range, between the limits from ATLAS searches in a four-lepton final state \cite{SUSY-2018-02} and an all-hadronic final state \cite{SUSY-2018-41}.
It also reaches higher \chionezero\ masses around chargino masses of 600~\GeV\ for the C1N2 model described in Table~\ref{tab:models}, between the limits from ATLAS searches in a three-lepton final state \cite{SUSY-2019-09} and the same electroweak all-hadronic search.
The Strong search has a gluino mass sensitivity similar to that of an ATLAS search in a single-lepton final state \cite{SUSY-2018-10}. The zero-lepton ATLAS search \cite{SUSY-2018-22} has sensitivity to gluinos a few hundred~\GeV\ heavier than in both of these searches, but the Strong search presented here has sensitivity to higher \chionezero\ masses.
A strong all-hadronic search also has sensitivity to squarks a few hundred~\GeV\ heavier than in the Strong search.
Unlike the EWK searches targeting the same model with different \ac{SM} boson decays, the decay chains of the gluinos and squarks differ between the various strong searches.
Thus each analysis complements each other by testing different assumptions of the SUSY particle spectra.
 
Similar searches have been performed by the CMS Collaboration with the full Run~2 dataset.
For electroweak production, limits on chargino--neutralino production (C1N2 model in Table~\ref{tab:models}) were presented in Ref.~\cite{CMS2LEWKStrong2020} and on the GMSB model in Ref.~\cite{CMS-PAS-SUS-19-012}, albeit with a different final state requiring same-charge leptons or more than two leptons.
For strong production, limits on similar models but with different parameterizations were presented in Ref.~\cite{CMS2LEWKStrong2020}.
 
This paper is organized with the common aspects of the searches preceding sections with the details specific to each search.
Section~\ref{sec:susy} describes the SUSY signal models targeted in this paper.
Section~\ref{sec:detector} describes the ATLAS detector.
Section~\ref{sec:datamc} describes the data and simulated samples used to guide the analysis strategy and estimate background and signal yields.
Section~\ref{sec:objects} describes the event reconstruction and criteria used to identify the physics objects used in the searches.
Section~\ref{sec:selection} describes the event selections that define the various search regions in each search.
Section~\ref{sec:background} describes the background estimation for each search.
Section~\ref{sec:systs} describes the uncertainties.
Sections~\ref{sec:result} and \ref{sec:interpretation} present the results and interpretations of the results, respectively.
The conclusions are presented in Section~\ref{sec:conclusion}.


\FloatBarrier
\section{SUSY signal models}
\label{sec:susy}

Simplified models \cite{Alwall:2008ve,Alwall:2008ag,Alves:2011wf} inspired by SUSY are used to guide the search strategy and interpret the results.
Two classes of models that contain production of weakly interacting or strongly interacting SUSY particles are used.
Each model is scanned over a two-dimensional space, varying the masses or decay branching ratios of sparticles.
Table~\ref{tab:models} summarizes the simplified models considered for analysis.
As mentioned in Section~\ref{sec:intro}, electrons or muons are required to be in the final state, including those from leptonic decays of $\tau$-leptons.
However, these will often fail \ac{SR} requirements related to boson mass compatibility.

\begin{table}[htbp]
\begin{center}
\caption{
Summary of the simplified signal model topologies used in this paper.
Here $x$ and $y$ denote the $x$--$y$ plane across which the signal model masses, or branching fraction $B$, are varied to construct the signal grid.
A dash ($-$) signifies the column is irrelevant for that model.
For the gluino--slepton model, the masses of the superpartners of the left-handed leptons are given by $[m(\chitwozero)+m(\chionezero)]/2$,
while the superpartners of the right-handed leptons are decoupled.
The gluino and squark decays have equal branching fractions for $q=u,d,c,s,b$.
}
\begin{tabular}{lcccc}
\hline \\ [-1em]
Model                               & Production mode  &  $x$ & $y$ & $[x+y]/2$ \\
\hline \\ [-1em]
C1N2            & \chionepm\chitwozero    & $m(\chionepm)=m(\chitwozero)$ & $m(\chionezero)$ & $-$  \\
GMSB            & \chionezero\chionezero  & $m(\chionezero)$  & $B(\chionezero\rightarrow h \tilde G)$ & $-$ \\[0.5ex]
\hline
Gluino--slepton    & $\gluino\gluino$       & $m(\gluino)$                        &  $m(\chionezero)$        & $m(\chitwozero)$          \\[0.5ex]
Gluino--$Z^{(*)}$  & $\gluino\gluino$       & $m(\gluino)$                        &  $m(\chionezero)$   & $m(\chitwozero)$           \\[0.5ex]
Squark--$Z^{(*)}$  & $\tilde{q}\tilde{q}$   & $m(\tilde{q})$               &   $m(\chionezero)$                         & $m(\chitwozero)$       \\[0.5ex]
\hline
\end{tabular}
\label{tab:models}
\end{center}
\end{table}
 
\begin{figure}[htbp]
\centering
\subfloat[]{\includegraphics[width=0.27\textwidth]{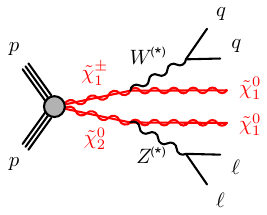}\label{subfig:c1n1}}
\hspace{1cm}
\subfloat[]{\includegraphics[width=0.27\textwidth]{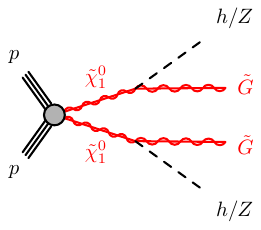}\label{subfig:gmsb}}\\
\subfloat[]{\includegraphics[width=0.27\textwidth]{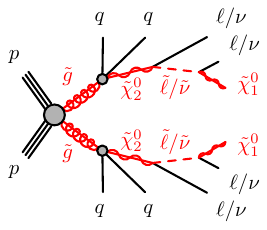}\label{subfig:ggsln}}
\hspace{1cm}
\subfloat[]{\includegraphics[width=0.27\textwidth]{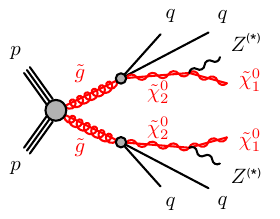}\label{subfig:ggzn}}
\hspace{1cm}
\subfloat[]{\includegraphics[width=0.27\textwidth]{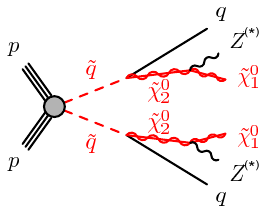}\label{subfig:sszn}}
\caption{Example decay topologies for the (top row) two electroweak SUSY models and (bottom row) three strong SUSY models considered in the analysis.
Model (a) shows the process $\tilde{\chi}_{2}^{0} \tilde{\chi}_{1}^{\pm} \rightarrow W^{(*)}(jj)Z^{(*)}(\ell\ell)+\met$ and model (b) shows a \acl{GMSB} model with a higgsino \acl{NLSP} and gravitino \acl{LSP}.
For the strong-production models, the left two decay topologies involve gluino pair production,
with the gluinos following an effective three-body decay for $\gluino\to q \bar{q} \tilde{\chi}_2^0$,
with $\tilde{\chi}_{2}^{0} \to \tilde{\ell}^{\mp}\ell^{\pm} / \tilde{\nu}\nu$ for the gluino--slepton model (c) and $\tilde{\chi}_2^0\rightarrow Z^{(*)} \tilde{\chi}_1^0$ in the gluino--$Z^{(*)}$ model (d).
The diagram (e) illustrates the squark--$Z^{(*)}$ model, where squarks are pair-produced,
followed by the decay $\tilde{q}\to q \tilde{\chi}_2^0$, with $\tilde{\chi}_2^0\rightarrow Z^{(*)} \tilde{\chi}_1^0$.
}
\label{fig:models}
\end{figure}

Two models of electroweak sparticle production are considered.
The first is the production of a chargino (\chionepm) and the second-lightest neutralino (\chitwozero), henceforth the C1N2 model, which decay via a $W$ boson and $Z$ boson respectively into \acp{LSP}, \chionezero.
The \chionepm\ and \chitwozero\ are assumed to have equal masses and always decay into a $W$ and $Z$ boson respectively.
A diagram of this model is shown in Figure~\ref{subfig:c1n1}.
The second is the pair production of higgsino neutralinos (\chionezero), which decay into a Higgs or $Z$ boson and a nearly massless gravitino $(\tilde{G})$.
This model is inspired by \ac{GMSB} \cite{Dine:1981gu,AlvarezGaume:1981wy,Nappi:1982hm} and referred to as the GMSB model.
A diagram of this model is shown in Figure~\ref{subfig:gmsb}.
The branching ratio of the \chionezero\ to a Higgs boson, alternatively a $Z$ boson, is varied from 0 to 100\%.
 
Three models of strong sparticle pair production are considered.
The choice of mass parameterization, namely setting intermediate particles halfway between the parent and child, enhances the topological differences between these simplified models and others with fewer intermediate particles in the decay chains~\cite{wacker} or very small mass differences between particles.
Changing this assumption matters when interpreting the particle spectrum of a new signal, but does not affect the sensitivity of the analysis to generic kinematic endpoint features.
In all three models, the gluino (or squark) and \chionezero\ masses are varied to produce a two-dimensional grid of signal models for interpretation.
The gluino and squark decays have equal branching fractions for $q=u,d,c,s,b$.
The first strong-production model, shown in Figure~\ref{subfig:ggsln}, is referred to as the gluino--slepton model, and assumes that the sleptons are lighter than the \chitwozero.
The gluino decays via a \chitwozero, which subsequently decays via a slepton or sneutrino $(\tilde{\nu})$ to the LSP (\chionezero).
The two decay channels, $\tilde{\chi}_{2}^{0} \to \tilde{\ell}^{\mp}\ell^{\pm}$ with $\tilde{\ell} \to \ell\tilde{\chi}_{1}^{0}$
or $\tilde{\chi}_{2}^{0} \to \tilde{\nu}\nu$ with $\tilde{\nu} \to \nu\tilde{\chi}_{1}^{0}$, have equal probability.
Only the superpartners of the left-handed leptons are allowed in the decays and they are taken to be mass degenerate, with a mass equal to the average of the \chitwozero\ and \chionezero\ masses.
The superpartners of the right-handed leptons are decoupled.
In this model, a kinematic endpoint in the dilepton invariant mass forms near half of the difference between the gluino and \chionezero\ masses, given by $m_{\ell\ell} = \sqrt{ (m^2_{\chitwozero}-m^2_{\tilde{\ell}})(m^2_{\tilde{\ell}}-m^2_{\chionezero}) / m^2_{\tilde{\ell}}}$ \cite{Paige:1996nx}.
This endpoint serves as an unambiguous signature of new physics that probes a wide variety of signal models, which can produce excesses over the entire \mll\ spectrum.
 
The second model, shown in Figure~\ref{subfig:ggzn}, is referred to as the gluino--\zstar\ model, where the \chitwozero\ from the gluino decay then decays as $\tilde{\chi}_{2}^{0} \to Z^{(*)}\tilde{\chi}_{1}^{0}$.
The \chitwozero\ mass is set to the average of the gluino and \chionezero\ masses.
 
The final model, shown in Figure~\ref{subfig:sszn}, is referred to as the squark--\zstar\ model, where the squark decays through a \chitwozero\ to a $Z$ boson and the LSP (\chionezero).
Similarly to the gluino--\zstar\ model, the \chitwozero\ mass is set to the average of the \squark\ and \chionezero\ masses.
Both of these models with decays to a $Z$ boson result in a kinematic endpoint in the dilepton invariant mass if the mass-splitting between the \chitwozero\ and \chionezero\ is smaller than the $Z$ boson mass.
If the mass-splitting is larger than the $Z$ boson mass, the models result in a peak in \mll\ at the $Z$ boson mass.
The superpartners of $q=u,d,c,s,b$ are all set to the same mass, with the superpartner of the $t$-quark decoupled.


\FloatBarrier
\section{ATLAS detector}
\label{sec:detector}

The ATLAS detector~\cite{PERF-2007-01} is a multipurpose particle detector with a forward--backward symmetric cylindrical geometry and a near \(4\pi\) coverage in solid angle.\footnote{
ATLAS uses a right-handed coordinate system with its origin at the nominal interaction point (IP) in the centre of the detector and the \(z\)-axis along the beam pipe.
The \(x\)-axis points from the IP to the centre of the LHC ring, and the \(y\)-axis points upwards.
Cylindrical coordinates \((r,\phi)\) are used in the transverse plane, \(\phi\) being the azimuthal angle around the \(z\)-axis.
The pseudorapidity is defined in terms of the polar angle \(\theta\) as \(\eta = -\ln \tan(\theta/2)\).
Angular distance is measured in units of \(\Delta R \equiv \sqrt{(\Delta\eta)^{2} + (\Delta\phi)^{2}}\).}  
It consists of an inner tracking detector surrounded by a thin superconducting solenoid providing a \SI{2}{\tesla} axial magnetic field, electromagnetic and hadronic calorimeters, and a muon spectrometer.
The inner tracking detector covers the pseudorapidity range \(|\eta| < 2.5\).
It consists of silicon pixel, silicon microstrip, and transition radiation tracking detectors.
An additional layer of silicon pixels, the  insertable B-layer~\cite{ATLAS-TDR-19,PIX-2018-001}, was installed before Run~2.
Lead/liquid-argon (LAr) sampling calorimeters provide electromagnetic (EM) energy measurements with high granularity.
A steel/scintillator-tile hadron calorimeter covers the central pseudorapidity range (\(|\eta| < 1.7\)).
The endcap and forward regions are instrumented with LAr calorimeters for both the EM and hadronic energy measurements up to \(|\eta| = 4.9\).
The muon spectrometer surrounds the calorimeters and is based on three large superconducting air-core toroidal magnets with eight coils each.
The field integral of the toroids ranges between \num{2.0} and \SI{6.0}{\tesla\metre} across most of the detector.
The muon spectrometer includes a system of precision chambers for tracking and fast detectors for triggering.
A two-level trigger system is used to select events.
The first-level trigger is implemented in hardware and uses a subset of the detector information to accept events at a rate below \SI{100}{\kHz}.
This is followed by a software-based trigger that reduces the accepted event rate to \SI{1}{\kHz} on average depending on the data-taking conditions.
An extensive software suite~\cite{ATL-SOFT-PUB-2021-001} is used in the reconstruction and analysis of real and simulated data, in detector operations, and in the trigger and data acquisition systems of the experiment.


\FloatBarrier
\section{Data and simulated samples}
\label{sec:datamc}

The LHC $pp$ collision data used in this analysis were collected by the ATLAS detector during 2015--2018 at a centre-of-mass collision energy of 13~\TeV.
After imposing requirements for beam and detector conditions and data quality \cite{DAPR-2018-01}, the dataset corresponds to an integrated luminosity of 139~\ifb.
The uncertainty in the combined 2015--2018 integrated luminosity is 1.7\% \cite{ATLAS-CONF-2019-021}, obtained using the LUCID-2 detector \cite{LUCID2} for the primary luminosity measurements.
 
Data events were collected using dilepton triggers with \pT\ thresholds of 7--26~\GeV\ varying with lepton flavour and data-taking period \cite{TRIG-2018-01,TRIG-2018-05}.
In 2017 and 2018, the asymmetric electron--muon trigger had thresholds of 26 and 8~\GeV\ respectively, where 26~\GeV\ is above the lepton \pt\ requirement used in the analysis.
The \pt\ range below the trigger threshold is covered by an asymmetric muon--electron trigger with thresholds of 24 and 7~\GeV\ respectively.
The use of the electron--muon trigger compensates for trigger inefficiencies due to the first-level muon trigger as it is seeded only by the first-level electromagnetic trigger.
The rest of the triggers used have thresholds of at most 24~\GeV.
Typical efficiencies for the muon part of a dilepton trigger requiring a single muon, including a Level-1 accept, is between 75--85\% depending on the muon \pt\ and $\eta$.
For electrons, the typical efficiency for a single part of a dilepton trigger is between 85--97\% depending on the electron  \pt\ and $\eta$.
 
Simulated event samples are used to help estimate the SM backgrounds, validate the analysis techniques, optimize the event selection, and provide predictions of the SUSY signal processes.
The majority of the \ac{SM} process samples were generated with \Sherpa[2.2] \cite{Bothmann:2019yzt} or with \POWHEGBOX[v2]~\cite{Nason:2004rx,Frixione:2007vw,Alioli:2010xd} and \Pythia[8] \cite{Sjostrand:2014zea} for the simulation of the \ac{PS}, hadronization, and underlying event.
The details of the \ac{ME} generator, \ac{PS} and parameter values (tune), \ac{PDF} choice, and cross-section for the \ac{SM} processes are listed in Table~\ref{tab:MC}.
The \ttbar\ and $Wt$ processes are referred to as `Top' events.
The $VV$ processes, where $V=W$ or $Z$, are referred to as `Diboson' events.
The remaining smaller backgrounds, except \zjets, are referred to as `Other' events.
The Other events category also includes the \ac{FNP} lepton background estimated from data as described in Section~\ref{sec:background}.
 
\begin{table*}[htbp]
\begin{center}
\caption{Simulated background event samples used in this analysis with the corresponding matrix element and parton shower generators,
cross-section order in perturbative QCD (pQCD) used to normalize the event yield, underlying-event tune and PDF set.
$V$ represents either a $W$ or $Z$ boson.
}
\scriptsize
\begin{tabular}{l l l l l l }
\hline
Physics process &  Generator  & Parton & Cross-section & Tune & PDF set\\
&             & shower & order in pQCD &      & \\
\noalign{\smallskip}\hline\noalign{\smallskip}
\ttbar~\cite{ATL-PHYS-PUB-2016-004}         & \POWHEGBOX[v2] \cite{Nason:2004rx,Frixione:2007vw,Alioli:2010xd}    & \Pythia[8.230] \cite{Sjostrand:2014zea} & NNLO+NNLL \cite{ttbarxsec1,ttbarxsec2,ttbarxsec3,ttbarxsec4,ttbarxsec5}                   & A14 \cite{ATL-PHYS-PUB-2014-021}   & \textsc{NNPDF2.3lo} \cite{Ball:2012cx}\\
$Wt$~\cite{ATL-PHYS-PUB-2016-004}  & \POWHEGBOX[v2]    & \Pythia[8.230] & NNLO+NNLL \cite{Kidonakis:2010b,Kidonakis:2010tc,Kidonakis:2011wy,Frederix:2012dh}        &A14    & \textsc{NNPDF3.0nlo} \cite{Ball:2014uwa}\\
$VV$~\cite{ATL-PHYS-PUB-2016-002} & \Sherpa[2.2.2] \cite{Bothmann:2019yzt} & \Sherpa[2.2.2] & NLO \cite{diboson1,diboson2} & \textsc{Sherpa}\ default & \textsc{NNPDF3.0nnlo} \\
$Z/\gamma^{*}(\rightarrow \ell \ell)$\,+\,jets~\cite{ATL-PHYS-PUB-2016-003}& \Sherpa[2.2.1]           & \Sherpa[2.2.1]  &NNLO \cite{DYNNLO1,DYcrosssection}       & \textsc{Sherpa}\ default     &\textsc{NNPDF3.0nnlo}\\
Higgs ($t\bar{t}H$) & \POWHEGBOX[v2]    & \Pythia[8.230] & NLO \cite{deFlorian:2016spz} & A14    & \textsc{NPDF2.3lo}\\
Higgs (VBF, $Vh$) & \POWHEGBOX[v2]     & \Pythia[8.212] & NNLO \cite{deFlorian:2016spz} & AZNLO \cite{STDM-2012-23} &  \textsc{CTEQ6L1} \cite{Pumplin:2002vw} \\
Higgs (ggF) & \POWHEGBOX[v2]     & \Pythia[8.212] & NNNLO \cite{deFlorian:2016spz} & AZNLO & \textsc{CTEQ6L1} \\
$ttt$, $tttt$, $t\bar{t}$\,+\,$WW$ & \textsc{MG5\_aMC@NLO} \cite{Alwall:2014hca} & \Pythia[8.186] \cite{Skands:2014pea} & NLO \cite{Frederix:2017wme}, \cite{Alwall:2014hca} & A14 & \textsc{NNPDF2.3lo}\\
$t\bar{t}$\,+\,$V$~\cite{ATL-PHYS-PUB-2016-005,Garzelli:2012bn}& \textsc{MG5\_aMC@NLO}        & \Pythia[8.210] & NLO \cite{Campbell:2012,Lazopoulos:2008,Alwall:2014hca} & \textsc{A14} & \textsc{NNPDF2.3lo}\\
$VVV$~\cite{ATL-PHYS-PUB-2016-002} & \Sherpa[2.2.2] & \Sherpa[2.2.2] &  NLO \cite{Bothmann:2019yzt} & \textsc{Sherpa}\ default     & \textsc{NNPDF3.0nnlo}\\
\noalign{\smallskip}\hline\noalign{\smallskip}
\end{tabular}
\label{tab:MC}
\end{center}
\end{table*}

The signal samples were generated using \textsc{MG5\_aMC@NLO}\,2.2.3 \cite{Alwall:2014hca} interfaced to \Pythia[8.186] with the A14 tune for the modelling of the \ac{PS}, hadronization and underlying event. The \ac{ME} calculation was performed at tree level and includes the emission of up to two additional partons. The \ac{PDF} set used for event generation was \textsc{NNPDF2.3lo}.
The ME--PS matching used the CKKW-L prescription, with a matching scale set to a quarter of the mass of the initial SUSY particle.
 
For the C1N2 and GMSB signal models, cross-sections are calculated to next-to-leading order in the strong coupling constant, adding the resummation of soft gluon emission at next-to-leading-logarithm accuracy (NLO+NLL)~\cite{Beenakker:1999xh,Debove:2010kf,Fuks:2012qx,Fuks:2013vua,Fiaschi:2018hgm}.
The nominal cross-section and its uncertainty are taken from an envelope of cross-section predictions using different \ac{PDF} sets and factorization and renormalization scales, as described in Ref.~\cite{Borschensky:2014cia}.
 
For the gluino and squark signal models, cross-sections are calculated to approximate next-to-next-to-leading order in the strong coupling constant, adding the resummation of soft gluon emission at next-to-next-to-leading-logarithm accuracy (approximate NNLO+NNLL)~\cite{Beenakker:2016lwe,Beenakker:2014sma,Beenakker:2013mva,Beenakker:2011sf,Beenakker:2009ha,Kulesza:2009kq,Kulesza:2008jb,Beenakker:1996ch}.
The nominal cross-section and its uncertainty are derived using the \PDFforLHC[15\_mc] PDF set, following the recommendations of Ref.~\cite{Butterworth:2015oua}.

The \ac{SM} background \ac{MC} samples were passed through a full simulation of the ATLAS detector~\cite{SOFT-2010-01} using \GEANT \cite{Agostinelli:2002hh}.
A fast simulation~\cite{SOFT-2010-01} was used for the signal MC samples; it relies on a parameterization of the lateral and longitudinal shower shapes of single particles in the  calorimeters, and on \GEANT elsewhere.
The effect of multiple $pp$ interactions per bunch crossing (pile-up) and detector-response effects due to interactions in neighbouring bunch crossings were included by overlaying the simulated hard-scattering events with additional inelastic $pp$ collision events generated with \Pythia[8.186].


\FloatBarrier
\section{Analysis object identification and selection}
\label{sec:objects}

Leptons and jets selected for analysis are categorized as `baseline' or `signal' objects by using various quality and kinematic requirements.
Baseline objects are used in the computation of the missing transverse momentum (\ptmiss) and its magnitude (\etmiss), to resolve ambiguities between closely spaced analysis objects, and to ensure orthogonality to other analyses.
The signal objects used in the analysis selection are required to pass more stringent requirements.
 
Electron candidates are reconstructed using energy clusters in the electromagnetic calorimeter matched to inner-detector tracks.
Baseline electrons are required to have $\pt>4.5$~\GeV, satisfy the `loose likelihood' criteria described in Ref.~\cite{EGAM-2018-01}, and reside within the region $|\eta|<2.47$.
Additionally, the baseline-electron tracks must pass within $|z_0\sin\theta| = 0.5$~mm of the primary vertex,
defined as the vertex with the largest sum of track \pTsq,
where $z_0$ is the longitudinal impact parameter with respect to the primary vertex.
Signal electrons are required to satisfy the `medium likelihood' criteria of Ref.~\cite{EGAM-2018-01} and to have $\pt>25$~\GeV.
The transverse-plane distance of closest approach of the signal electron to the beamline, divided by the corresponding uncertainty, must satisfy $|d_0/\sigma_{d_0}|<5$.
These electrons must also be isolated from other objects in the event according to a \pt-dependent isolation requirement based on calorimeter ($\text{E}_\text{T}^{\text{cone20}}/\pt<0.06$) and tracking information ($\pt^{\text{varcone20}}/\pt < 0.06$).
In RJR search regions with more than two leptons (only used for \acp{VR} and \acp{CR}), the electron $\pt$ requirement is lowered to 20~\GeV\ for the third and fourth electrons.
Measured in $Z\rightarrow ee$ events, the `medium' electron identification efficiency increases from 75\% at a \pt\ of 20~\GeV\ to 91\% at 100~\GeV.
The isolation requirement is approximately 70\% efficient for `medium likelihood' electrons at a \pt\ of 20~\GeV, rising to above 98\% at 60~\GeV\ and beyond \cite{EGAM-2018-01}.
 
Baseline muons are reconstructed from either inner-detector tracks matched to muon track segments in the muon spectrometer or combined tracks formed in the inner detector and muon spectrometer.
They are required to satisfy the `medium' selection criteria described in Ref.~\cite{MUON-2018-03}, $p_{\text{T}}>3$~\GeV, and $|\eta|<2.7$.
For consistency with the previous analysis, the RJR search limits the muon acceptance to $|\eta|<2.4$.
Similarly to electrons, the muon must pass within $|z_0\sin\theta| = 0.5$~mm of the primary vertex.
Signal muons are required to be isolated with $\text{E}_\text{T}^{\text{topocone20}}/\pt<0.15$ and $\pt^{\text{varcone30}}/\pt < 0.04$, and they must have $|d_0/\sigma_{d_0}|<3$, $|\eta|<2.6$, and $\pT>25$~\GeV.
In RJR search regions with more than two leptons (only used for \acp{VR} and \acp{CR}), the muon $\pt$ requirement is lowered to 20~\GeV\ for the third and fourth muon.
Measured in $Z\rightarrow\mu\mu$ events, the `medium' muon identification efficiency exceeds 98\% in regions with muon spectrometer coverage \cite{MUON-2018-03}.
The isolation requirement efficiency increases from approximately 85\% at a \pt\ of 20~\GeV\ to more than 99\% at 60~\GeV\ and above.
Proximity to a jet, which can occur in the Strong search, lowers the muon isolation efficiency by up to 10\%.
 
Jets are reconstructed from topological clusters of energy~\cite{PERF-2014-07} in the calorimeter using the anti-$k_{t}$ algorithm~\cite{Cacciari:2008gp,Cacciari:2005hq} with a radius parameter of 0.4 by making use of utilities within the \texttt{FastJet} package~\cite{Fastjet}.
The reconstructed jets are then calibrated to parton level by the application of a jet energy scale (JES) derived from 13~\TeV\ data and simulation~\cite{JETM-2018-05}.
A residual correction applied to jets in data is based on studies of the $\pT$ balance between jets and well-calibrated objects in the MC simulation and data.
Baseline jets are defined as jet candidates that have $\pt>20$~\GeV\ and reside within the region $|\eta|<2.8$.
Additional track-based criteria designed to select jets from the hard scatter and reject those originating from pile-up are applied to jets with $\pt<120$~\GeV\ and $|\eta|<2.5$.
These are imposed by using the `medium' working point of the \ac{JVT} described in Refs.~\cite{PERF-2014-03,PERF-2016-04}.
For jets with $\pT<50~\GeV$ in the region $|\eta|>2.5$, outside of the inner-detector acceptance, the tighter working point of forward \ac{JVT}, $\mathrm{fJVT}<0.4$ as defined in Ref.~\cite{PERF-2016-06}, uses shape and topological information to suppress jets originating from pile-up.
Signal jets are further required to have $\pt>30$~\GeV.
To be consistent with the previous analysis, the RJR search limits the jet acceptance to $|\eta|<2.4$.
Finally, events containing a baseline jet that does not pass jet quality requirements are vetoed in order to remove events impacted by detector noise and non-collision backgrounds~\cite{DAPR-2012-01,ATLAS-CONF-2015-029}.

The MV2C10 boosted decision tree algorithm~\cite{FTAG-2018-01} identifies jets containing $b$-hadrons by using quantities such as
the impact parameters of associated tracks and the positions of any good reconstructed secondary vertices.
A selection that provides 77\% efficiency for tagging jets from $b$-quarks in simulated \ttbar\ events is used.
These tagged jets are called $b$-jets.
The corresponding rejection factors against jets originating from $c$-quarks and light quarks in the same sample for this selection are 5 and 110 respectively \cite{FTAG-2018-01,ATLAS-CONF-2018-006}.
 
Photon candidates, used only in the computation of \ptmiss, are reconstructed using energy clusters in the electromagnetic calorimeter without matching inner-detector tracks, or with tracks consistent with having originated from a photon conversion vertex.
Photons are required to satisfy the `tight' selection criteria described in Ref.~\cite{EGAM-2018-01}, have $\pt>25~\GeV$, and reside within the region $|\eta|< 2.37$, excluding the transition region ($1.37<|\eta|<1.52$) between the barrel and endcaps of the electromagnetic calorimeter.

To avoid the duplication of analysis objects and resolve ambiguities, an overlap removal procedure is applied to baseline jets and signal leptons in the following order.
Electron candidates that share an inner-detector track with a baseline muon are rejected to remove electrons originating from photons radiated from muons.
Any baseline jet within $\Delta R=0.2$ of a baseline electron is removed.
Any electron that lies within $\Delta R=\mathrm{ min } (0.04+(10~\GeV)/\pt ,0.4)$ of a baseline jet is removed in order to suppress electrons from heavy-flavour decays.
If a baseline muon either resides within $\Delta R=0.2$ of, or has a track associated with, a remaining baseline jet, that jet is removed.
Muons are removed in favour of baseline jets with the same \pt-dependent $\Delta R$ requirement as electrons.
For the purpose of computing \ptmiss, photon candidates sharing their calorimeter cluster with an electron are removed.
 
The \ptmiss\ is defined as the negative vector sum of the transverse momenta of all baseline electrons,
muons, jets, and photons~\cite{PERF-2016-07}.
Low-momentum contributions from particle tracks originating from the primary vertex that are not associated with reconstructed analysis objects are also included in the calculation of \ptmiss.

All MC samples have corrections applied to take into account small differences between data and MC simulation in the identification, reconstruction and trigger efficiencies.
The $\pt$ values of leptons in MC samples are smeared to match the momentum resolution in data.


\FloatBarrier
\section{Event selection}
\label{sec:selection}

The searches are carried out in \acp{SR} designed to be sensitive to heavy new particles inspired by SUSY.
Auxiliary measurements are performed in \acp{CR}, which are orthogonal to their associated SRs, designed to be enriched in a particular background process and have low contamination from signal processes.
To validate the background estimation in the SRs, \acp{VR} are defined to be similar, but orthogonal, to the SRs and CRs.
The VRs typically have signal contaminations of less than 10\% near the exclusion limits, but in a few regions can be larger, close to 20\%.
 
All three searches share some common event requirements.
Events entering the SRs must have exactly two \ac{OS}~\ac{SF} signal leptons (electrons or muons) with $\pt>25$~\GeV, without any additional baseline leptons.
This requirement rejects events with a third, low-momentum lepton, such as from $WZ$ production.
This also means that for signal models which may result in more than two leptons, those decay modes are suppressed in favour of hadronic decays of bosons or additional neutrinos in the final state.
The majority of the search regions require events to have at least two jets with $\pt>30$~\GeV.
Analysis objects are ordered by decreasing \pt\ for further selections.
Sections \ref{sec:discVars} and \ref{sec:rjr_reco} describe variables used to construct the regions in the searches.
The RJR, EWK, and Strong search-specific selections are described in Sections~\ref{sec:rjr_selection}, \ref{sec:ewk_selection}, and \ref{sec:strong_selection} respectively.
The regions defined in the searches, in the same order, have a suffix -RJR, -EWK, or -STR.
Except in the case of the RJR search, which replicates the selection of the previous search, the event selections are optimized for the different final states depending on the type of sparticles and sparticle masses in the targeted models.
 
\subsection{Additional event selection variables}
\label{sec:discVars}
 
Signal models with large hadronic activity are targeted by placing additional requirements on the quantity \HT, defined as the scalar sum of the $\pT$ values
of all signal jets.
For the purpose of rejecting \ttbar\ background events, the \mttwo\ \cite{Lester:1999tx,Barr:2003rg} variable is used, defined as an extension of the transverse mass $m_{\text{T}}$ for the case of two missing particles:
\begin{equation}
m_{\text{T}}^2\left(\,\vec{p}_{\text{T},a},\vec{p}_{\text{T}}^{\text{miss}}\right) = 2 \times \left( p_{\text{T},a} \times \met - \vec{p}_{\text{T} a} \cdot \vec{p}_{\text{T}}^{\text{miss}} \right),
\label{eq:mt}
\end{equation}
\begin{equation}
\mttwo^2 = \min_{\vec{x}_{\text{T,1}}+\vec{x}_{\text{T,2}}=\vec{p}_{\text{T}}^{\text{miss}}} \left[ \max \{ m_{\text{T}}^2\left(\,\vec{p}_{\text{T} 1},\vec{x}_{\text{T,1}}\right) , m_{\text{T}}^2\left(\,\vec{p}_{\text{T} 2},\vec{x}_{\text{T,2}}\right) \} \right],
\end{equation}
where $\vec{p}_{\text{T}, a}$ is the transverse-momentum vector of the highest-\pt ($a=1$) or second-highest-\pt ($a=2$) lepton, and $\vec{x}_{\text{T,b}}$ ($b=1,2$) are two vectors representing the possible momenta of the invisible particles that minimize the \mttwo\ in the event.
The masses of the invisible particles are set to zero in the calculation.
The \mttwo\ variable tends to have an endpoint near the mass of the parent particle, e.g.\ the $W$ boson mass for \ttbar\ events when considering the leptons and \met.
Thus it can be used to separate signal models with larger mass-splittings from typical \ttbar\ events.
 
The \met\ significance \metsig\ quantifies how consistent the \met\ is with mismeasurements of objects in events without any genuine source of \met.
It is used to select events where it is more likely  that invisible particles are contributing to the \met.
The \met\ significance is defined \cite{METSig} as
\begin{equation*}
\mathcal{S}^2 = \frac{|\met|^2}{\sigma_\text{L}^2 (1-\rho_\text{LT}^2)},
\label{eq:metsig}
\end{equation*}
where $\sigma_\text{L}^2$ is the total variance in the longitudinal direction along \ptmiss, with the resolutions of all of the objects taken into account, and $\rho_\text{LT}$ is the correlation between the longitudinal and transverse resolutions of the objects.
 
\subsection{Recursive-jigsaw reconstruction}
\label{sec:rjr_reco}
 
The \ac{RJR} technique~\cite{RJR} is a method for decomposing measured properties event by event to provide a basis of kinematic variables.
This is achieved by approximating the rest frames of intermediate particle states in each event.
This reconstructed view of the event gives rise to a natural basis of kinematic observables, calculated by evaluating the momentum and energy of different objects in these rest frames.
Backgrounds are reduced by testing whether  each event exhibits the anticipated properties of the imposed decay tree under investigation while only applying minimal selection criteria to visible-object momenta and missing momenta.
The RJR technique is described in detail in Refs.~\cite{SparticlesInMotion,RJR,SantoniEWK}, and has been used in previous ATLAS searches~\cite{SUSY-2016-15,SUSY-2016-07,SUSY-2016-16}.

\begin{figure}[htbp]
\begin{center}
\subfloat[]{\includegraphics[width=0.32\textwidth]{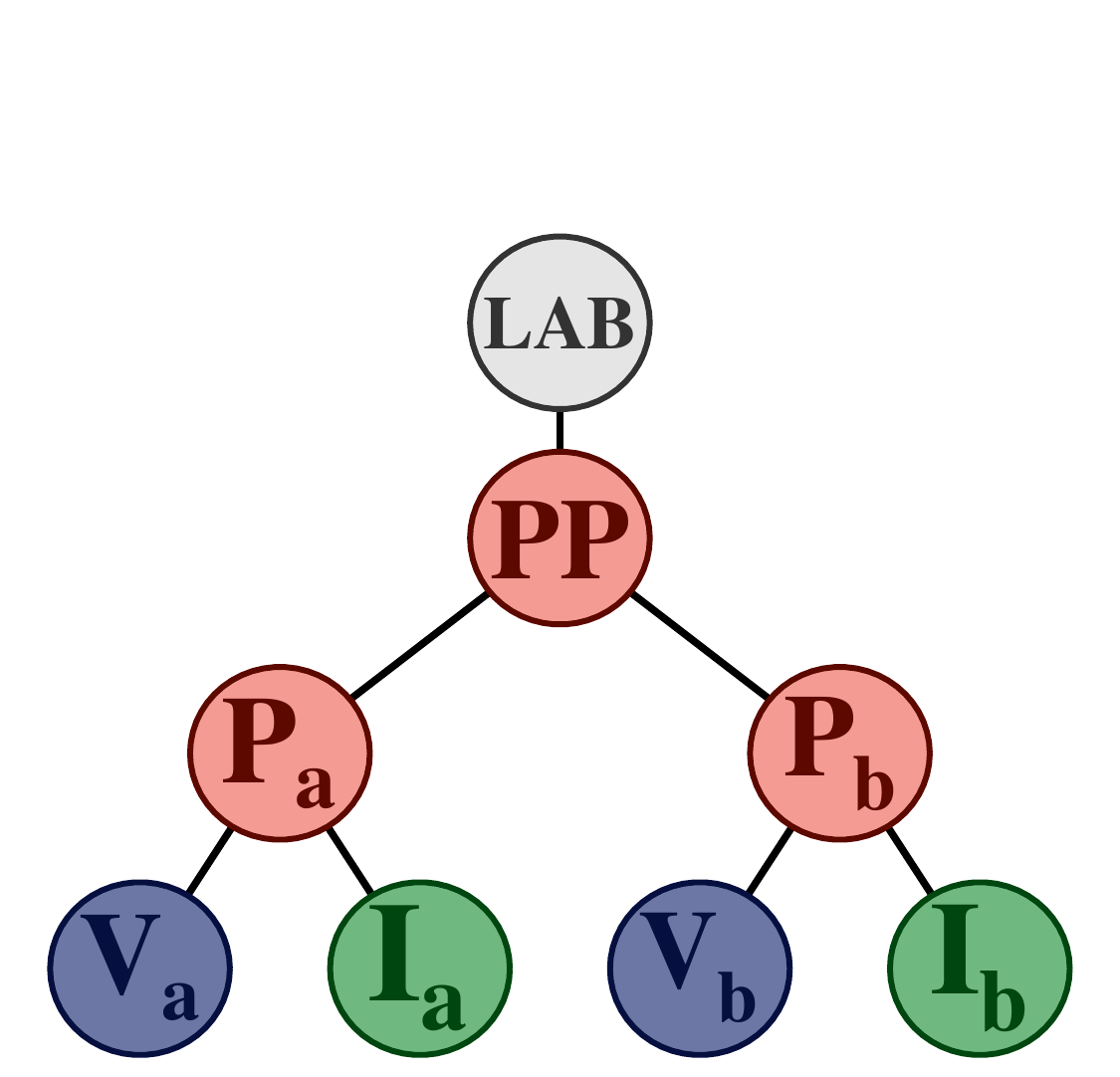}\label{subfig:rjr_treeA}}
\subfloat[]{\includegraphics[width=0.32\textwidth]{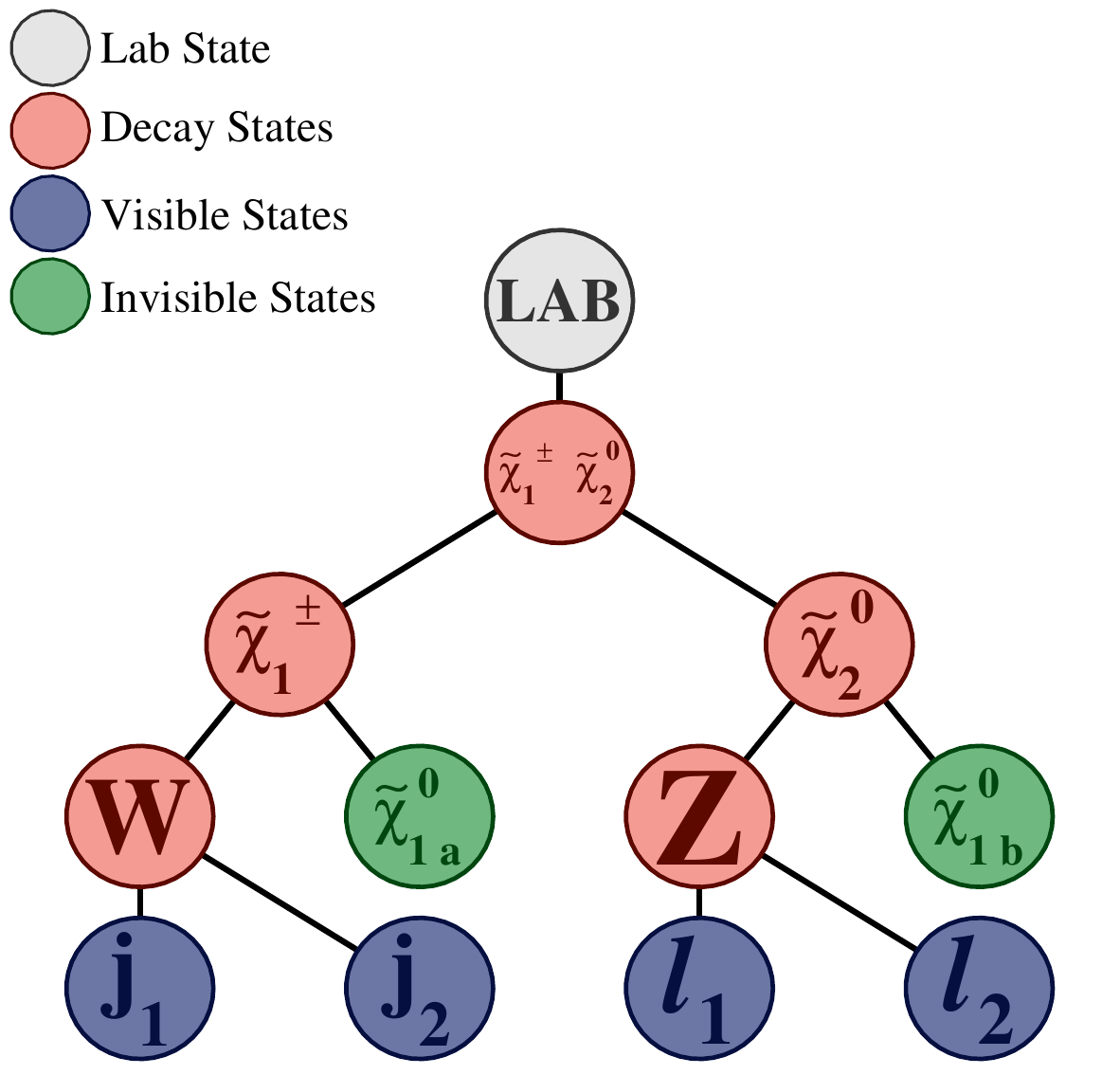}\label{subfig:rjr_treeB}}
\subfloat[]{\includegraphics[width=0.32\textwidth]{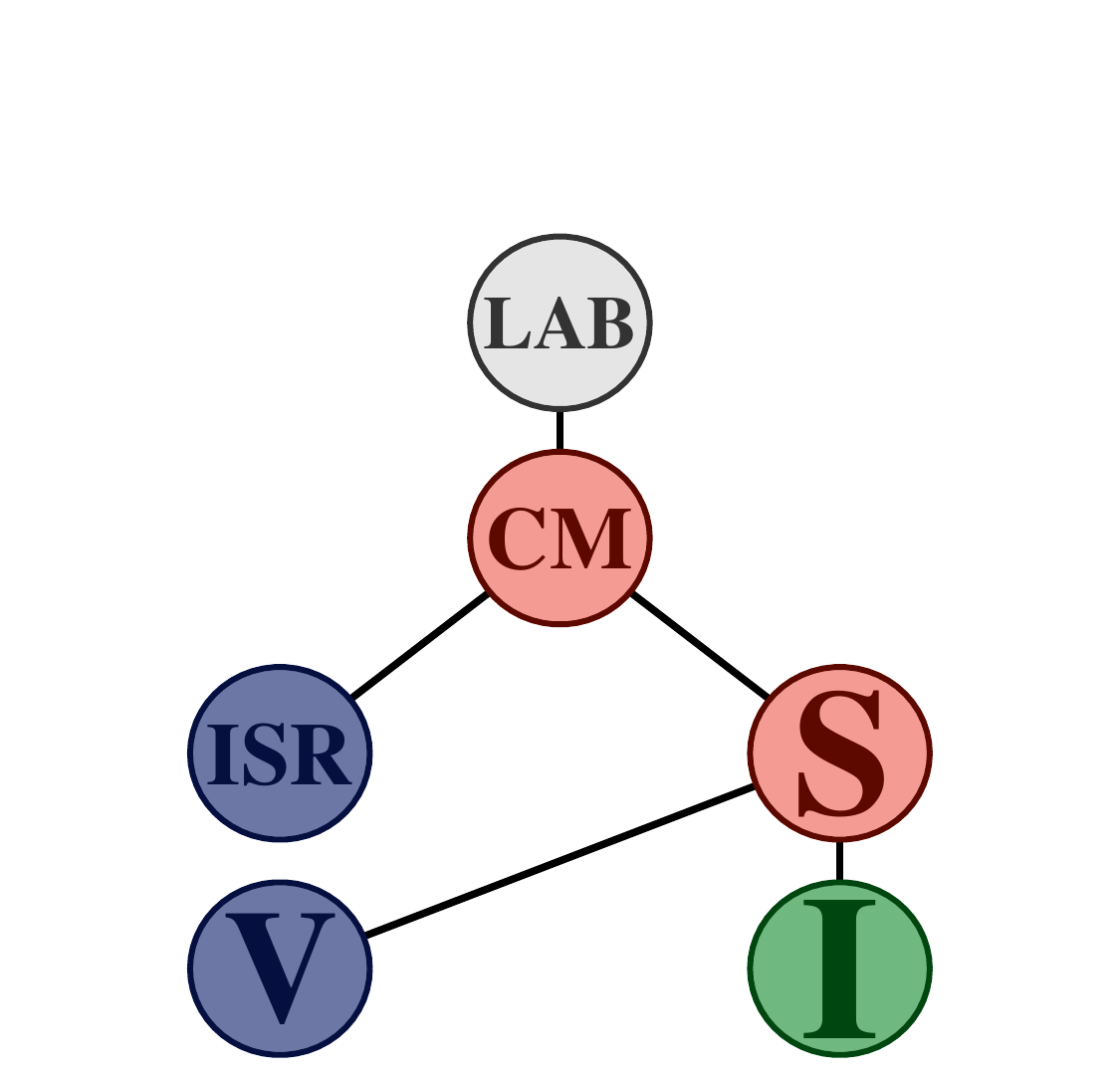}\label{subfig:rjr_treeC}}
\end{center}
\caption{\label{fig:InclusiveTree}
(a) The standard decay tree applied to pair-produced (PP) sparticles (parent objects), P$_{\textrm{a,b}}$, decaying to visible states V$_{\textrm{a,b}}$ and invisible states I$_{\textrm{a,b}}$.
(b) Decay tree for the C1N2 model with decays to the $2\ell$\,+\,$2$-jets final state.
(c) The decay tree for compressed scenarios with an initial-state radiation jet (ISR).
The signal sparticle system S recoils against the jet-radiation system ISR.
CM denotes the centre-of-mass frame.
}
\end{figure}

Electrons, muons, jets, and \ptmiss\ are used as input to the RJR algorithm.
A decay tree for the $2\ell$\,+\,$2j$ final state in this search, shown in Figure~\ref{subfig:rjr_treeB}, is constructed following the canonical process shown in Figure~\ref{subfig:rjr_treeA}.
These decay trees are motivated by the pair production of sparticles in $R$-parity-conserving models.
Each event is evaluated as if two sparticles (labelled PP) were produced, assigned to two hemispheres (P$_{\textrm{a}}$ and P$_{\textrm{b}}$), and subsequently decay into the particles observed in the detector, with V denoting the visible objects and I the invisible objects.
The benchmark signal models probed in this search give rise to signal events with at least two weakly interacting particles associated with two systems of invisible particles (shown in green). 
Both of the leptons must be assigned to either V$_{a}$ or V$_{b}$, not split across both. The jets must also be assigned to either of the two visible frames.
 
After partitioning the visible objects, the remaining unmeasured quantities related to the invisible particles (masses, longitudinal momenta, and contribution to \ptmiss) are associated with the two collections of invisible particles $\text{I}_{\text{a/b}}$.
The RJR algorithm determines values for the unmeasured particles' four-momenta, while requiring the masses of the invisible particles to be positive. In cases with multiple solutions, the solution with the smallest invariant mass of the visible system is chosen~\cite{RJR}. Once all measured and unmeasured momenta are defined, a set of variables can be constructed, such as multi-object invariant masses and angles between objects.
The primary energy-scale-sensitive observables used in the search presented here are a suite of variables denoted by $H$.
As shown in Eq.~(\ref{eq:Hdefinition}), the $H$ variables are constructed using different combinations of object momenta, including contributions from the invisible particles' four-momenta, and are not necessarily evaluated in the lab frame, nor only in the transverse plane.
The $H$ variables are labelled with a superscript F and two subscripts $n$ and $m$:
 
\begin{equation}
\label{eq:Hdefinition}
H_{n,m}^{\textrm{F}} = \sum_{i=1}^{n} |\vec{p}_{\textrm{vis},\ i}^{\textrm{~F}}| + \sum_{j=1}^{m} |\vec{p}_{\textrm{inv},\ j}^{\textrm{~F}}|.
\end{equation}

The F represents the rest frame in which the momenta are evaluated.
In this analysis, this may be the lab frame, the proxy for the sparticle--sparticle frame PP, or the proxy for the rest frame of an individual sparticle, P.
The subscripts $n$ and $m$ represent the number of visible and invisible momentum vectors considered, respectively. For events with fewer than $n$ visible objects, the sum only runs over the available momenta.
Only the leading $n-n_{\ell}$ jets are considered, where $n_{\ell}$ is the number of reconstructed leptons in the event.
An additional subscript `T' denotes a transverse version of the variable, where the transverse plane is defined in a frame F as follows: the Lorentz transformation relating F to the lab frame is decomposed into a boost along the beam axis, followed by a subsequent transverse boost.
The transverse plane is defined to be perpendicular to the longitudinal boost.
In this analysis, it is the plane transverse to the beamline.
 
The following variables are used in the definition of the \acp{SR}.
The value of $n$ depends on the number of visible objects; for the $2\ell$\,+\,$2$-jets final state, $n=4$.
 
\begin{itemize}
\item $H_{4,1}^{~\textrm{PP}}$ is a scale variable as described above that behaves similarly to the effective mass, $\meff$ (defined as the scalar sum of the transverse momenta of the visible objects and $\met$), used in previous ATLAS SUSY searches.
\item $H_{1,1}^{~\textrm{PP}}/H_{4,1}^{~\textrm{PP}}$ provides additional information when testing the balance between the two scale variables.
This provides excellent discrimination against unbalanced events where the large scale is dominated by a particular object's \pt\ or by large $\met$.
It behaves similarly to $\met/\meff$ and is used to reduce the \zjets\ background in cases where one high-\pt\ jet dominates.
\item $p_{\textrm{T}~\textrm{PP}}^{\textrm{lab}}/(p^{\textrm{lab}}_{\textrm{T}~\textrm{PP}} + H_{\textrm T~4,1}^{~\textrm{PP}})$ compares the magnitude of the vector sum of the transverse momenta of all objects assigned to the PP system in the lab frame ($p^{\textrm{lab}}_{\textrm{T}~\textrm{PP}}$) with the overall transverse scale variable considered.
This quantity tests for a significant boost in the transverse direction.
For signal events this quantity peaks sharply towards zero while for background processes the distribution is broader.
It is a test of how much a given process resembles the imposed PP system in the decay tree.
\item $\textrm{min}(H^{\textrm{P}_{\textrm{a}}}_{1,1},H^{\textrm{P}_{\textrm{b}}}_{1,1})/\textrm{min}(H^{\textrm{P}_{\textrm{a}}}_{2,1},H^{\textrm{P}_{\textrm{b}}}_{2,1})$ compares the scale due to one visible object and \met\ ($H^{\textrm{P}_{\textrm{a}}}_{1,1}$ and $H^{\textrm{P}_{\textrm{b}}}_{1,1}$ in their respective production frames) with the scale due to two visible objects ($H^{\textrm{P}_{\textrm{a}}}_{2,1}$ and $H^{\textrm{P}_{\textrm{b}}}_{2,1}$).
The numerator and denominator are each defined by finding the smaller of the values of these quantities.
This variable tests against a single object taking a large portion of the hemisphere momentum.
It is particularly useful in discriminating against \zjets\ events.
\item $\Delta\phi_{\textrm{V}}^{\textrm{P}}$ is the azimuthal opening angle between the visible system V in frame P and the direction of the boost from the PP frame to the P frame.
Standard Model backgrounds from diboson, top and \zjets\ processes peak towards zero and $\pi$ due to their topologies not obeying the imposed decay tree while signal events tend to have a flat distribution in this variable.
\end{itemize}
 
For selections involving three charged leptons, the $W$ boson transverse mass, $m_{\text{T}}^{W}$, is used.
It is derived as in Eq.~(\ref{eq:mt}) from \ptmiss\ and the transverse momentum of the charged lepton not associated with the $Z$ boson.
The three-lepton regions are only used for CRs and VRs.
 
For sparticle spectra with smaller mass-splittings and lower intrinsic \met, it can be useful to define a decay tree with a partially resolved sparticle system recoiling against a high-\pt\ \ac{ISR} jet, instead of the fully resolved decay tree described above.
This is shown Figure~\ref{subfig:rjr_treeC}.
This tree is simpler and attempts to identify visible (V) and invisible (I) systems that are the result of an intermediate state corresponding to the system of sparticles and their decay products (S).
Since the \met\ is used to identify which jets come from \ac{ISR}, a transverse view of the reconstructed event is used which ignores the longitudinal momentum of the jets and leptons, as described in Ref.~\cite{SparticlesInMotion}.
The reference frames appearing in the decay tree shown in Figure~\ref{subfig:rjr_treeC}, such as the \ac{CM} frame of the whole interaction, are approximations in this transverse projection.
The variables which are derived from this process leverage the relationship between the \ac{ISR} and the total S system:
 
\begin{itemize}
\item $p_{\mathrm{T\ ISR}}^{~\textrm{CM}}$ is the magnitude of the vector-summed transverse momenta of all jets assigned to the ISR system.
\item $p_{\mathrm{T\ I}}^{~\textrm{CM}}$ is the magnitude of the vector-summed transverse momenta of the invisible system, which behaves similarly to $\met$.
\item $p_{\mathrm{T}}^{~\textrm{CM}}$ is the magnitude of the vector-summed transverse momenta of the CM system.
\item $R_{\textrm{ISR}} \equiv \vec{p}_{\textrm{I}}^{~\textrm{CM}}\cdot \hat{p}_{\mathrm{T}~\textrm{S}}^{~\textrm{CM}}/p_{\mathrm{T}~\textrm{S}}^{~\textrm{CM}}$ serves as an estimate of $m_{\tilde{\chi}^{0}_{1}}/m_{\tilde{\chi}^{0}_{2}/\tilde{\chi}^{\pm}_{1}}$.
This corresponds to the fraction of the momentum of the S system that is carried by its invisible system I, with momentum $\vec{p}_{\textrm{I}}^{~\textrm{CM}}$ in the CM frame.
As $p_{\mathrm{T}~\textrm{S}}^{~\textrm{CM}}$ grows it becomes increasingly hard for backgrounds to possess a large value in this ratio -- a feature exhibited by compressed signals~\cite{SparticlesInMotion}.
\item $N_{\textrm{jet}}^{\textrm{S}}$ is the number of jets assigned to the signal system S.
\item $N_{\textrm{jet}}^{~\textrm{ISR}}$ is the number of jets assigned to the ISR system.
\item $\Delta\phi_{~\textrm{ISR},\textrm{I}}^{~\textrm{CM}}$ is the azimuthal opening angle between the ISR system and the invisible system in the CM frame.
\item $m_\text{T}^Z$ is the transverse mass of the dilepton pair assigned to the signal system.
\item $m_\text{T}^J$ is the transverse mass of the jet system assigned to the signal system.
\end{itemize}

\subsection{Recursive-jigsaw reconstruction search selections}
\label{sec:rjr_selection}

The RJR search is a follow-up to the $36.1~\mathrm{fb}^{-1}$ search \cite{SUSY-2017-03} which observed 1.4$\sigma$ and 2.0$\sigma$ excesses in regions named SR$2\ell$-Low-RJR and SR$2\ell$-ISR-RJR respectively.
Thus the selection is kept the same and not optimized for the full Run~2 dataset.
These \acp{SR} were designed to target C1N2 models with \chionepm--\chionezero\ mass-splittings of approximately $100~\GeV$.
The selections begin with the common selections defined in Section~\ref{sec:selection}, and include requirements on the lepton multiplicity ($n_{\mathrm{leptons}}$), the jet multiplicity ($n_{\mathrm{jets}}$), the $b$-tagged jet multiplicity ($n_{b\textrm{-tag}}$), the transverse momenta of the leading $(p^{\ell_{1}}_\mathrm{T},p^{j_{1}}_\mathrm{T})$ and subleading $(p^{\ell_{2}}_\mathrm{T},p^{j_{2}}_\mathrm{T})$ leptons and jets, as well as the invariant masses of the dilepton $m_{\ell\ell}$ and dijet $m_{jj}$ systems.
The selections used to define SR$2\ell$-Low-RJR, and the CRs and VRs associated with it, are summarized in Tables~\ref{tab:StandardRJPreselections}~and~\ref{tab:StandardRJSelections}.
Selections which enforce orthogonality between the CRs and VRs are in boldface.
CRs are defined to extract data-driven normalization factors for the main background processes: diboson and $\ttbar$\,+\,$Wt$.
Details of the background estimation are described in Section~\ref{sec:rjr_bkg}.
 
Most of the regions are defined to have exactly two \ac{OS} \ac{SF} leptons with transverse momentum greater than $25~\GeV$, and a dilepton invariant mass consistent with originating from a $Z$ boson.
The diboson CR (CR2$\ell$-VV-RJR) is an exception and requires three or four leptons, which ensures orthogonality to the SRs while also enriching the sample with diboson events.
The \ac{OS} \ac{SF} dilepton pair with an invariant mass closest to the $Z$ boson mass is chosen as the $Z$ boson candidate.
For the purpose of RJR calculations, the third and fourth leptons are treated as invisible objects contributing to \ptmiss.
An additional requirement on $m_{\text{T}}^{W}$ is applied, which ensures orthogonality to the $3\ell$ searches using the $36.1~\mathrm{fb}^{-1}$ dataset.
Both the top CR (CR$2\ell$-$\mathrm{Top}$-RJR) and VR (VR$2\ell$-$\mathrm{Top}$-RJR) require a $b$-tagged jet.
Inverting the \mll\ requirement makes the VR orthogonal to the CR.
The SRs require $m_{jj}$ to be consistent with a $W$ boson, whereas VR$2\ell$-VV-RJR selects events outside the $W$ boson mass window.
The min~\mindphimj\ variable corresponds to the azimuthal angle between the jets and \ptmiss\ and is applied to suppress \zjets\ contributions to SR$2\ell$-Low-RJR.
VR$2\ell$-VV is the only region with an $H_{1,1}^{\mathrm{PP}}$ requirement, and it suppresses the \zjets\ contribution.
 
SR$2\ell$-ISR-RJR requires three or four jets, which makes it orthogonal to SR$2\ell$-Low-RJR.
All CRs and VRs for SR$2\ell$-ISR-RJR require at least one jet to be assigned to the ISR system ($N_{J}^{\mathrm{ISR}}$), and at least two to the signal system ($N_{J}^{\mathrm{S}}$).
The assignment of the jets is determined by the configuration that minimizes the mass of both S and ISR systems.
Both CR$2\ell$-ISR-VV-RJR and VR$2\ell$-ISR-VV-RJR require three or four leptons.
To increase the number of events in VR$2\ell$-ISR-VV-RJR, the transverse momentum requirement for jets is relaxed to 20~\GeV, compared to 30~\GeV\ in other regions.
The ISR regions are further defined by a series of requirements based on the variables from the ISR decay tree, described in Section~\ref{sec:rjr_reco}.
These requirements are listed in Tables~\ref{tab:CompressedRJPreselections2L}~and~\ref{tab:2L2J_compressedCRs}.
Selections that enforce orthogonality to the SR are in boldface.
SR$2\ell$-ISR-RJR requires a highly energetic ISR jet system which recoils against the signal system in the CM frame.
In VR$2\ell$-ISR-VV-RJR the $m_\text{T}^Z$ requirement is inverted in order to be orthogonal to CR$2\ell$-ISR-VV-RJR.
The top CR (CR$2\ell$-ISR-Top-RJR) and VR (VR$2\ell$-ISR-Top-RJR) both require a $b$-tagged jet and have broader $m_\text{T}^Z$ and $m_\text{T}^J$ requirements.
These regions are orthogonal due to the inversion of the $p_\mathrm{T}^{~\mathrm{CM}}$ requirement.

\begin{table}
\centering
\scriptsize
\caption{Preselection criteria for the standard-decay-tree $2\ell$ SR (SR$2\ell$-Low-RJR) and the associated CRs and VRs. Selections which enforce orthogonality between the CRs and VRs are in boldface.}
\begin{tabular}{l cccccccc}
\noalign{\smallskip}\hline\noalign{\smallskip}
\textbf{Region}                    &   $\bolden{n_{\textrm{leptons}}}$   &        $\bolden{n_{\textrm{jets}}}$  &  $\bolden{n_{b\textrm{-tag}}}$ & $\bolden{\pt^{\ell_{1},\ell_{2}}}$ & $\bolden{\pt^{j_{1},j_{2}}}$  & $\bolden{m_{\ell\ell}}$ & $\bolden{m_{jj}}$ & $\bolden{m_{\textrm{T}}^{W}}$\\
&    &     &   & \textbf{[\GeV]} & \textbf{[\GeV}]  & \textbf{[\GeV}] & \textbf{[\GeV}] & \textbf{[\GeV}]\\
\noalign{\smallskip}\hline\noalign{\smallskip}
CR$2\ell$-VV-RJR              &  $\bolden{\in[3,4]}$  & $\geq 2$ & $=0$    & $>25$ & $>30$ & $\in(80,100)$& $>20$~ & $\in(70,100)$\\
&    &  &    &  &  & &  & if $n_{\textrm{leptons}} =3$\\
CR$2\ell$-Top-RJR             &    $=2$     & $\geq 2$ & $\bolden{=1}$    & $>25$ & $>30$ & $\bolden{\in(20,80)}$~~ &~~$\in(40,250)$& $-$\\
&          &          &      &       &       & \textbf{or} $\bolden{>100}$    &                & $-$\\

\noalign{\smallskip}\hline\noalign{\smallskip}
VR$2\ell$-VV-RJR              &    $=2$     & $\geq 2$ & $=0$    & $>25$ & $>30$ & $\in(80,100)$&$\bolden{\in(40,70)}$& $-$\\
&          &          &     &       &       &              &\textbf{or} $\bolden{\in(90,500)}$~~~~& $-$\\
VR$2\ell$-Top-RJR             &    $=2$     & $\geq 2$ & $\bolden{=1}$ & $>25$ & $>30$ & $\in(80,100)$& ~~$\in(40,250)$ & $-$\\
\noalign{\smallskip}\hline\noalign{\smallskip}
SR$2\ell$-Low-RJR            &    $=2$     & $=2$     & $=0$    & $>25$ & $>30$ & $\in(80,100)$& $\in(70,90)$    & $-$\\
\noalign{\smallskip}\hline\noalign{\smallskip}
\end{tabular}
\label{tab:StandardRJPreselections}
\end{table}
 
\begin{table}
\centering
\scriptsize
\caption{Selection criteria for the standard-decay-tree $2\ell$ SR (SR$2\ell$-Low-RJR) and the associated CRs and VRs. }
\begin{tabular}{l ccccccc}
\noalign{\smallskip}\hline\noalign{\smallskip}
\textbf{Region} & $\bolden{H_{4,1}^{\mathrm{PP}}}$ & $\bolden{H_{1,1}^{\mathrm{PP}}}$ & $\bolden{\frac{p^{\mathrm{lab}}_{\mathrm{T}~\mathrm{PP}}}{p^{\mathrm{lab}}_{\mathrm{T}~\mathrm{PP}}+H^{\mathrm{PP}}_{\mathrm{T}\ 4,1}}}$ & $\bolden{\frac{\mathrm{min}(H^{\mathrm{P}_{\mathrm{a}}}_{1,1},H^{\mathrm{P}_{\mathrm{b}}}_{1,1})}{\mathrm{min}(H^{\mathrm{P}_{\mathrm{a}}}_{2,1},H^{\mathrm{P}_{\mathrm{b}}}_{2,1})}}$ & $\bolden{\frac{H_{1,1}^{\textrm{PP}}}{H_{4,1}^{\textrm{PP}}}}$ & $\bolden{\Delta\phi_{\textrm{V}}^{\textrm{P}}}$  & \bolden{min \mindphimj}\\
& \textbf{[\GeV}] & \textbf{[\GeV}] &  &  &   & \\
\noalign{\smallskip}\hline\noalign{\smallskip}
CR$2\ell$-VV-RJR              &$>200$&$-$   &$<0.05$& $>0.2$     &$-$             &$\in(0.3,2.8)$&$-$ \\
CR$2\ell$-Top-RJR             &$>400$&$-$   &$<0.05$& $>0.5$     &$-$             &$\in(0.3,2.8)$&$-$ \\
\noalign{\smallskip}\hline\noalign{\smallskip}
VR$2\ell$-VV-RJR              &$>400$&$>250$&$<0.05$& $\in(0.4,0.8)$&$-$             &$\in(0.3,2.8)$&$-$ \\
VR$2\ell$-Top-RJR             &$>400$&$-$   &$<0.05$& $>0.5$     &$-$             &$\in(0.3,2.8)$&$-$ \\
\noalign{\smallskip}\hline\noalign{\smallskip}
 
SR$2\ell$-Low-RJR            &$>400$&$-$   &$<0.05$ &  $-$      &$\in(0.35,0.60)$&$-$           &$>2.4$\\
\noalign{\smallskip}\hline\noalign{\smallskip}
\end{tabular}
\label{tab:StandardRJSelections}
\end{table}
 
\begin{table}
\centering
\scriptsize
\caption{Preselection criteria for the ISR-decay-tree $2\ell$ SR (SR$2\ell$-ISR-RJR) and the associated CRs and VRs. Selections which enforce orthogonality between the CRs and VRs are in boldface.}
\begin{tabular}{l ccccccc}
\noalign{\smallskip}\hline\noalign{\smallskip}
\textbf{Region}                &   $\bolden{n_{\textrm{leptons}}}$   & $\bolden{N_{\textrm{jet}}^{~\textrm{ISR}}}$ & $\bolden{N_{\textrm{jet}}^{\textrm{S}}}$ & $\bolden{n_{\textrm{jets}}}$  &  $\bolden{n_{b\textrm{-tag}}}$ & $\bolden{\pt^{\ell_{1},\ell_{2}}}$ & $\bolden{\pt^{j_{1},j_{2}}}$\\
&    & &  &  & & \textbf{[\GeV]} & \textbf{[\GeV]}\\
\noalign{\smallskip}\hline\noalign{\smallskip}
CR$2\ell$-ISR-VV-RJR     &  $\bolden{\in[3,4]}$  & $\geq 1$  & $\geq 2$   & $> 2$        & $=0$    & $>25$ & $>30$\\
CR$2\ell$-ISR-Top-RJR    &    $=2$     & $\geq 1$  & $=2$       & $\in[3,4]$   & $\bolden{=1}$ & $>25$ & $>30$\\
\noalign{\smallskip}\hline\noalign{\smallskip}
VR$2\ell$-ISR-VV-RJR     &  $\bolden{\in[3,4]}$  & $\geq 1$  & $\geq 2$  & $\geq 3$      & $=0$    & $>25$ & $>20$ \\
VR$2\ell$-ISR-Top-RJR    &    $=2$     & $\geq 1$  & $= 2$     & $\in[3,4]$    & $\bolden{=1}$ & $>25$ & $>30$ \\
\noalign{\smallskip}\hline\noalign{\smallskip}
SR$2\ell$-ISR-RJR        &    $=2$     & $\geq 1$  & $= 2$     &$\in[3,4]$     & $=0$    & $>25$ & $>30$ \\
\noalign{\smallskip}\hline\noalign{\smallskip}
\end{tabular}
\label{tab:CompressedRJPreselections2L}
\end{table}
 
\begin{table}
\centering
\scriptsize
\caption{Selection criteria for the ISR-decay-tree $2\ell$ SR (SR$2\ell$-ISR-RJR) and the associated CRs and VRs. Selections which enforce orthogonality between the CRs and VRs are in boldface.}
\begin{tabular}{l ccccccc}
\noalign{\smallskip}\hline\noalign{\smallskip}
\textbf{Region}                 & $\bolden{m_\text{T}^Z}$ & $\bolden{m_\text{T}^J}$ & $\bolden{\Delta\phi_{\textrm{ISR,I}}^{\textrm{CM}}}$ & $\bolden{R_{\textrm{ISR}}}$        & $\bolden{p^{\textrm{CM}}_{\textrm{T\ ISR}}}$ & $\bolden{p^{\textrm{CM}}_{\textrm{T\ I}}}$  & $\bolden{p^{\textrm{CM}}_{\textrm{T}}}$  \\
& \textbf{[\GeV}] & \textbf{[\GeV}] & &       &  \textbf{[\GeV}] & \textbf{[\GeV}] & \textbf{[\GeV}] \\
\noalign{\smallskip}\hline\noalign{\smallskip}
CR$2\ell$-ISR-VV-RJR      & $\in(80,100)$     &   $>20$           &  $>2.0$               &$\in(0.0,0.5)$~~    & ~~$>50$               &  ~~$>50$            &    $<30$         \\
CR$2\ell$-ISR-Top-RJR     & $\in(50,200)$     &   $\in(50,200)$   &  $>2.8$               &$\in(0.4,0.75)$   & $>180$              &  $>100$           &    $<20$         \\
\noalign{\smallskip}\hline\noalign{\smallskip}
VR$2\ell$-ISR-VV-RJR      &  $\bolden{\in(20,80)}$~~     &   $>20$           & $>2.0$               & $\in(0.0,1.0)$~~   & ~~$>70$               & ~~$>70$             & $<30$            \\
&  \,\textbf{or} $\bolden{>100}$           &                   &                      &    &         &          &         \\
VR$2\ell$-ISR-Top-RJR     &  $\in(50,200)$    &   $\in(50,200)$   &$>2.8$               & $\in(0.4,0.75)$  & $>180$              & $>100$             & $\bolden{>20}$           \\
\noalign{\smallskip}\hline\noalign{\smallskip}
SR$2\ell$-ISR-RJR         &  $\in(80,100)$    &  $\in(50,110)$ & $>2.8$               & $\in(0.4,0.75)$  & $>180$              & $>100$            & $<20$                 \\
\noalign{\smallskip}\hline\noalign{\smallskip}
\end{tabular}
\label{tab:2L2J_compressedCRs}
\end{table}


\FloatBarrier
\subsection{Electroweak search selections}
\label{sec:ewk_selection}

The EWK search uses 13 orthogonal \acp{SR} designed to cover different regions of the C1N2 and GMSB models' parameter spaces.
In addition to the use of new kinematic variables, the strategy from the 36~\ifb\ search~\cite{SUSY-2016-24} is extended by optimizing binned \acp{SR} to maximize the model-dependent search sensitivity.
The \acp{SR}  labelled -OffShell, -Low, -Int, and -High target increasing  \ac{NLSP}--\ac{LSP} mass-splittings.
The \ac{SR} labelled -$\ell\ell bb$ targets the GMSB model with either a Higgs or $Z$ boson decaying into two $b$-quarks.
The selections defining each region in the search are summarized in Tables~\ref{tab:ewkRegionDefinitionsHigh}--\ref{tab:ewkRegionDefinitionsOffShell}, along with the control and validation regions used for the estimation and validation of the SM backgrounds.
The acceptance times efficiency for several example signal models are listed in Table~\ref{tab:ewk_acceff}.
 
Lepton pair and jet pair mass windows are used to select events with a $ZW$, $Zh$, or $ZZ$ topology.
In all regions, except for the OffShell regions, the dilepton invariant mass is required to be on the $Z$ boson mass peak, and the mass of the jet system is required to be consistent with a hadronically decaying boson.
Most regions require the jet system mass to be around the $W$ or $Z$ boson mass, $60<m_X<110$~\GeV, where $m_X$ refers to the dijet system for all regions except for SR-1J-High-EWK, where it is the single-jet mass.
The $\ell\ell bb$ region expands this window to $60<\mbb<150$~\GeV\ in order to additionally account for Higgs boson decays in the GMSB model.
 
When performing model-dependent fits, SR-High-EWK uses a two-dimensional binning in $\metsig$ with boundaries $(18,21,\infty)$ and $\Delta R_{jj}$ with boundaries $[0,0.8,1.6)$, labelled as $\Delta R_{X}$ in Tables~\ref{tab:ewkRegionDefinitionsHigh} and \ref{tab:ewkRegionDefinitionsLow}.
The variable $\Delta R_{X}$ is sensitive to signal events where the leptons or jets are expected to be closer together due to the boost of the decay system, which is increased by the large mass splittings targeted by these regions.
SR-Int-EWK and SR-Low-EWK are binned only in $\metsig$, with boundaries $(12,15,18)$ and $(6,9,12)$ respectively.
SR-OffShell-EWK is split into ranges of 12--40 and 40--71~\GeV\ in \mll.
The \ac{SR} binning was optimized by checking the performance of a few variables, e.g.\ binning in jet \pt\ did not perform as well as $\Delta R_{jj}$, and binning the main discriminant, \metsig, so that at least one background event is expected in each bin.
All remaining \acp{SR} are treated as single bins.
For all of the \acp{SR}, \acp{VR} are defined using the jet system mass sidebands or by inverting criteria for kinematic variables defining the \acp{SR}.
The criteria ensuring orthogonality to the corresponding \acp{SR} are highlighted in boldface in Tables~\ref{tab:ewkRegionDefinitionsHigh}--\ref{tab:ewkRegionDefinitionsOffShell}.
\acp{CR} are defined in order to extract data-driven normalization factors for the main background processes: diboson, \ttbar, \zjets, and low-mass off-shell \zjets.
These are discussed further in Section~\ref{sec:ewk_bkg}.

\begin{table}[htbp]
\centering
\caption{Overview of the High and Intermediate signal, control, and validations regions of the electroweak search.
The main requirements that distinguish the control and validation regions from the signal regions are highlighted in boldface.
The bracketed requirements denote the binning used in model-dependent fits, as described in the text.
}
\resizebox{1\textwidth}{!}{
\begin{tabular}{lcccccccc}
\noalign{\smallskip}\hline\noalign{\smallskip}
\textbf{Region} & $\bolden{\njets}$ & $\bolden{\nbjets}$  & $\bolden{\metsig}$ & $\bolden{\mll}$ & $\bolden{m_{\!X}}$  & $\bolden{\mttwo}$ & $\bolden{\Delta R_{X}}$ & $\bolden{p_\text{T}^{j_1}}$ \\
& & & & \bolden{[\GeV]} & \bolden{[\GeV]} & \bolden{[\GeV]} & & \bolden{[\GeV]} \\
\noalign{\smallskip}\hline\noalign{\smallskip}
SR-High-EWK   & $\geq2$ & $\leq 1$ & $(18,21,\infty)$ &  71--111 & $60<\mjj<110$  & $>80$ & $\Rjj\in(0,0.8,1.6)$ & $-$ \\
VR-High-Sideband-EWK   & $\geq2$ & $\leq 1$ & $>18$ &  71--111 & $\bolden{20<\mjj<60\cup\mjj>110}$  & $>80$ & $\Rjj <1.6$ & $-$ \\
VR-High-R-EWK & $\geq2$ & $\leq 1$ & $>18$ &  71--111 & $\bolden{\mjj>20}$  & $>80$ & $\bolden{\Rjj >1.6}$ & $-$ \\
SR-1J-High-EWK & ~~~~$1$ & $\leq 1$ & $>12$ &  71--111 & $60<\mj<110$  & $>80$ & $-$ & $-$ \\
VR-1J-High-Sideband-EWK & ~~~~$1$ & $\leq 1$ & $>12$ &  71--111 & $\bolden{20<\mj<60\cup\mj>110}$  & $>80$ & $-$ & $-$ \\
SR-$\ell\ell bb$-EWK & $\geq2$ & $\geq 2$ & $>18$ &  71--111 & $60<\mbb<150$ & $>80$ & $-$ & $-$\\
VR-$\ell\ell bb$-EWK & $\geq2$ & $\geq 2$ & \textbf{12--18}~~ &  71--111 & $60<\mbb<150$ & $>80$ & $-$ & $-$ \\
\noalign{\smallskip}\hline\noalign{\smallskip}
SR-Int-EWK & $\geq2$ & ~~~~$0$ & $(12,15,18)$ &  81--101 & $60<\mjj<110$  & $>80$ & $-$ & $>60$ \\
VR-Int-EWK & $\geq2$ & ~~~~$0$ & 12--18~~ &  81--101 & $60<\mjj<110$  & $>80$ & $-$ & $\bolden{<60}$ \\
CR-VZ-EWK & $\geq2$ & ~~~~$0$ & 12--18~~ &  81--101 & $\bolden{20<\mjj<60\cup\mjj>110}$  & $>80$ & $-$ & $\bolden{-}$ \\
CR-tt-EWK & $\geq2$ & $\bolden{\geq1}$ & \textbf{9--12} &  81--101 & $\bolden{\mjj>20}$  & $>80$ & $-$ & $>60$ \\
\noalign{\smallskip}\hline\noalign{\smallskip}
\end{tabular}
}
\label{tab:ewkRegionDefinitionsHigh}
\end{table}
 
\begin{table}[htbp]
\centering
\caption{Overview of the Low signal, control, and validations regions of the electroweak search.
The main requirements that distinguish the control and validation regions from the signal regions are highlighted in boldface.
The bracketed requirements denote the binning used in model-dependent fits, as described in the text.
}
\resizebox{1\textwidth}{!}{
\begin{tabular}{lcccccccc}
\noalign{\smallskip}\hline\noalign{\smallskip}
\textbf{Region} & $\bolden{\njets}$ & $\bolden{\nbjets}$  & $\bolden{\metsig}$ & $\bolden{\mll}$ & $\bolden{m_{\!X}}$  & $\bolden{\mttwo}$ & $\bolden{\Delta R_{X}}$ & $\bolden{\Delta\phi(\ptll, \ptmiss)}$ \\
& & & & \bolden{[\GeV]} & \bolden{[\GeV]} & \bolden{[\GeV]} &  &  \\
\noalign{\smallskip}\hline\noalign{\smallskip}
SR-Low-EWK & $2$ & $0$ & $(6,9,12)$ &  81--101 & $60<\mjj<110$ & $>80$ & $\Rll<1$~~~ & $-$ \\
VR-Low-EWK & $2$ & $0$ & 6--12~~ &  81--101 & $60<\mjj<110$ & $>80$ & $\bolden{1<\Rll<1.4}$ & $-$ \\
SR-Low-2-EWK & $2$ & $0$ & 6--9 &  81--101 & $60<\mjj<110$ & $<80$ & $\Rll<1.6$ &$<0.6$ \\
VR-Low-2-EWK & $2$ & $0$ & 6--9 &  81--101 & $\bolden{20<\mjj<60\cup\mjj>110}$ & $<80$ & $\Rll<1.6$ &  $<0.6$ \\
CR-Z-EWK     & $2$ & $0$ & 6--9 &  81--101 & $\bolden{20<\mjj<60\cup\mjj>110}$ & $\bolden{>80}$ & $\bolden{-}$ &  $\bolden{-}$ \\
\noalign{\smallskip}\hline\noalign{\smallskip}
\end{tabular}
}
\label{tab:ewkRegionDefinitionsLow}
\end{table}
 
\begin{table}[htbp]
\centering
\caption{Overview of the OffShell signal, control, and validations regions of the electroweak search.
The main requirements that distinguish the control and validation regions from the signal regions are highlighted in boldface.
The bracketed requirements denote the binning used in model-dependent fits, as described in the text.
}
\begin{tabular}{lccccccc}
\noalign{\smallskip}\hline\noalign{\smallskip}
\textbf{Region} & $\bolden{\njets}$ & $\bolden{\nbjets}$  & $\bolden{\metsig}$ & $\bolden{\mll}$  & $\bolden{\mttwo}$ & $\bolden{p_\text{T}^{j_1}}$  & $\bolden{\Delta\phi(p_\text{T}^{j_1}, \ptmiss)}$\\
& & & & \bolden{[\GeV]}  & \bolden{[\GeV]}  & \bolden{[\GeV]}  & \\
\noalign{\smallskip}\hline\noalign{\smallskip}
SR-OffShell-EWK & $\geq2$ & $0$ & $>9$ &  $(12,40,71)$   & $>100$  & $>100$ & $>2$\\
VR-OffShell-EWK & $\geq2$ & $0$ & $>9$ &  12--71   & \textbf{80--100}~~  & $>100$ & $>2$\\
CR-DY-EWK       & $\geq2$ & $0$ & \textbf{6--9} &  12--71   & $>100$  & $\bolden{-}$ & $\bolden{-}$\\
\noalign{\smallskip}\hline\noalign{\smallskip}
\end{tabular}
\label{tab:ewkRegionDefinitionsOffShell}
\end{table}
 
\begin{table}[htbp]
\centering
\caption{Acceptance times efficiency in the summed EWK signal regions for example C1N2 and GMSB signal models.
}
\begin{tabular}{ccl}
\hline
\multicolumn{3}{c}{C1N2}   \\
$m_{\chionepm/\chitwozero}$ [\GeV] & $m_{\chionezero}$ [\GeV] & $A\times\epsilon$ [\%]    \\[1.5ex]
\hline
100 & ~~60 & 0.0003       \\
150 & ~~70 & 0.0007       \\
250 & 100 & 0.042       \\
400 & 200 & 0.21     \\
650 & 250 & 0.64     \\
\hline
\end{tabular}
\hspace{1cm}
\begin{tabular}{ccl}
\hline
\multicolumn{3}{c}{GMSB}  \\
$m_{\chionezero}$  [\GeV] & $B(\chionezero\rightarrow h \tilde G)$ [\%] & $A\times\epsilon$ [\%]  \\[1.5ex]
\hline
200 & 50 & 0.02  \\
500 & 80 & 0.21   \\
700 & ~~0  & 1.6   \\
\hline
\end{tabular}
\label{tab:ewk_acceff}
\end{table}


\FloatBarrier
\subsection{Strong search selections}
\label{sec:strong_selection}

The \acp{SR} targeting production of gluinos and squarks start with the common selection described at the beginning of Section~\ref{sec:selection}.
The four overlapping \acp{SR} designed for a kinematic endpoint, or `edge' feature, are binned in the dilepton invariant mass, \mll.
They are named SRC-STR, SRLow-STR, SRMed-STR, and SRHigh-STR in order of sensitivity to gluino--\chionezero\ mass splittings from those that are compressed, or small, to those that are large.
\acp{SR} designed for an excess of events near the $Z$ boson mass, `\onz', are a single bin in the window $81<\mll<101$~\GeV.
They share the same naming as the edge \acp{SR}, with the addition of a `Z' in the name.
The selection used for each region in the analysis is summarized in Table~\ref{tab:strongRegionDefinitions}.
The bin boundaries in \mll\ used for interpretations were chosen such that there is a finer division in the mass region targeted by each SR, while keeping enough events per bin for the background estimates and a bin around the $Z$ mass where possible.
The boundaries in units of \GeV\ are as follows:
\begin{itemize}
\item SRC-STR: 12, 31, 46, 61, 71, 81, 101, 201;
\item SRLow-STR: 12, 41, 61, 81, 101, 141, 201, 301, 501;
\item SRMed-STR: 12, 81, 101, 201, 301, 601;
\item SRHigh-STR: 12, 101, 301, 1001.
\end{itemize}
The acceptance times efficiency for the simulated simplified models in the various regions depends on the model and region.
SRHigh-STR has an acceptance times efficiency of 10\% for the gluino-slepton model with $m_{\gluino}=2000~\GeV$ and $m_{\chionezero}=300~\GeV$.
SRC-STR has an acceptance times efficiency of 0.02\% for the  gluino-$Z^{(*)}$ model with $m_{\gluino}=800~\GeV$ and $m_{\chionezero}=700~\GeV$.
This is very small due to the compressed mass-splittings between sparticles, resulting in low-momentum decay products, and the $Z$ boson branching fraction to leptons.
SRZMed-STR has an acceptance times efficiency of 1\% for the squark-$Z^{(*)}$ model with $m_{\squark}=1200~\GeV$ and $m_{\chionezero}=700~\GeV$.

As described before, all of the SRs require exactly two \ac{OS} \ac{SF} signal leptons with $\pt>25$~\GeV, without any additional baseline leptons.
This also applies to the Strong search CRs and VRs, except for VR3L-STR, which requires exactly three signal leptons.
All signal models studied are expected to have several quarks in the final state, so events are further required to have at least two jets with $\pt>30$~\GeV.
Since the SR selection requires exactly two leptons, it is likely that events passing the selection from models with $Z$ bosons contain both a leptonically and hadronically decaying $Z$ boson.
Thus the \onz\ \acp{SR} require at least four jets with $\pt>30$~\GeV\ to take this into account.
All regions require $\mll>12$~\GeV\ in order to reduce contributions from low-mass resonances.
Several variables, such as \met\ and \HT, are used to isolate SUSY-like events from the background.
Most regions require the angular separation in $\phi$ between the \ptmiss\ and the two leading jets, $\mindphimj$, to be greater than 0.4 to remove events with \met\ arising from jet mismeasurements.
CRs for \zjets\ are defined, identified with a `-Z', by inverting this requirement, adding an \mll\ window of 81--101~\GeV, and otherwise keeping the selections the same.
CRs for \ac{FS} processes, such as \ttbar\ and $WW$, are defined (labelled with `-FS') by requiring the leptons to be \ac{DF}.
VRs are defined for each edge SR, below their \met\ requirement, by $150<\met<250$~\GeV.
Additional VRs targeting $WZ$ and \ac{FNP} leptons are defined by requiring three leptons and \ac{SS} leptons respectively; the \met\ requirement in these regions is lowered in order to increase the number of events.
 
The requirements chosen for each SR are based on an optimization performed with a few test points with different mass-splittings between the \gluino\ or \squark\ and the \chionezero\ from each model.
One of the main changes with larger mass-splittings is the increase in jet activity.
The requirement on \HT\ varies from $>250$~\GeV\ in the more compressed regions to $>800$~\GeV\ in the regions targeting large splittings.
The lower bound on the dilepton system's momentum, $\ptll>40$~\GeV, was imposed to allow use of the $\gamma$\,+\,jets process to estimate the \zjets\ background, but this method was not used due to poor modelling.
After all of the other requirements this removes less than one expected event from each SR.
The more compressed signal models tend to have small values of \ptll, so an upper bound is imposed to reduce the \zjets\ contribution.
The requirement is relaxed for \acp{SR} targeting larger mass-splittings.
The \met\ requirement removes a large portion of the SM backgrounds and is set to $>250$~\GeV\ for the C and Low regions and $>300$~\GeV\ for the Med and High regions.
Requirements on \metsig\ and \mttwo\ are used to reduce the contribution from \ttbar\ in the signal regions.

\begin{table}[htbp]
\centering
\caption{Overview of all of the signal, control, and validation regions of the Strong search.
The flavour of the lepton pair is denoted by `SF' for same-flavour and `DF' for different-flavour.
All regions require exactly two opposite-charge signal leptons with $\pt>25$~\GeV, with the exception of VR3L which requires exactly three signal leptons (3L).  
The main requirements that distinguish the control and validation regions from the signal regions are highlighted in boldface.
The validation regions have an equivalent set of CRs with the lepton flavour and $\mindphimj$ flipped, but are omitted here for brevity.
}
\resizebox{1\textwidth}{!}{
\begin{tabular}{lccccccccc}
\noalign{\smallskip}\hline\noalign{\smallskip}
\textbf{Region} & $\bolden{\njets}$         & $\bolden{\HT}$    & $\bolden{\met}$ & $\bolden{\mttwo}$ & $\bolden{\metsig}$ & $\bolden{\ptll}$ & $\bolden{\mindphimj}$ & \textbf{SF/DF} & $\bolden{\mll}$ \\
& & \bolden{[\GeV]} & \bolden{[\GeV]} & \bolden{[\GeV]} & & \bolden{[\GeV]} & & & \bolden{[\GeV]} \\
\noalign{\smallskip}\hline\noalign{\smallskip}
Signal regions \\
\noalign{\smallskip}\hline\noalign{\smallskip}
SRC-STR                     & $\geq2$  & $>250$ & $>250$ & $>90$ & $>10$ & 40--100  & $>0.4$ & SF & $>12$ \\
SRLow-STR                   & $\geq2$ & $>250$ & $>250$ & $>100$~~ & $-$ & 40--500 & $>0.4$ & SF & $>12$   \\
$\hookrightarrow$SRZLow-STR & $\geq4$  & $>250$ & $>250$ & $>100$~~ & $-$ & 40--500 & $>0.4$& SF & 81--101   \\
SRMed-STR                   & $\geq2$  & $>500$ & $>300$ & $>75$ & $-$ & 40--800 & $>0.4$&  SF & $>12$\\
$\hookrightarrow$SRZMed-STR & $\geq4$  & $>500$ & $>300$ & $>75$ & $-$ & 40--800 & $>0.4$&  SF &  81--101   \\
SRHigh-STR                   & $\geq2$  & $>800$ & $>300$ & $>75$ & $-$ & $>40$    & $>0.4$ & SF &  $>12$   \\
$\hookrightarrow$SRZHigh-STR & $\geq4$  & $>800$ & $>300$ & $>75$ & $-$ & $>40$ & $>0.4$& SF & 81--101  \\
\noalign{\smallskip}\hline\noalign{\smallskip}
Control regions\\
\noalign{\smallskip}\hline\noalign{\smallskip}
CRC-FS-STR   & $\geq2$  & $>250$ & $>250$ & $>90$ & $>10$ & 40--100  & $>0.4$ & \textbf{DF} & $>12$ \\
CRLow-FS-STR & $\geq2$  & $>250$ & $>250$ & $>100$~~ & $-$ & 40--500 & $>0.4$ & \textbf{DF} & $>12$   \\
$\hookrightarrow$CRZLow-FS-STR & $\geq4$  & $>250$ & $>250$ & $>100$~~ & $-$ & 40--500 & $>0.4$& \textbf{DF} & \textbf{61--121}   \\
CRMed-FS-STR & $\geq2$  & $>500$ & $>300$ & $>75$ & $-$ & 40--800 & $>0.4$&  \textbf{DF} & $>12$\\
$\hookrightarrow$CRZMed-FS-STR & $\geq4$  & $>500$ & $>300$ & $>75$ & $-$ & 40--800 & $>0.4$&  \textbf{DF} &  \textbf{61--121}  \\
CRHigh-FS-STR & $\geq2$  & $>800$ & $>300$ & $>75$ & $-$ & $>40$    & $>0.4$ & \textbf{DF} &  $>12$   \\
$\hookrightarrow$CRZHigh-FS-STR & $\geq4$  & $>800$ & $>300$ & $>75$ & $-$ & $>40$ & $>0.4$& \textbf{DF} & \textbf{61--121} \\
CRC-Z-STR   & $\geq2$  & $>250$ & $>250$ & $>90$ & $>10$ & 40--100  & $\bolden{<0.4}$ & SF & \textbf{81--101} \\
CRLow-Z-STR & $\geq2$  & $>250$ & $>250$ & $>100$~~ & $-$ & 40--500 & $\bolden{<0.4}$ & SF & \textbf{81--101}   \\
$\hookrightarrow$CRZLow-Z-STR & $\geq4$  & $>250$ & $>250$ & $>100$~~ & $-$ & 40--500 & $\bolden{<0.4}$& SF & 81--101   \\
CRMed-Z-STR & $\geq2$  & $>500$ & $>300$ & $>75$ & $-$ & 40--800 & $\bolden{<0.4}$&  SF & \textbf{81--101}\\
$\hookrightarrow$CRZMed-Z-STR & $\geq4$  & $>500$ & $>300$ & $>75$ & $-$ & 40--800 & $\bolden{<0.4}$&  SF &  81--101   \\
CRHigh-Z-STR & $\geq2$  & $>800$ & $>300$ & $>75$ & $-$ & $>40$    & $\bolden{<0.4}$ & SF &  \textbf{81--101}   \\
$\hookrightarrow$CRZHigh-Z-STR & $\geq4$  & $>800$ & $>300$ & $>75$ & $-$ & $>40$ & $\bolden{<0.4}$& SF & 81--101  \\
\noalign{\smallskip}\hline\noalign{\smallskip}
Validation regions\\
\noalign{\smallskip}\hline\noalign{\smallskip}
VRC-STR    & $\geq2$  & $>250$ & \textbf{150--250} & $>90$ & $>10$ & 40--100 & $>0.4$ & SF & $>12$ \\
VRLow-STR  & $\geq2$  & $>250$ & \textbf{150--250} & $>100$~~ & $-$ & 40--500 & $>0.4$ & SF & $>12$   \\
VRMed-STR  & $\geq2$  & $>500$ & \textbf{150--250} & $>75$ & $-$ & 40--800 & $>0.4$ & SF & $>12$   \\
VRHigh-STR & $\geq2$  & $>800$ & \textbf{150--250} & $>75$ & $-$ & $>40$    & $>0.4$ & SF & $>12$   \\
VR3L-STR   & $\geq2$  & $>250$ & ~~~~$\bolden{>200}$ & $>100$~~ & $-$ & $>40$& $>0.4$ & \textbf{3L}  & $>12$   \\ 
\noalign{\smallskip}\hline\noalign{\smallskip}
\end{tabular}
}
\label{tab:strongRegionDefinitions}
\end{table}



\FloatBarrier
\section{Background estimation}
\label{sec:background}

This section describes the methods used to estimate the contributions from \ac{SM} processes to each of the search regions.
An overview of how the various processes are estimated in each search is shown in Table~\ref{tab:bkg_est_breakdown}.
Several processes are estimated using data-driven methods in order to estimate both the yield and distribution from data.
Depending on the search and \ac{SR}, some processes are estimated using orthogonal \acp{CR} to normalize the yield to data while taking the shape from \ac{MC} simulation.
The choice of background estimation method depends on the dominant backgrounds, expected yields, and whether the shape of the background is important.
For the RJR search, the background estimation methods are repeated from the 36~\ifb\ analysis.
The use of $b$-tagging in the EWK search requirements allows for a \ttbar\ CR with inverted $b$-jet requirements.
While the Strong search is inclusive in $b$-jets and modelling the shape of the \mll\ spectrum is important, thus a method that also estimates the shape of the backgrounds from data is used.
Finally, processes with smaller expected yields such as triboson production ($VVV$), Higgs production, and rare top processes (`Other Top') such as $t\bar{t}$\,+\,$V$ production are estimated directly from \ac{MC} simulation.
The common estimate of FNP leptons is described, and then methods specific to each search are described in Sections~\ref{sec:rjr_bkg}--\ref{sec:strong_bkg}.
 
\begin{table}[htbp]
\caption{Breakdown of background estimation methods for the three searches.
`MC' processes are estimated directly from a Monte Carlo (MC) simulated sample, `CR' processes are MC samples normalized to data in a control region (CR), and `DD' processes are estimated via a data-driven method where both the shape and normalization are obtained from data.
Entries with multiple categories depend on the specific region or process, e.g.\ \ttbar\ and $Wt$.}
\resizebox{1\textwidth}{!}{
\begin{tabular}{lccccccc}
\hline
Process & \zjets and Drell--Yan & Top & Diboson & FNP & $VVV$ & Other Top & Higgs \\
\hline
RJR     & DD & CR & CR & DD & MC & MC & MC \\
EWK     & CR & MC/CR & CR & DD & MC & MC & MC  \\
Strong  & CR & DD & MC/DD & DD & MC & MC & MC \\
\hline
\end{tabular}
}
\label{tab:bkg_est_breakdown}
\end{table}

All three searches estimate events with fake, misidentified and non-prompt leptons using the matrix method \cite{ATLAS-CONF-2014-058}.
This method inverts a system of equations relating the number of observed baseline and signal leptons to the estimated number of real and \ac{FNP} leptons via measured real- and FNP-lepton efficiencies.
These events come mostly from semileptonic \ttbar, $W\rightarrow \ell\nu$, and single top ($s$- and $t$-channel) decays, which enter the dilepton channel when one hadron, photon, or non-prompt lepton from a heavy-flavour decay is misidentified as a signal lepton.
In most \acp{SR}, the contribution from this background is less than 5\%.
However, in a few regions it is up to 20\% of the background yield.

A control sample is constructed for each region in the searches with the same selection but removing the requirement of two signal leptons, requiring only two baseline leptons instead.
The number of leptons that pass or fail the stricter signal lepton requirements is counted.
A system of equations is constructed with the measured efficiencies for real and FNP leptons to pass these requirements and then inverted to solve for the number of FNP leptons passing the signal lepton requirement.
In the case of a single-lepton selection, the number of FNP-lepton events in a given region would be estimated according to:
\begin{equation*}
N_{\text{pass}}^{\text{FNP}} = \frac{N_{\text{fail}} - (1/\epsilon^{\text{real}} - 1) \times N_{\text{pass}} }{1/\epsilon^{\text{FNP}} - 1/\epsilon^{\text{real}}},
\end{equation*}
where $\epsilon^{\text{real}}$ is the efficiency for a real, prompt lepton to pass the signal lepton requirements, and $\epsilon^{\text{FNP}}$ is the same for FNP leptons.
The real-lepton efficiency is obtained from \ac{MC} simulation which has been corrected to match the efficiency found in data and uses the \ac{MC} simulation particle-level information to select prompt leptons.
The FNP-lepton efficiency is measured using a tag-and-probe method.
It is measured separately for misidentified leptons originating from light-flavour jets, heavy-flavour jets, and from the conversion of photons to electrons.
The heavy-flavour and photon-conversion efficiencies are measured with data in regions with a $b$-jet or $\mu\mu e$ events on the $Z$ boson mass peak respectively.
The light-flavour efficiency is measured in \ac{SS} events with \ac{MC} simulation since it is difficult to construct a region pure in FNP leptons from light-flavour jets.
The three sources are combined according to their expected relative contributions to the search regions, measured using \ac{MC} simulation with selections similar to those for each search.

\subsection{Recursive-jigsaw reconstruction search backgrounds}
\label{sec:rjr_bkg}

In the RJR search, the diboson and top backgrounds are normalized in CRs, the \zjets\ background is estimated using the so-called ABCD method, the \ac{FNP}-lepton background is estimated with the data-driven approach previously described, and the remaining backgrounds are estimated via MC simulation.
 
The ABCD method requires two SR selections that are uncorrelated with respect to the \zjets\ background.
These selections are used to construct three CRs: A, B, and D (region C is the SR), which are adjacent to the SR.
Table~\ref{tab:rjr:abcdregions} lists the requirements on these variables and their adjacency to the SRs for both cases.
As the selections are uncorrelated, there is a linear relation between the regions, $N_\text{SR} \approx N_\text{D}\cdot N_\text{A}/N_\text{B}$, where $N$ is the fitted \zjets\ estimate in each region.
This provides a background estimate in each of the SRs.
 
For SR$2\ell$-Low-RJR, the CRs are defined with \mjj\ and $H_{1,1}^{\mathrm{PP}}/H_{4,1}^{\mathrm{PP}}$, and for SR$2\ell$-ISR-RJR, $m_\text{T}^{J}$ and $p_{\mathrm{T\ I}}^{~\textrm{CM}}$ are used to define the CRs.
For $\mathrm{A}_{\mathrm{Low}}$, $\mathrm{B}_{\mathrm{Low}}$, and $\mathrm{D}_{\mathrm{Low}}$, the \zjets\ purities are 77\%, 98\%, and 96\% respectively.
For $\mathrm{A}_{\mathrm{ISR}}$, $\mathrm{B}_{\mathrm{ISR}}$, $\mathrm{D}_{\mathrm{ISR}}$, the \zjets\ purities are 36\%, 85\%, and 91\% respectively.
The ABCD regions are included in the simultaneous fits.
A cross-check \zjets\ estimate obtained by fitting the sidebands in \mjj, i.e.\ outside of the $W$ boson mass window, produces a compatible yield, but the ABCD method results in a lower overall uncertainty.
 
\begin{table}[htbp]
\centering
\caption{The selection criteria for the regions used in the ABCD \zjets\ background estimate for the RJR search.}
\resizebox{1.0\textwidth}{!}{
\begin{tabular}{l l l | l l l c}
\toprule
Standard Region & \quad$m_{jj}$ [\GeV]   & \quad$H_{1,1}^{\textrm{PP}}/H_{4,1}^{\textrm{PP}}$ & ISR Region & \quad$m_\text{T}^J$ [\GeV] & \quad$p_{\mathrm{T\ I}}^{~\textrm{CM}}$ [\GeV]\\ \hline \hline
SR$2\ell$-Low-RJR & $\in(70,90)$ & $\in(0.35,0.60)$  & SR$2\ell$-ISR-RJR & $\in(50,110)$ & \quad$ > 100$                \\ \midrule
$\mathrm{A}_{\mathrm{Low}}$  & $\in(20,70)$ & $\in(0.35,0.60)$ & $\mathrm{A}_{\mathrm{ISR}}$  & $\in(0,50)$~or~$>100$ & \quad$> 100$ \\
$\mathrm{B}_{\mathrm{Low}}$  & $\in(20,70)$ & $\in(0.20,0.35)$ & $\mathrm{B}_{\mathrm{ISR}}$  & $\in(0,50)$~or~$>100$ & \quad$< 100$\\
$\mathrm{D}_{\mathrm{Low}}$  & $\in(70,90)$ & $\in(0.20,0.35)$ & $\mathrm{D}_{\mathrm{ISR}}$  & $\in(50,110)$         & \quad$< 100$        \\ \bottomrule
\end{tabular}\label{tab:rjr:abcdregions}
}
\end{table}

The top and diboson backgrounds are normalized to data in the CRs defined in Section~\ref{sec:rjr_selection} with a separate simultaneous fit to each corresponding SR.
The \ttbar\ and $Wt$ processes share the same normalization factors of $1.03\pm0.08$ and $0.96\pm0.19$ in SR$2\ell$-Low-RJR and SR$2\ell$-ISR-RJR respectively.
The normalization factors derived for the diboson backgrounds are $1.09\pm0.10$ and $0.96\pm0.13$ in the same regions.
The yields in these CRs, along with the corresponding VRs, are listed in Tables~\ref{tab:standard_CRVRs_table_RJ}~and~\ref{tab:ISR_CRVRs_table_RJ} for the Low and ISR regions respectively.
Distributions in the Low and ISR top and diboson \acp{VR} are shown in Figure~\ref{fig:rjr:vrs}.
The yields and distributions for \zjets\ in the tables and figure are from MC simulation because the ABCD method is only constructed for the SRs.
The expected and observed distributions agree well in the VRs.

\begin{table}[htbp]
\caption{Breakdown of expected and observed yields in the diboson (VV) and top CRs and VRs for the standard-decay-tree regions associated with SR$2\ell$-Low-RJR after a simultaneous fit to the \aclp{CR}.
The yields for \zjets\ are from MC simulation.
The uncertainties include both the statistical and systematic sources.}
\label{tab:standard_CRVRs_table_RJ}
\begin{center}
\setlength{\tabcolsep}{0.0pc}{\small
\begin{tabular*}{\textwidth}{@{\extracolsep{\fill}}lP{-1}P{-1}P{-1}P{-1}}
\noalign{\smallskip}\hline\noalign{\smallskip}
& \multicolumn{1}{c}{CR2$\ell$-VV-RJR}            & \multicolumn{1}{c}{VR2$\ell$-VV-RJR}            & \multicolumn{1}{c}{CR2$\ell$-Top-RJR}            & \multicolumn{1}{c}{VR2$\ell$-Top-RJR}            \\[-0.05cm]
\noalign{\smallskip}\hline\noalign{\smallskip}
Observed events             & 216                   & 232                & 1759                  & 368                         \\
\noalign{\smallskip}\hline\noalign{\smallskip}
 
Total exp.\ bkg.\ events  & 216,16      & 285,78   & 1760,40     & 389,51                   \\
\noalign{\smallskip}\hline\noalign{\smallskip}
Diboson events   & 195,17      & 180,40         & 19.3,3.2          & 20.2,3.5                  \\
Top events       & $-$         & 47,8           & 1700,40           & 269,16                \\
\zjets\ events   & $-$         & 58,56           & 13,12        & 91,45                 \\
Other events     & 21,4        & 3.8,3.1        & 23,11             & 8.5,1.3                   \\
\noalign{\smallskip}\hline\noalign{\smallskip}
\end{tabular*}
}
\end{center}
\end{table}

\begin{table}[htbp]
\caption{Breakdown of expected an observed yields in the diboson (VV) and top CRs and VRs for the ISR-decay-tree regions associated with SR$2\ell$-ISR-RJR  after a simultaneous fit to the \aclp{CR}.
The yields for \zjets\ are from MC simulation.
The uncertainties include both the statistical and systematic sources.}
\label{tab:ISR_CRVRs_table_RJ}
\begin{center}
\setlength{\tabcolsep}{0.0pc}{\small
\begin{tabular*}{\textwidth}{@{\extracolsep{\fill}}lP{-1}P{-1}P{-1}P{-1}}
\noalign{\smallskip}\hline\noalign{\smallskip}
& \multicolumn{1}{c}{CR2$\ell$-ISR-VV-RJR}        & \multicolumn{1}{c}{VR2$\ell$-ISR-VV-RJR}       & \multicolumn{1}{c}{CR2$\ell$-ISR-Top-RJR}      & \multicolumn{1}{c}{VR2$\ell$-ISR-Top-RJR}              \\[-0.05cm]
\noalign{\smallskip}\hline\noalign{\smallskip}
Observed events                    & 149                    & 50                    & 289                   & 306                      \\
\noalign{\smallskip}\hline\noalign{\smallskip}
Total exp.\ bkg.\ events   & 149,14          & 63,9       & 289,19         & 350,60                \\
\noalign{\smallskip}\hline\noalign{\smallskip}
Diboson events    & 125,17       & 28,5        & 3.0,0.6             & 4.5,1.0                \\
Top events        & $-$          & $-$         & 254,18              & 330,60               \\
\zjets\ events    & $-$          & $-$         & 18,15              & 10,4                 \\
Other events      & 24,9         & 35,10       & 14.0,2.6            & 8.7,2.3 \\
\noalign{\smallskip}\hline\noalign{\smallskip}
\end{tabular*}
}
\end{center}
\end{table}

\begin{figure}[htbp]
\centering
\includegraphics[width=0.49\textwidth]{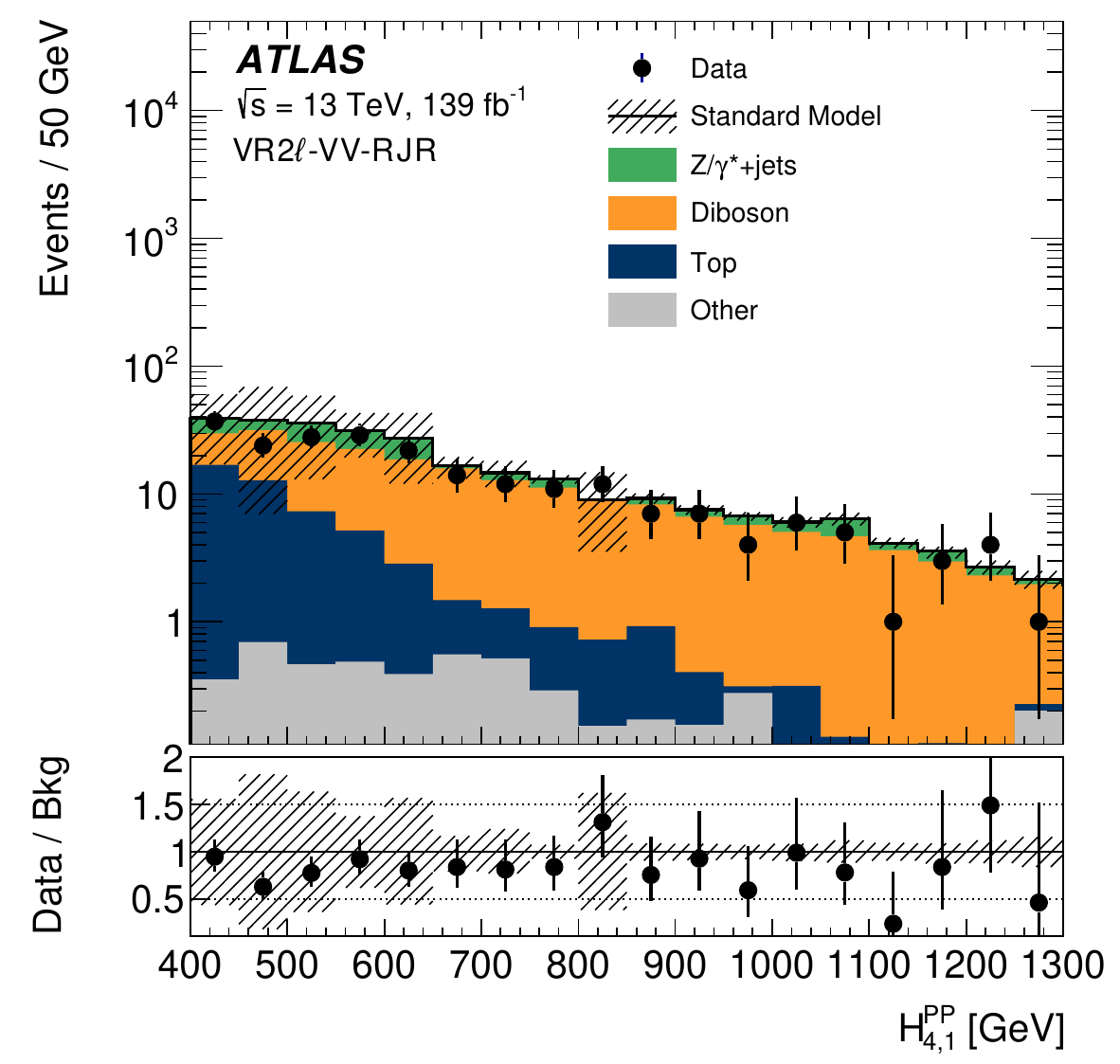}
\includegraphics[width=0.49\textwidth]{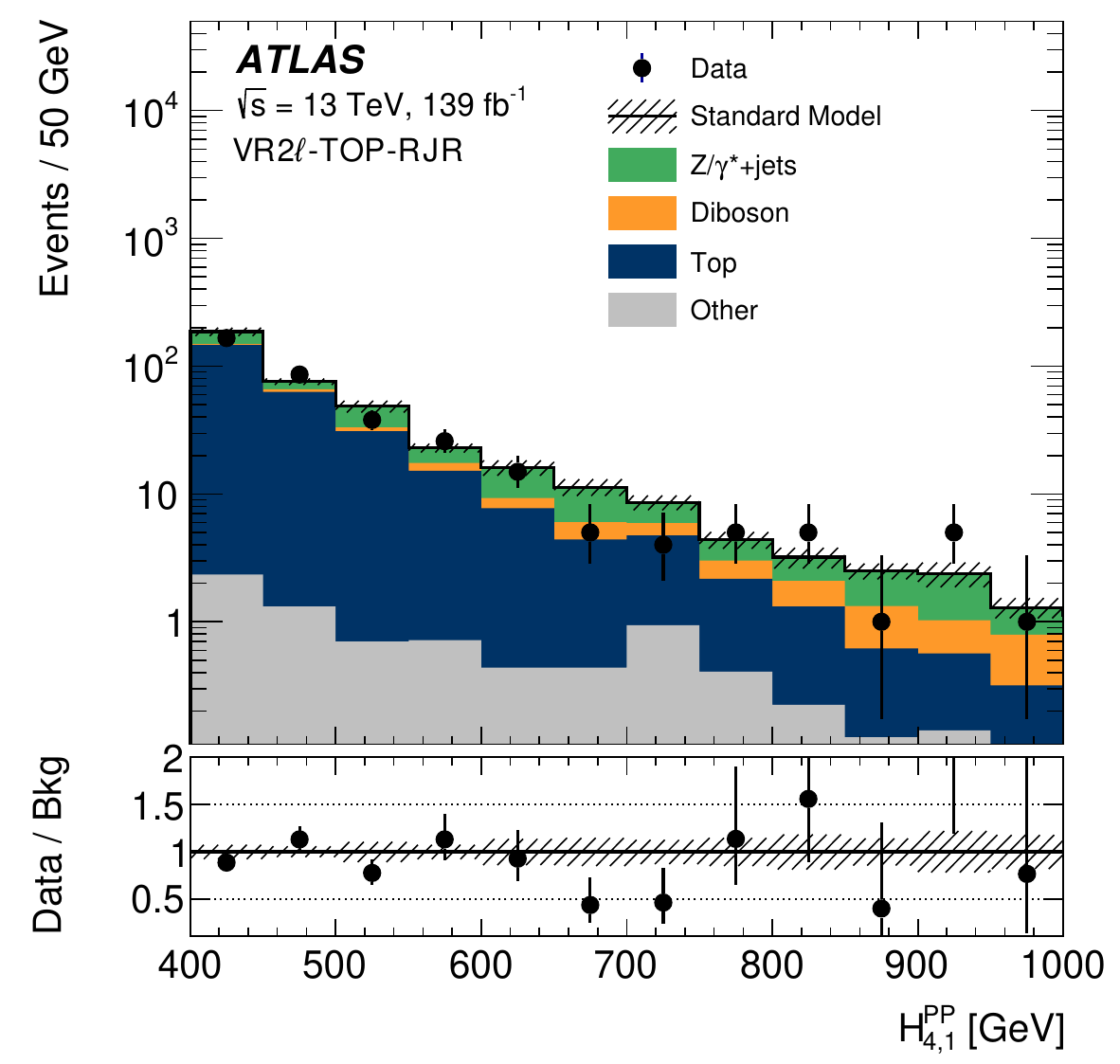}\\
\includegraphics[width=0.49\textwidth]{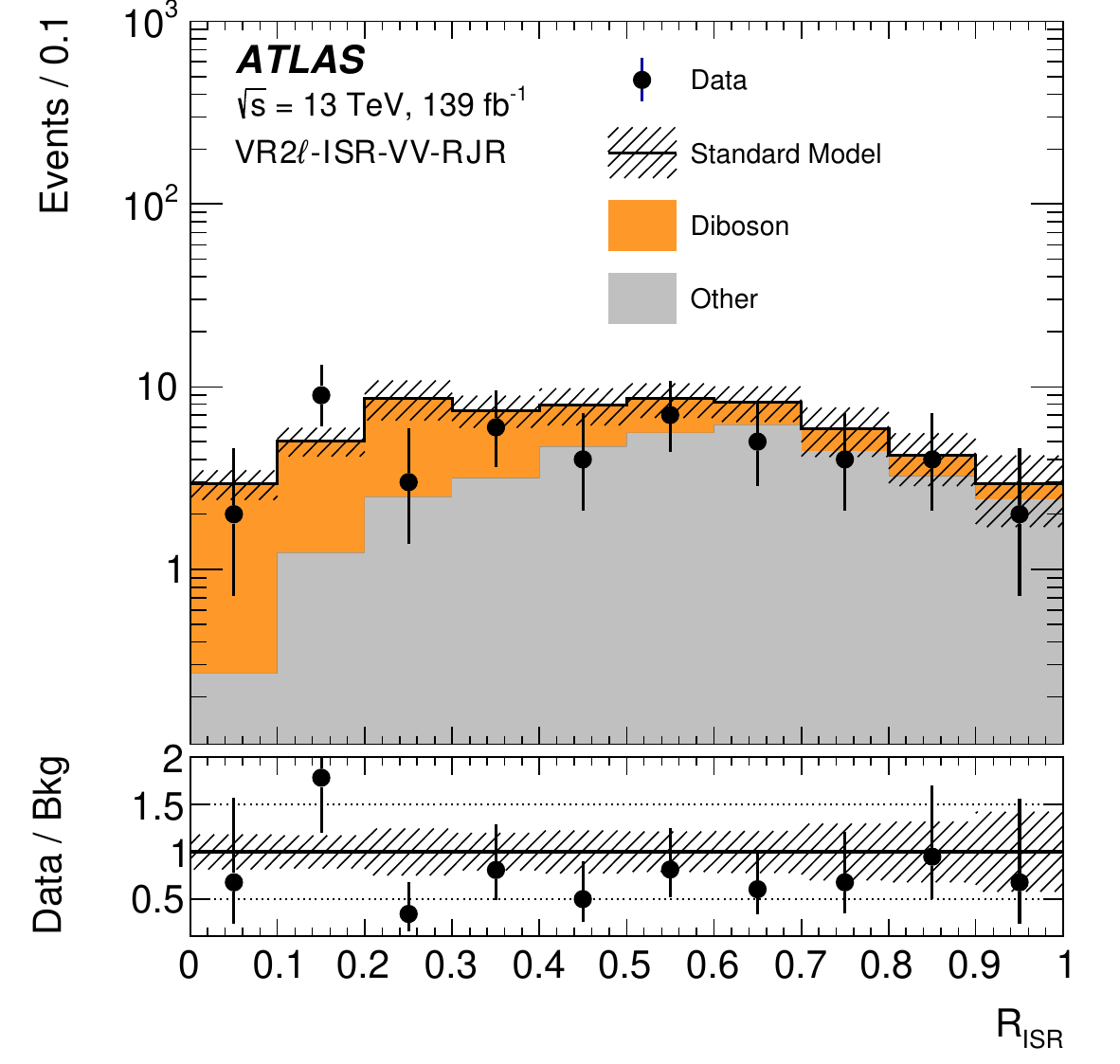}
\includegraphics[width=0.49\textwidth]{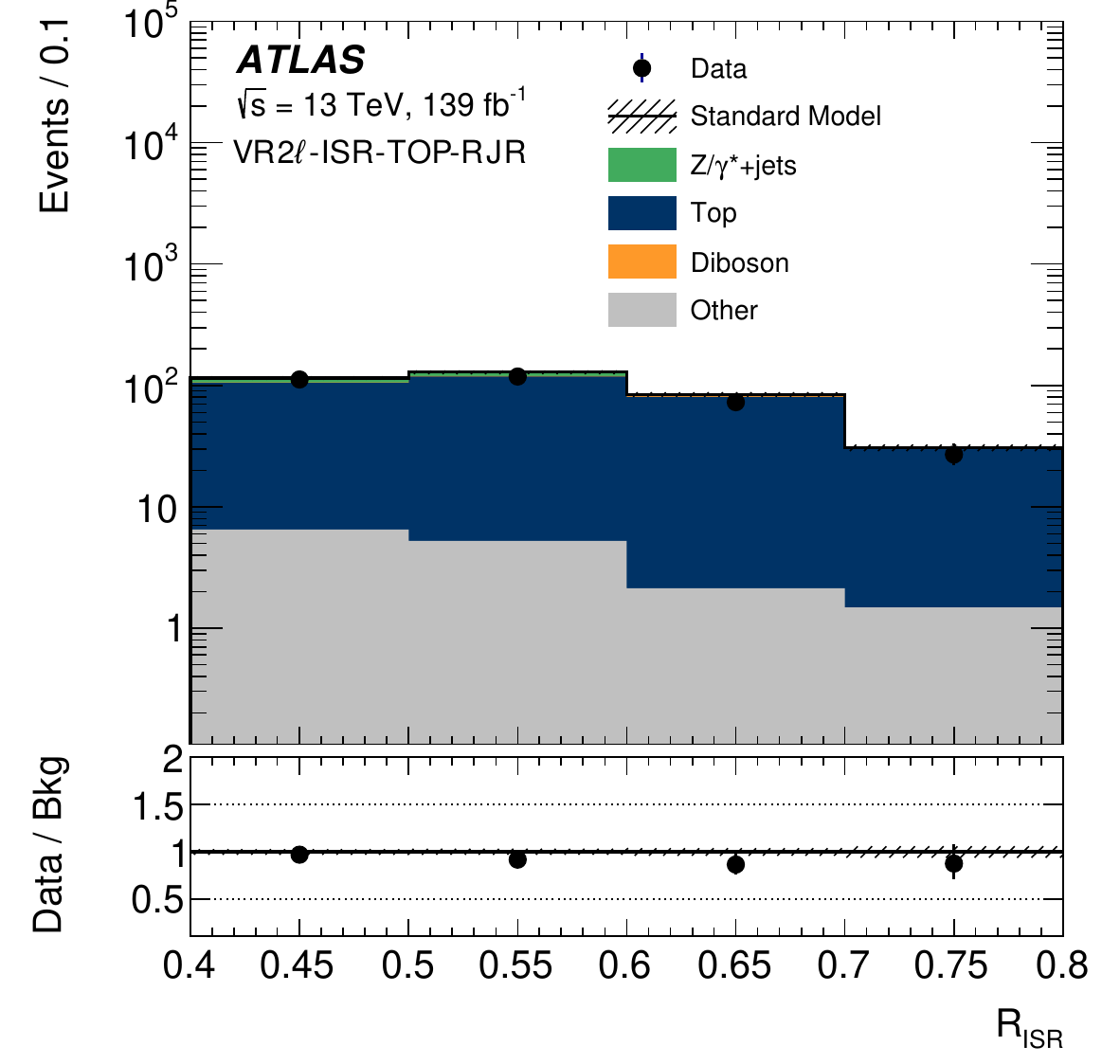}
\caption{Distributions of RJR variables in VR$2\ell$-VV-RJR (top-left), VR$2\ell$-Top-RJR (top-right), VR$2\ell$-ISR-VV-RJR (bottom-left), and VR$2\ell$-ISR-Top-RJR (bottom-right) from the RJR search.
The background estimate models the data well in all regions after a simultaneous fit of the \aclp{CR}.
The standard and ISR regions are fit separately.
The hatched band includes both the systematic and statistical uncertainties.
The last bin contains the overflow.}
 
\label{fig:rjr:vrs}
\end{figure}


\FloatBarrier

\subsection{Electroweak search backgrounds}
\label{sec:ewk_bkg}

The dominant backgrounds in the EWK \acp{SR} are $WZ$, $ZZ$, and \ttbar\ production.
Additionally, \zjets\ is an important background for the Low and OffShell regions.
Data-driven normalization factors for these backgrounds are extracted using a simultaneous likelihood fit to data in the \acp{SR} and \acp{CR} that are designed to be enriched in each background.
CR-tt-EWK targets \ttbar\ production and requires at least one $b$-jet and intermediate $\metsig \in [ 9,12 ]$ to ensure orthogonality to the \acp{SR}; unlike the RJR search, the resulting normalization factor is not applied to $Wt$, which is instead normalized to the theoretical cross-section.
CR-VZ-EWK targets $WZ$/$ZZ$ production and uses the sideband of the SR-Int-EWK $m_{jj}$ distribution, relaxing the $p_\text{T}^{j_1}$ requirement.
A common diboson normalization factor is applied to the \acp{SR}.
To account for different kinematics, two separate normalization factors are defined for \zjets.
CR-Z-EWK targets on-shell $Z$\,+\,jets production ($\mll > 71$~\GeV, which is the lowest edge of the $Z$ mass window) while CR-DY-EWK targets low-mass \zjets\ production ($12<\mll<71$~\GeV).
The resulting $Z$\,+\,jets normalization factor is applied everywhere, except for the OffShell regions, where the low-mass \zjets normalization factor is applied.
The definitions of the \acp{CR} are provided in Tables~\ref{tab:ewkRegionDefinitionsHigh}--\ref{tab:ewkRegionDefinitionsOffShell}.
The resulting normalization factors are $0.93\pm0.09$, $0.84\pm0.08$, $1.21\pm0.14$, and $0.86\pm0.25$ for \ttbar, diboson, $Z$\,+\,jets, and \zjets\ respectively.
 
To validate the normalization and modelling of the SM predictions, eight \acp{VR} are defined.
For the High \acp{SR}, two \acp{VR} use the sideband of the $m_{j_1}$ or $m_{jj}$ distribution, while VR-High-R-EWK provides additional validation of the diboson background modelling inside the $m_{jj}$ mass window by inverting the $\Delta R_{jj} $ requirement.
VR-$\ell\ell bb$-EWK validates the top background in a \metsig\ window of 12--18 below the SR-$\ell\ell bb$-EWK threshold.
Distributions of \metsig\ in the High and $\ell\ell bb$ \acp{VR}, with the normalizations from the simultaneous fit applied, are shown in Figure~\ref{fig:ewk:vrhigh_llbb}.
Good agreement between the data and the background prediction is observed.
 
VR-Int-EWK has the same requirements as SR-Int-EWK but with the requirement on $p_\text{T}^{j_1}$ inverted and is dominated by the diboson processes.
For the Low \acp{SR}, two \acp{VR} are defined.
VR-Low-EWK requires a larger value of $\Delta R_{\ell\ell}$ than SR-Int-EWK, and VR-Low-2-EWK inverts the $m_{jj}$ window.
For the OffShell regions, VR-OffShell-EWK maintains orthogonality by using an \mttwo\ window below that in the SR.
Distributions of \metsig\ in the Int, Low, and OffShell \acp{VR} are shown in Figure~\ref{fig:ewk:vrint_low_offshell} with the normalization from the simultaneous fit applied.
Good agreement is observed in all \acp{VR}.
 
\begin{figure}[htbp]
\centering
\includegraphics[width=0.49\textwidth]{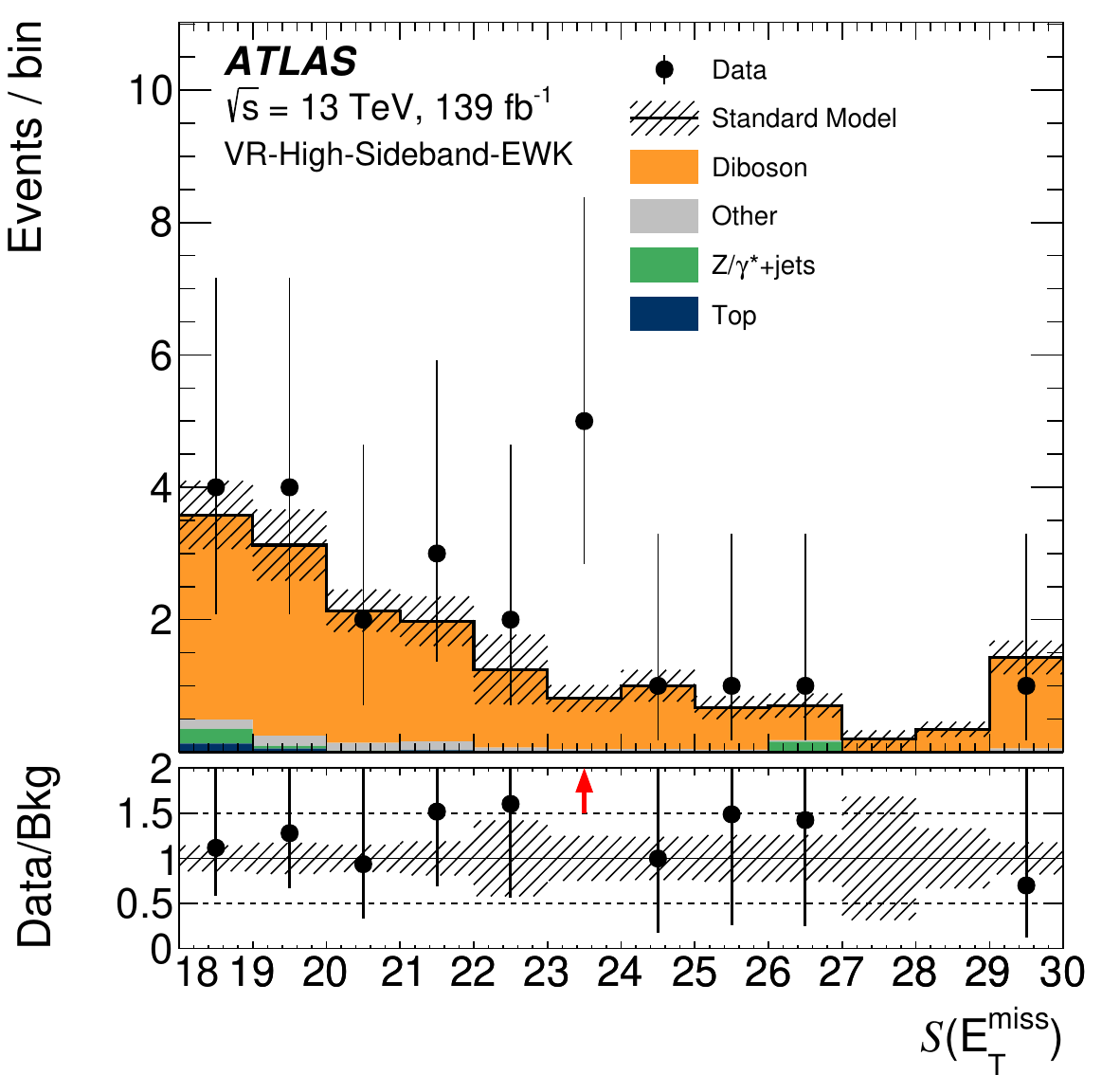}
\includegraphics[width=0.49\textwidth]{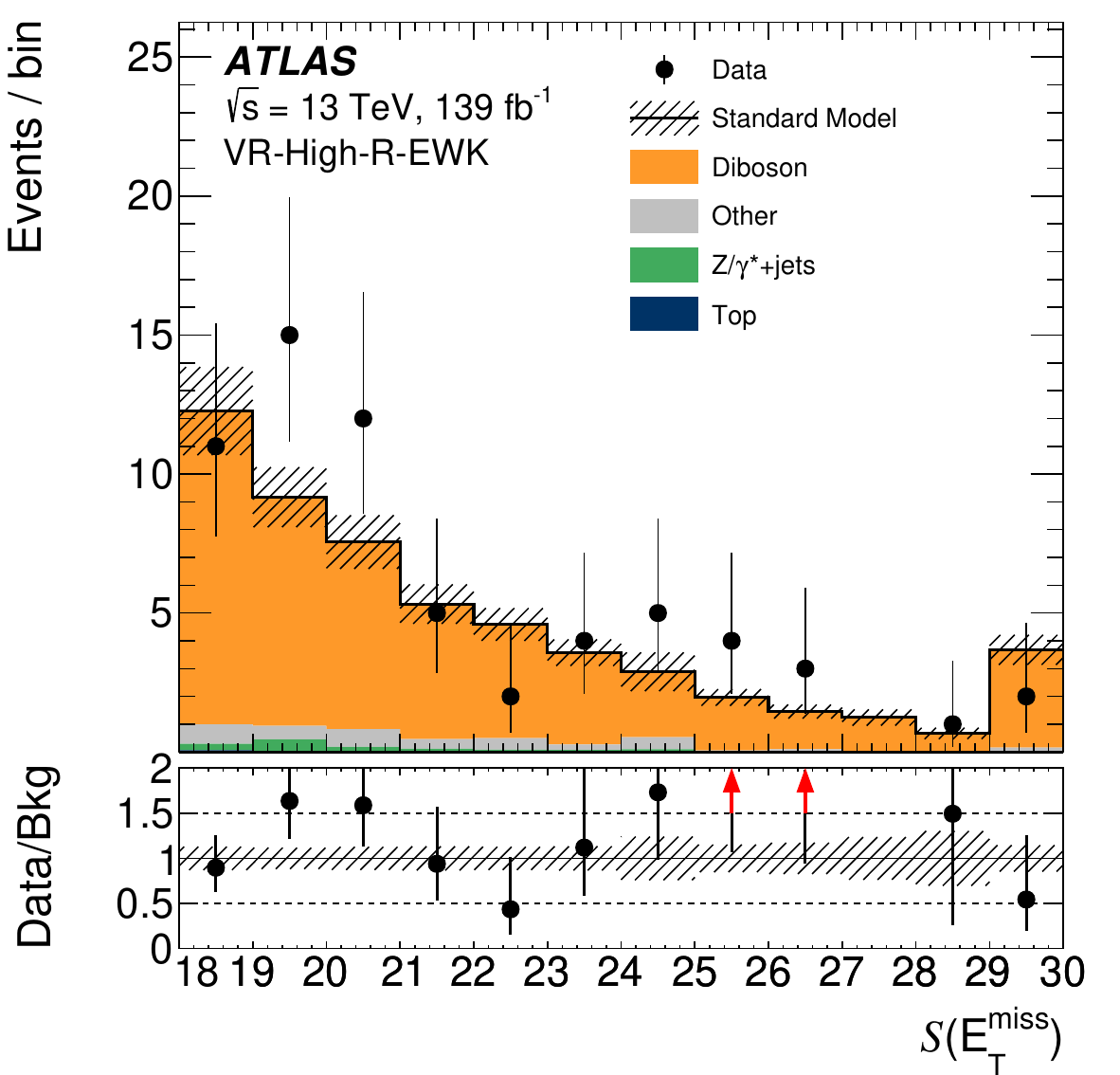}
\includegraphics[width=0.49\textwidth]{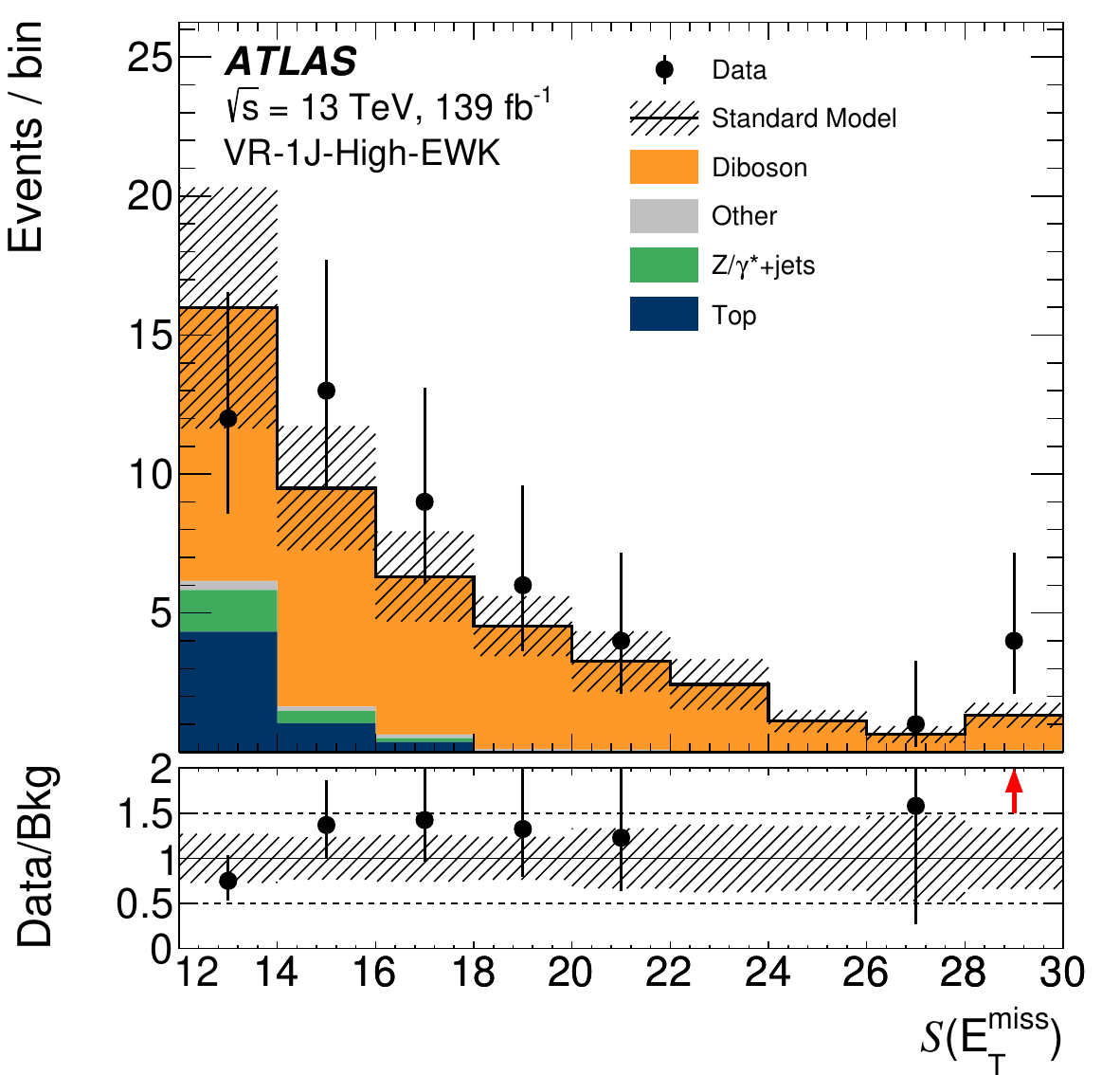}
\includegraphics[width=0.49\textwidth]{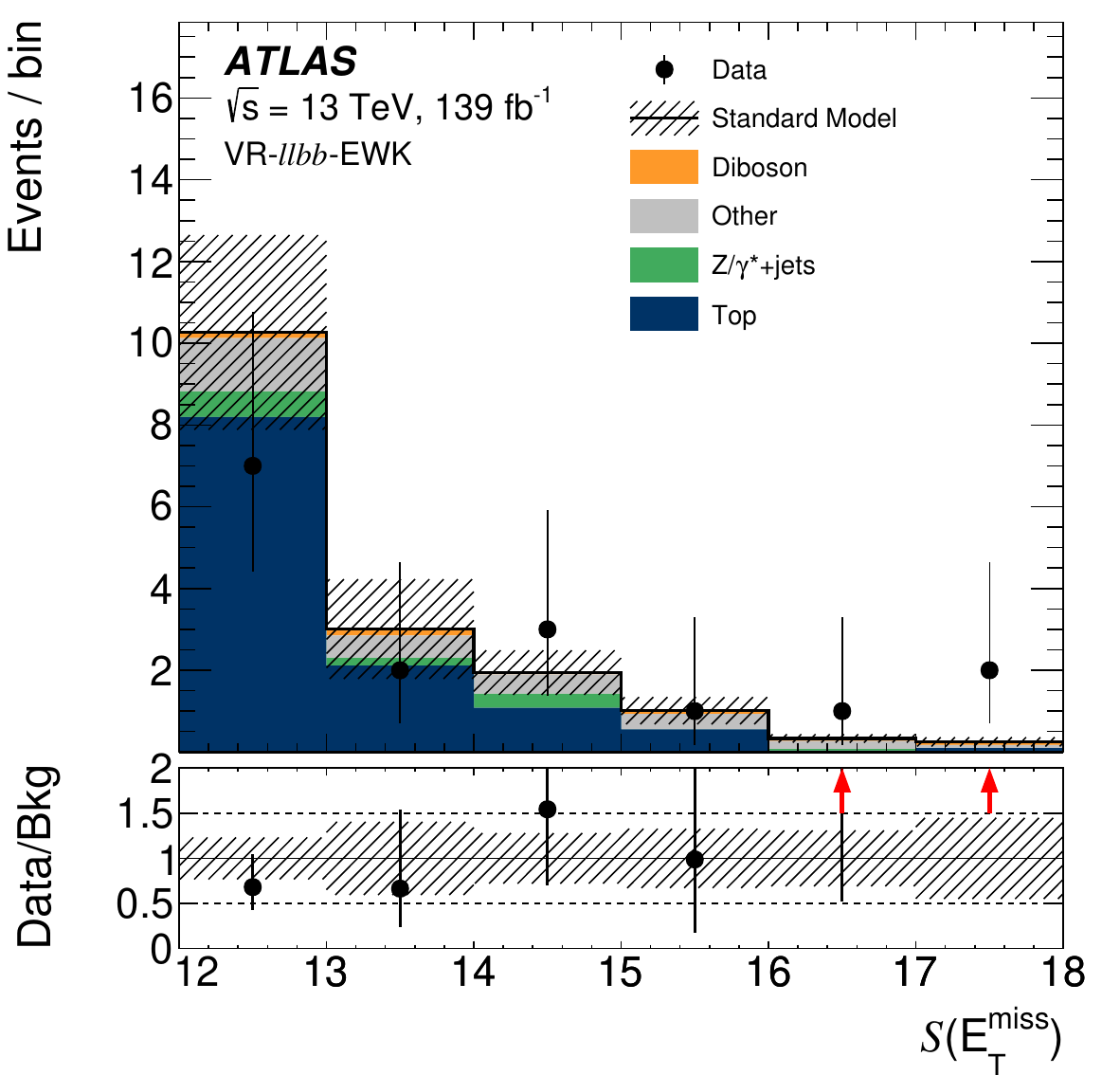}
\caption{Distributions of \metsig\ in VR-High-Sideband-EWK (top-left), VR-High-R-EWK (top-right), VR-1J-High-EWK (bottom-left), and VR-$\ell\ell bb$-EWK (bottom-right) from the EWK search  after a simultaneous fit of the \aclp{CR}.
The hatched band includes both the systematic and statistical uncertainties.
The last bin includes the overflow.}
\label{fig:ewk:vrhigh_llbb}
\end{figure}

\begin{figure}[htbp]
\centering
\includegraphics[width=0.49\textwidth]{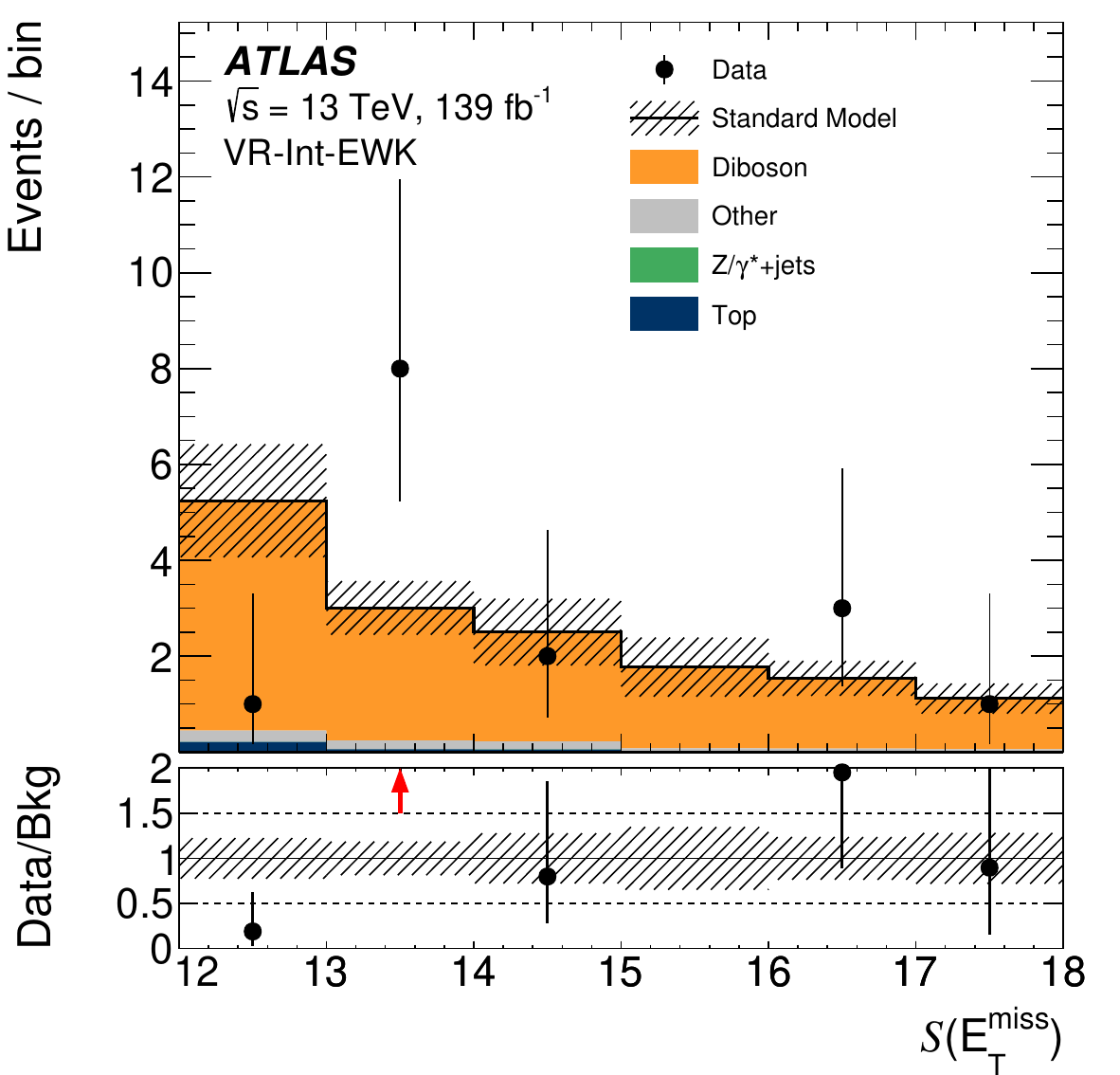}
\includegraphics[width=0.49\textwidth]{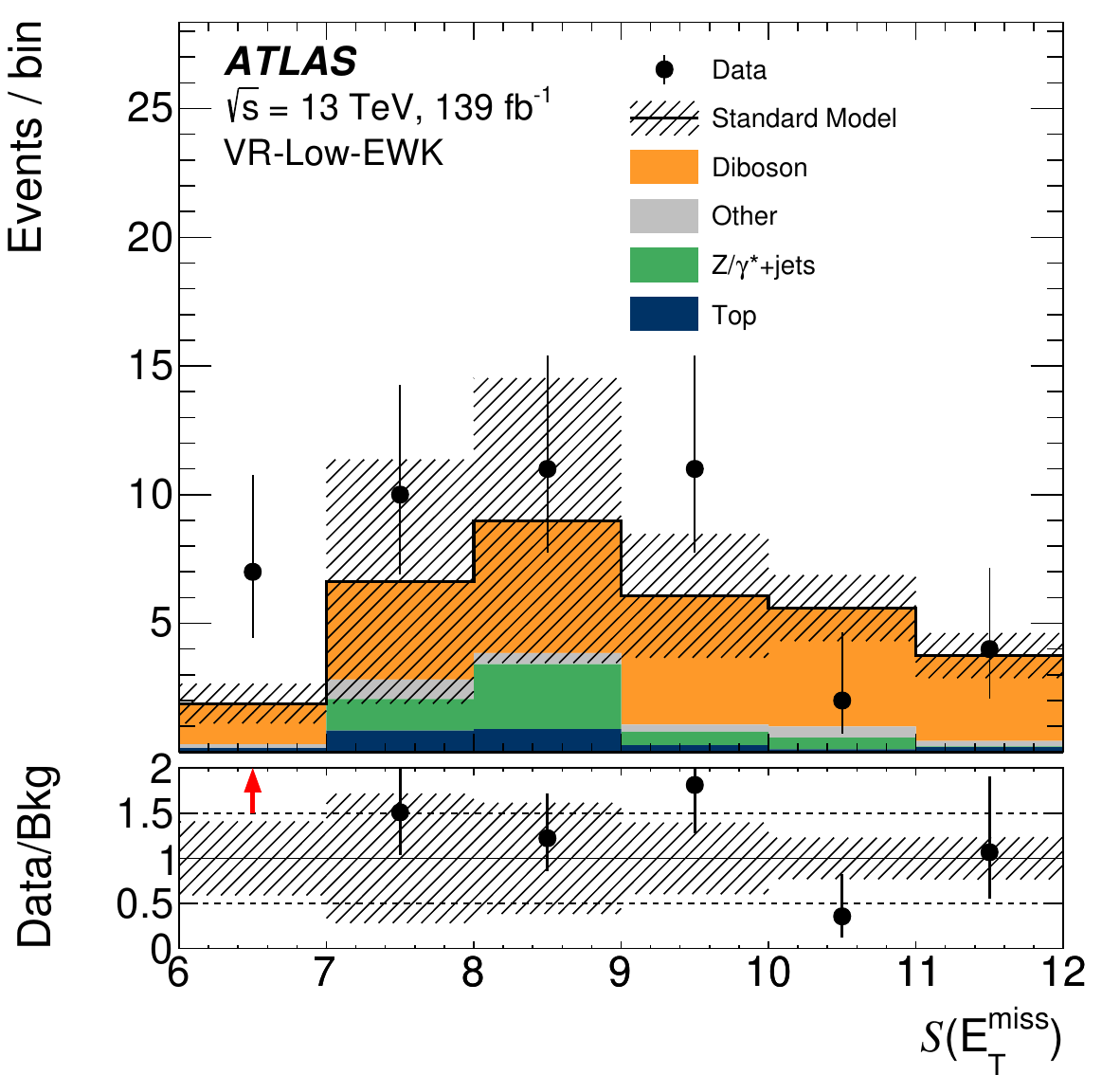}
\includegraphics[width=0.49\textwidth]{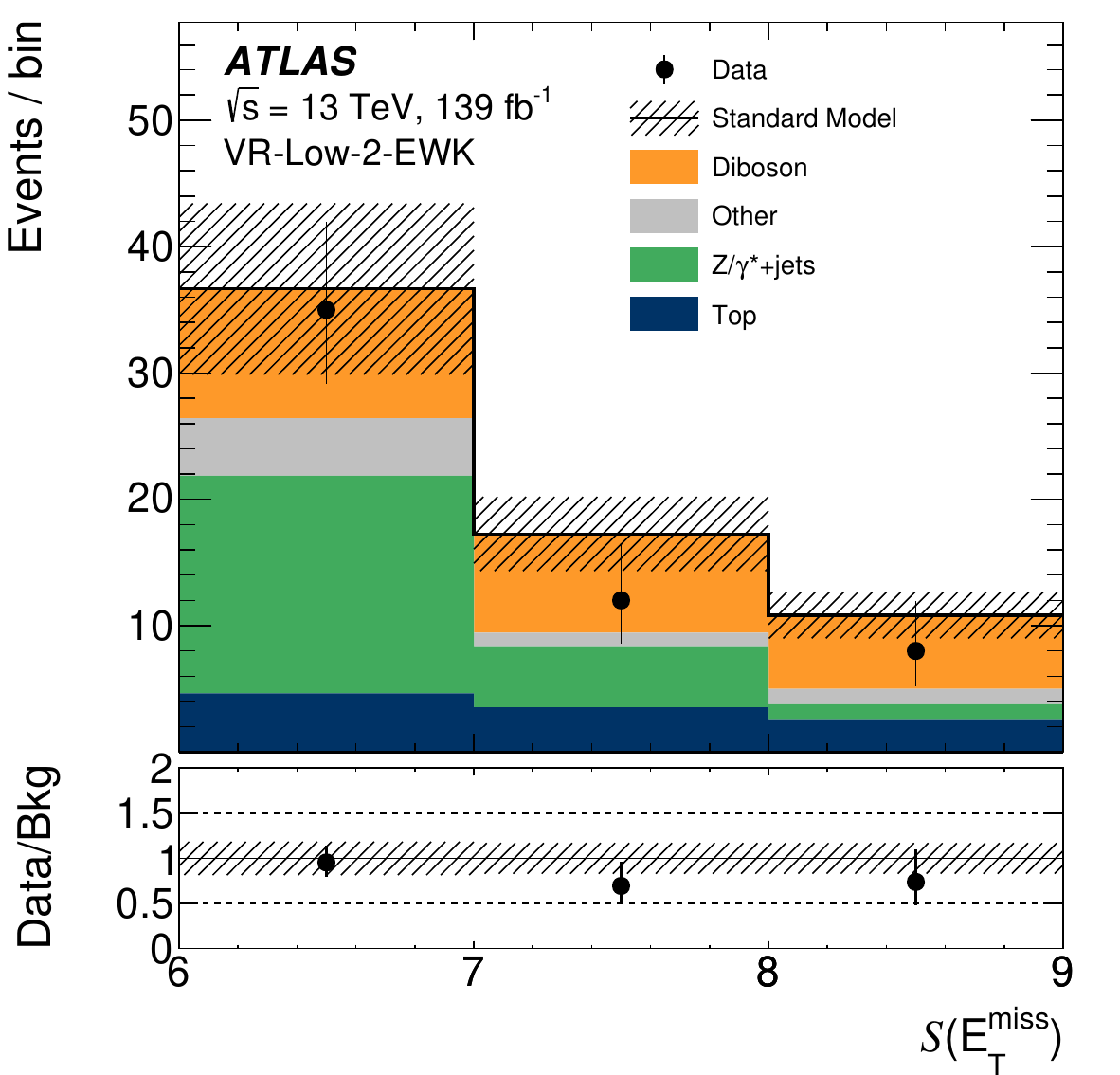}
\includegraphics[width=0.49\textwidth]{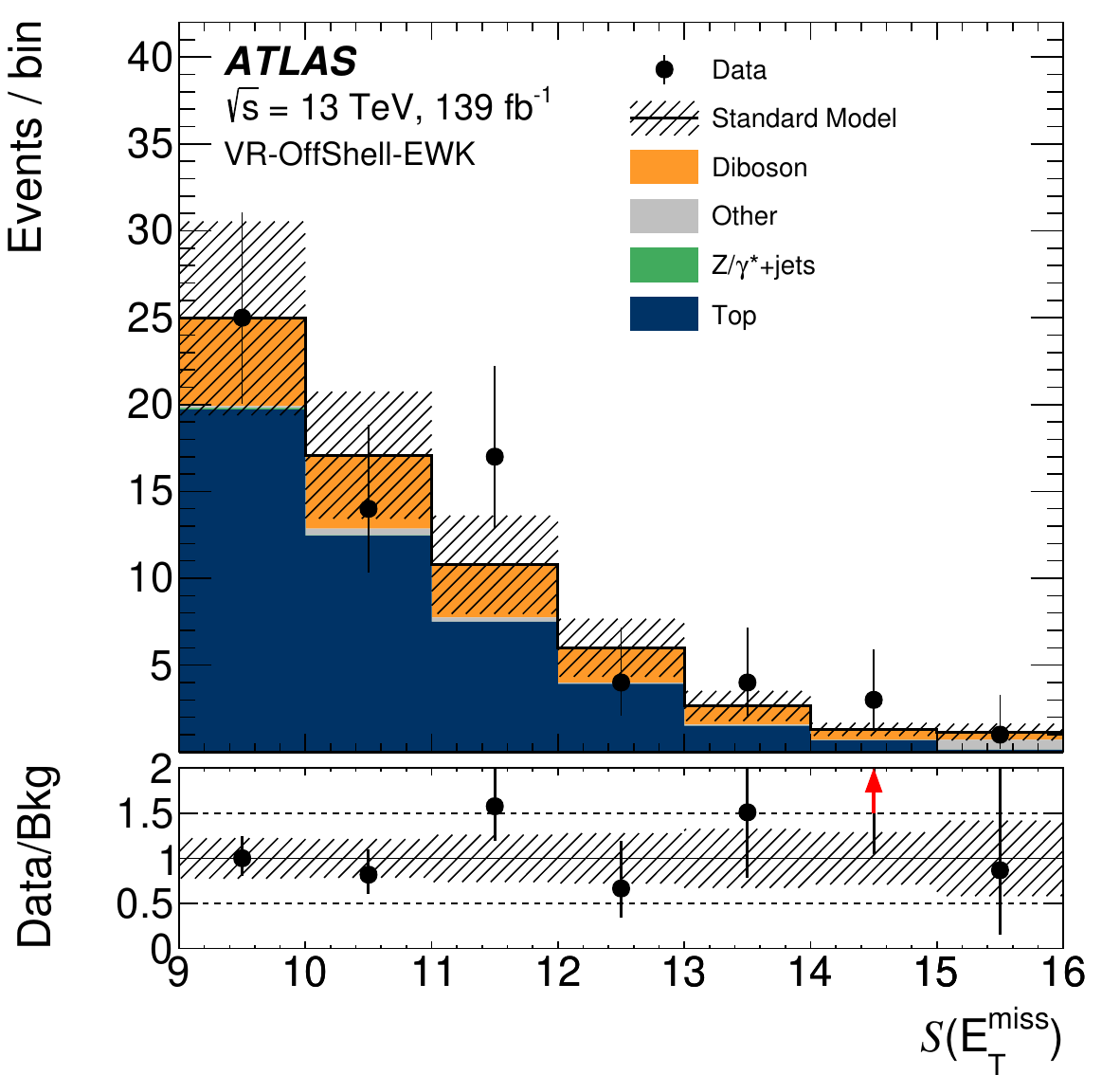}
\caption{Distributions of \metsig\ in VR-Int-EWK (top-left), VR-Low-EWK (top-right), VR-Low-2-EWK (bottom-left), and VR-OffShell-EWK (bottom-right) from the EWK search after a simultaneous fit of the \aclp{CR}.
The hatched band includes both the systematic and statistical uncertainties.
The last bin includes the overflow.}
\label{fig:ewk:vrint_low_offshell}
\end{figure}


\FloatBarrier
 
\subsection{Strong search backgrounds}
\label{sec:strong_bkg}

For the Strong search, the \zjets\ background is normalized in \acp{CR}, the \ac{FS} backgrounds are estimated from $e\mu$ data, the \ac{FNP}-lepton background is estimated in a data-driven way as previously described, and the remaining backgrounds are fully estimated via \ac{MC} simulation.
\ac{FS} processes include those with independent leptonic decays such that the expected number of $e\mu$ events is the same as the number $ee$\,+\,$\mu\mu$ events, e.g.\ \ttbar, $WW$, and $Z\rightarrow\tau\tau$.
\acp{VR} are constructed with an \met\ requirement, $150<\met<250~\GeV$, below the \ac{SR} requirement in order to validate the background estimation.
\acp{VR} with \ac{SS} instead of \ac{OS} leptons and with three instead of two leptons are also constructed in order to validate the \ac{FNP}-lepton and $WZ$/$ZZ$ backgrounds respectively.
 
The \zjets\ background, with decays to $ee$ or $\mu\mu$, mainly enter the \acp{SR} due to \met\ from the mismeasurement of jets or from contributions by neutrinos in heavy-flavour decays.
The requirement of $\mindphimj>0.4$ removes \zjets\ events where the \ptmiss\ is aligned with a jet, as is the case for mismeasured jets.
A \ac{CR} is constructed for each SR or VR by inverting this requirement, and roughly half of the yield in these CRs is from \zjets\ events.
The normalization of the \zjets\ \ac{MC} prediction is extracted with a simultaneous likelihood fit to the \ac{CR} and respective SR or VR.
The resulting factors are within one standard deviation of one for every region except SRC-STR, where the small MC simulation prediction is pulled up by approximately 1.5 standard deviations, and VRLow-STR, where the MC simulation prediction is pulled down by approximately 1.3 standard deviations.
 
The \ac{FS} method makes use of a \ac{CR} for each \ac{SF} region, with the lepton flavour requirement changed to select $e\mu$ data events.
Differences between the efficiencies to select events with electrons and events with muons are used to adjust the $e\mu$ data events to more accurately predict $ee$\,+\,$\mu\mu$ events.
For the \onz\ \acp{SR}, the $e\mu$ data yield is small.
In order to decrease the statistical uncertainty of the \ac{FS} prediction, the \mll\ window used to collect $e\mu$ events is widened to $61$--$121$~\GeV.
The yield is then scaled down by the ratio of the yields in the narrower and wider \mll\ windows obtained from \ac{MC} simulation.
The estimated $ee$\,+\,$\mu\mu$ yield $(N^\text{est})$ is obtained as:
\begin{align}
\label{eq:fs_est}
N^\text{est} = \frac{1}{2} \cdot
\left[ \sum^{N_{e\mu}^\text{data}}_{i}
\Bigl( k_{e}(\pT^{i,\mu}, \eta^{i,\mu}) + k_{\mu}(\pT^{i,e}, \eta^{i,e})
\Bigr) \cdot \alpha(\pT^{i,\ell_{1}}, \eta^{i,\ell_{1}}) \notag \right.\\
\left.- \sum^{N_{e\mu}^\text{MC}}_{i}
\Bigl( k_{e}(\pT^{i,\mu}, \eta^{i,\mu}) + k_{\mu}(\pT^{i,e}, \eta^{i,e})
\Bigr) \cdot \alpha(\pT^{i,\ell_{1}}, \eta^{i,\ell_{1}}) \right],
\end{align}
\noindent where $N_{e\mu}^\text{data}$ is the number of data events observed in a given \ac{CR}.
Events from non-FS processes, e.g.\ $WZ$/$ZZ$, are subtracted from the $e\mu$ data events using MC simulation.
This is the second term in Eq.~(\ref{eq:fs_est}), where $N_{e\mu}^\text{MC}$ is the number of events from non-FS processes in MC simulation in the CR.
The factor $\alpha(\pT^i, \eta^i)$ accounts for the different trigger efficiencies for $ee/\mu\mu$ and $e\mu$ events,
and $k_{e}(\pT^i, \eta^i)$ and $k_{\mu}(\pT^i, \eta^i)$ are the electron and muon selection efficiency factors for the kinematics of the lepton being replaced in event $i$.
For example, if an event with a muon and electron is used to predict a $\mu\mu$ event, the kinematics of the electron are used to evaluate $k_\mu$.
The trigger and selection efficiency correction factors are derived from data events in an inclusive region with two signal leptons and at least two signal jets,
according to:
\begin{eqnarray*}
k_{e}(\pT, \eta) = \sqrt{\frac{N_{ee}^{\text{meas}}(\pT, \eta)}{N_{\mu\mu}^{\text{meas}}(\pT, \eta)}}, \\
k_{\mu}(\pT, \eta) = \sqrt{\frac{N_{\mu\mu}^{\text{meas}}(\pT, \eta)}{N_{ee}^{\text{meas}}(\pT, \eta)}}, \\
\alpha(\pT, \eta) = \frac{\sqrt{\epsilon^\text{trig}_{ee}(\pt^{\ell_1},\eta^{\ell_1})\times\epsilon^\text{trig}_{\mu\mu}(\pt^{\ell_1},\eta^{\ell_1})}}{\epsilon^\text{trig}_{e\mu}(\pt^{\ell_1},\eta^{\ell_1})},
\end{eqnarray*}
 
\noindent where $\epsilon^\text{trig}_{ee/\mu\mu/e\mu}$ is the trigger efficiency as a function of the leading-lepton ($\ell_1$) kinematics and $N_{ee}^{\text{meas}}$ $(N_{\mu\mu}^{\text{meas}})$
is the number of $ee$ $(\mu\mu)$ data events in the inclusive region mentioned above.
The factors $k_{e}(\pT, \eta)$ and $k_{\mu}(\pT, \eta)$ are calculated separately for leading and subleading leptons.
The correction factors are typically within 20\% of unity, except in the region $|\eta|<0.1$, where they deviate by up to 40\% from unity because of a lack of coverage by the muon spectrometer.
 
Comparisons of the \mll\ distribution between the background estimate and data in the \acp{VR} corresponding to the four edge \acp{SR} are shown in Figure~\ref{fig:strong_VR_mll}.
Good agreement is observed in each of the regions.
Since the \onz\ \acp{SR} require at least four jets, their equivalent \acp{VR} correspond to the high-\njets\ regions in the distribution of the number of jets for each \ac{VR} shown in Figure~\ref{fig:strong_VR_njet}.
Good agreement is seen in each VR.
In order to validate the \ac{MC} simulation of $WZ$/$ZZ$ production, a three-lepton \ac{VR} is constructed by selecting events with  three signal leptons and applying requirements similar to those in the \acp{SR}, as shown near the bottom of Table~\ref{tab:strongRegionDefinitions}.
Figure~\ref{fig:strong_VR3L} shows the \mll\ distribution in this VR3L-STR with a 13\% theory uncertainty assigned to the $WZ$/$ZZ$ background.
Good agreement within the statistical uncertainty is observed.
 
\begin{figure}[htbp]
\centering
\includegraphics[width=0.49\textwidth]{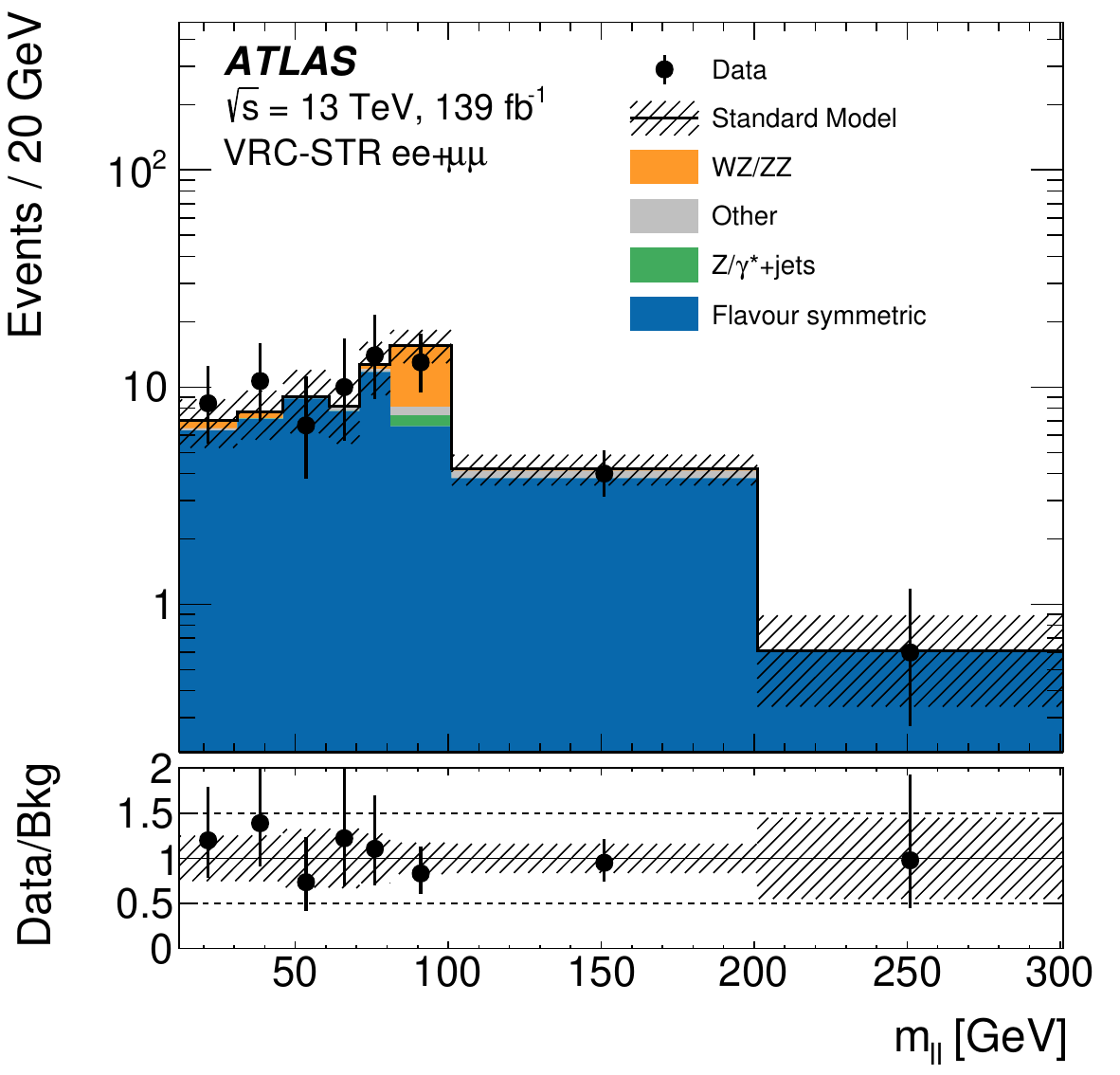}
\includegraphics[width=0.49\textwidth]{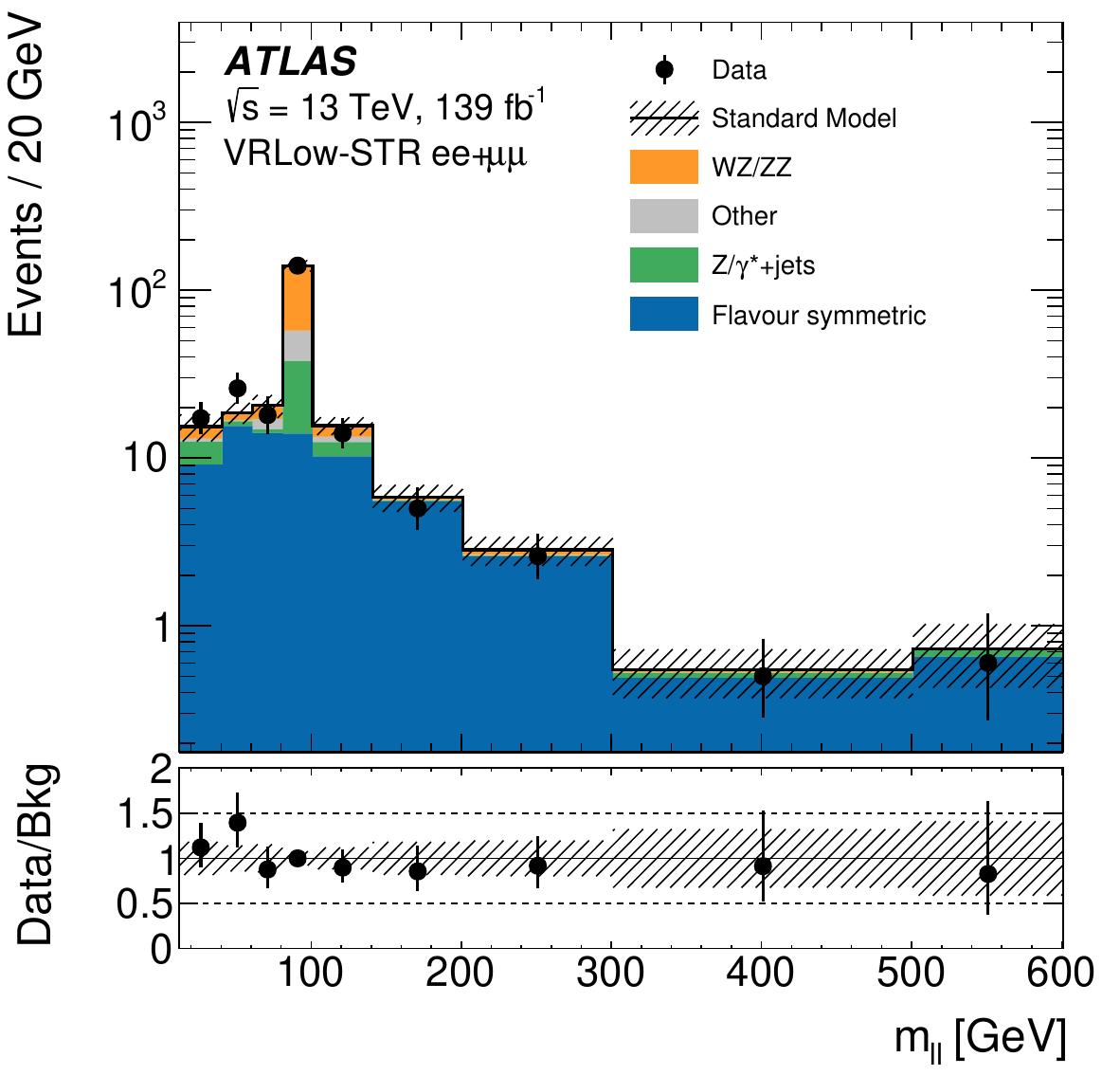}
\includegraphics[width=0.49\textwidth]{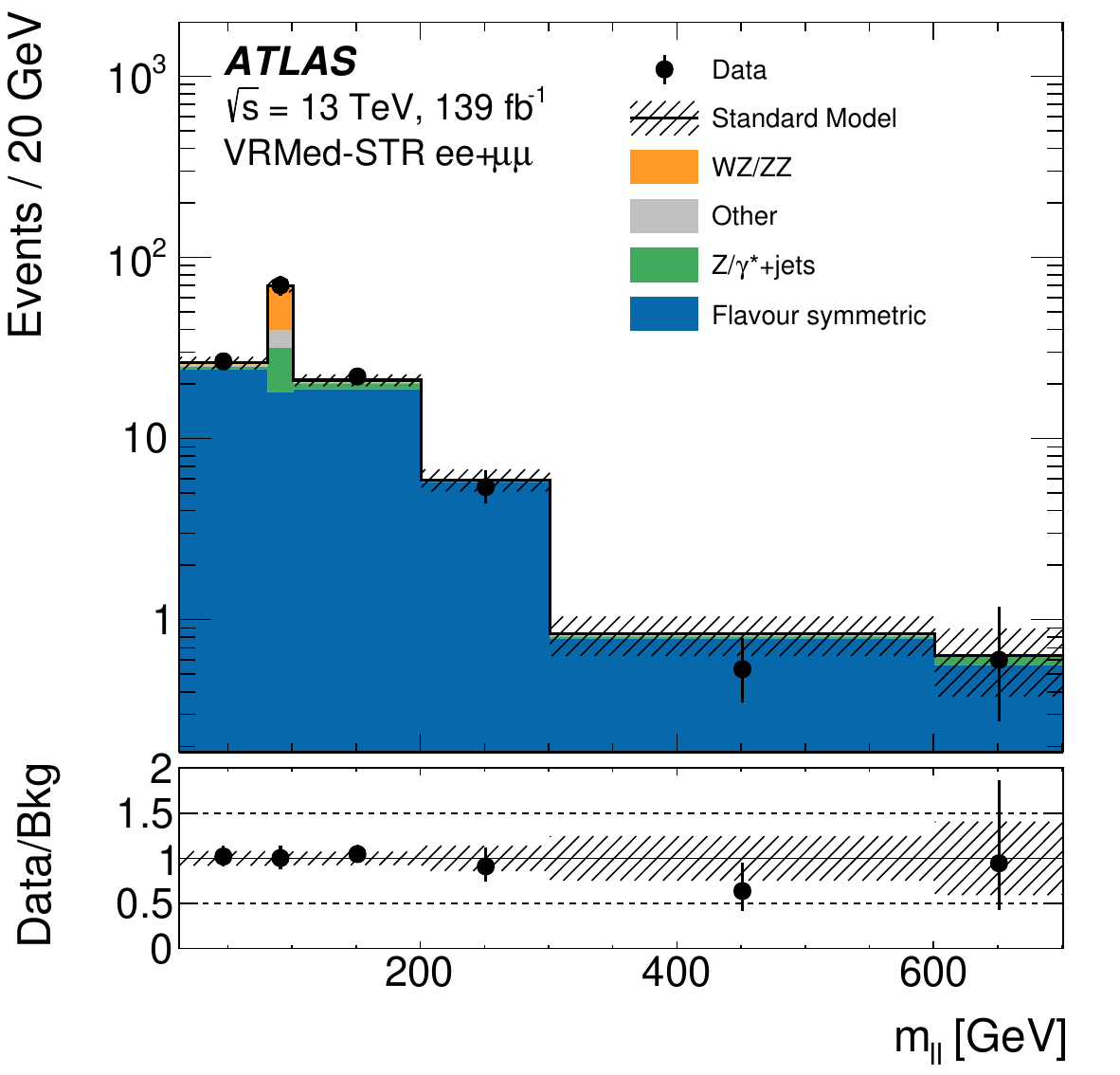}
\includegraphics[width=0.49\textwidth]{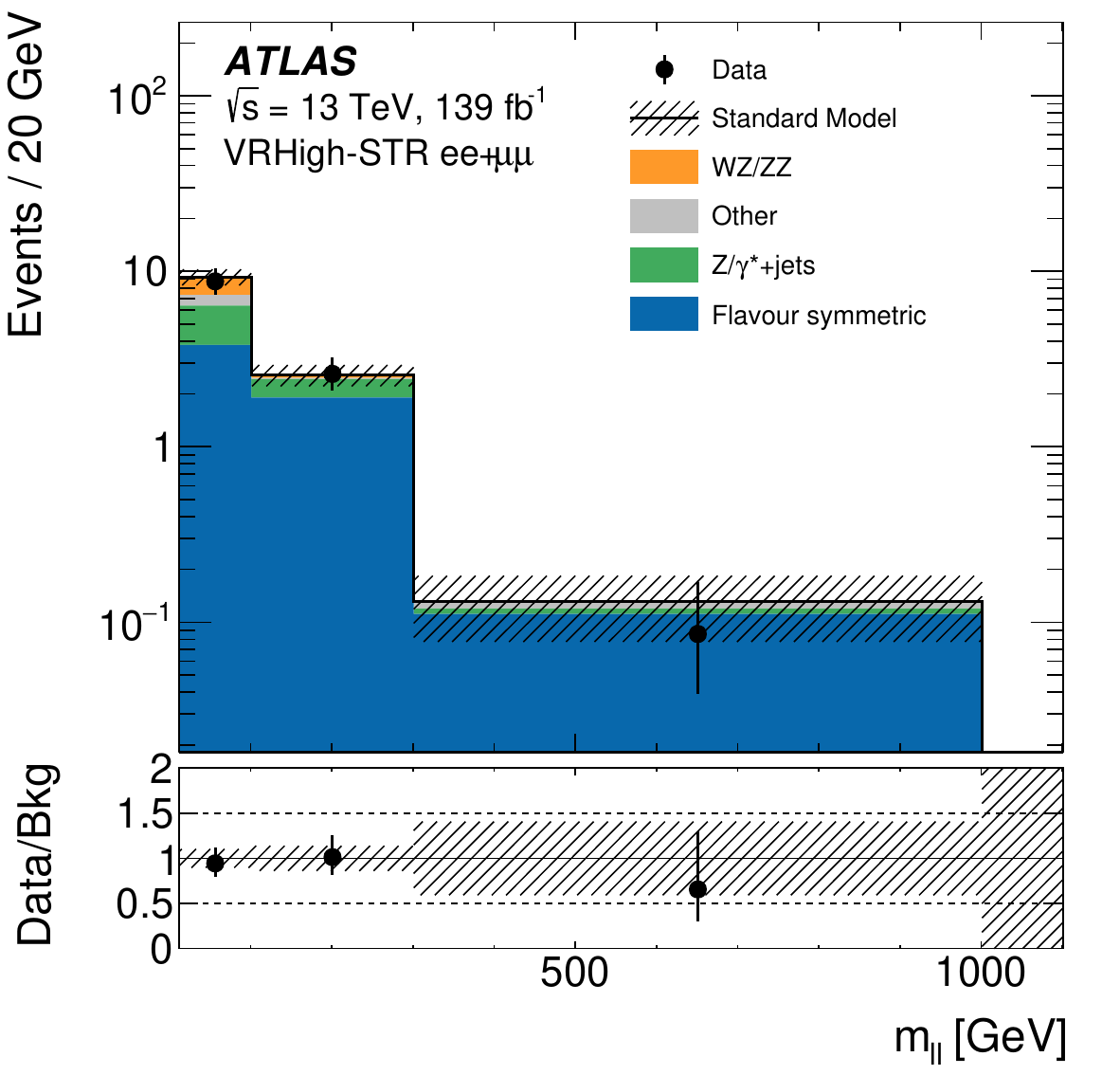}
\caption{Observed and expected dilepton mass distributions in VRC-STR (top-left), VRLow-STR (top-right), VRMed-STR (bottom-left), and VRHigh-STR (bottom-right).
Each \acl{VR} is fit separately with the corresponding \acl{CR}.
All statistical and systematic uncertainties are included in the hatched band.
The entries are normalized to the bin width, and the last bin is the overflow.
}
\label{fig:strong_VR_mll}
\end{figure}
 
\begin{figure}[htbp]
\centering
\includegraphics[width=0.49\textwidth]{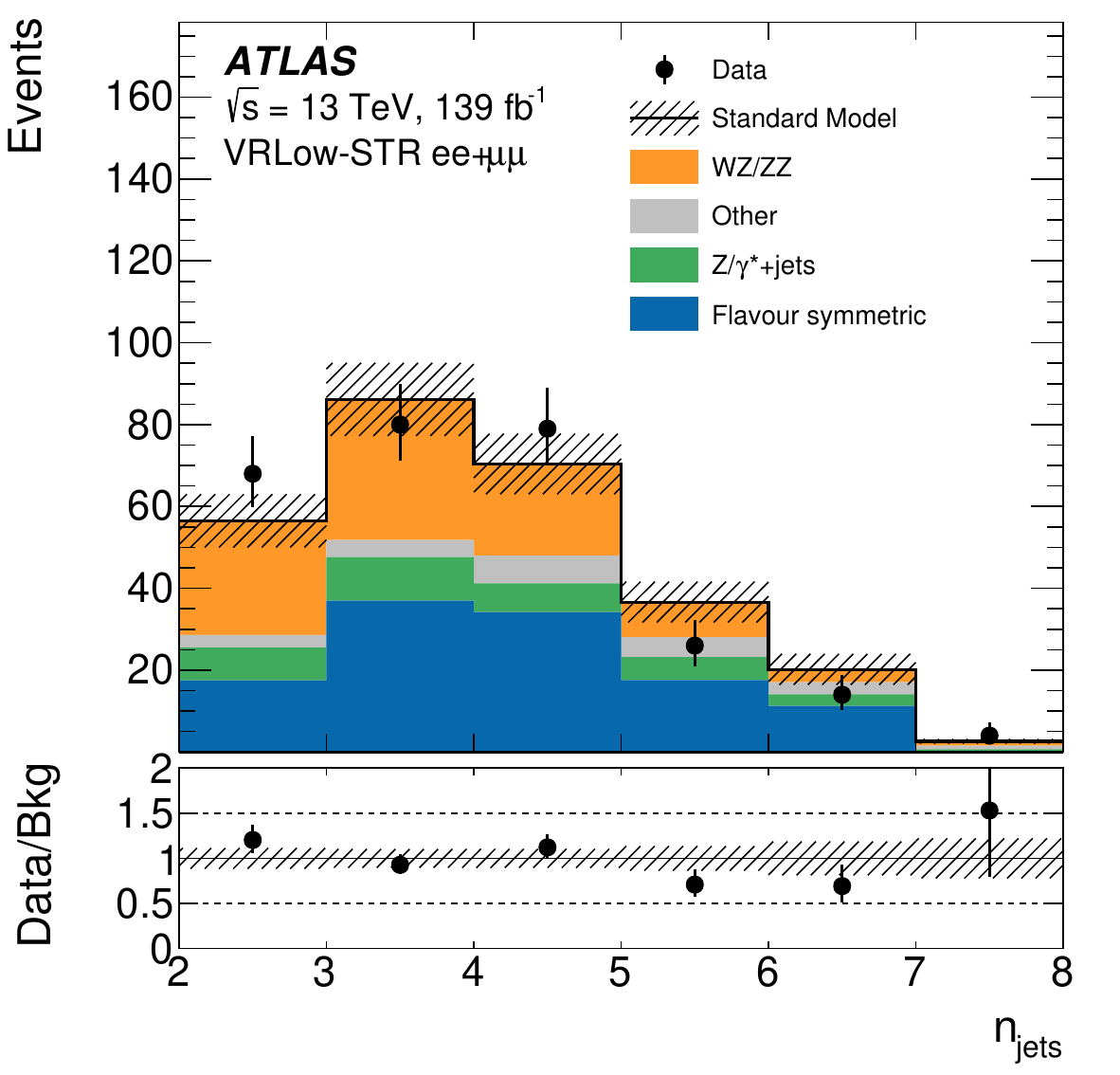}
\includegraphics[width=0.49\textwidth]{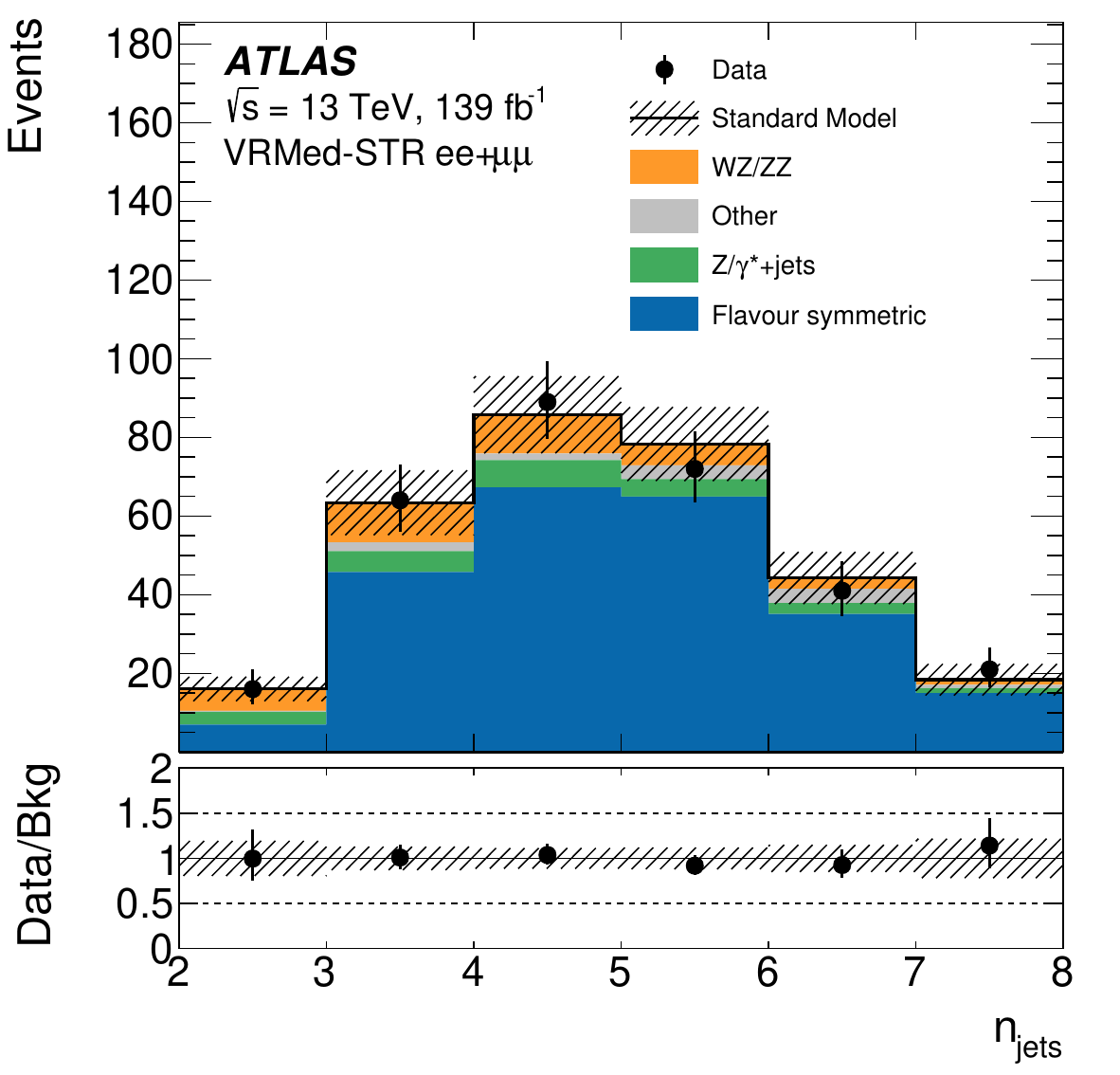}
\includegraphics[width=0.49\textwidth]{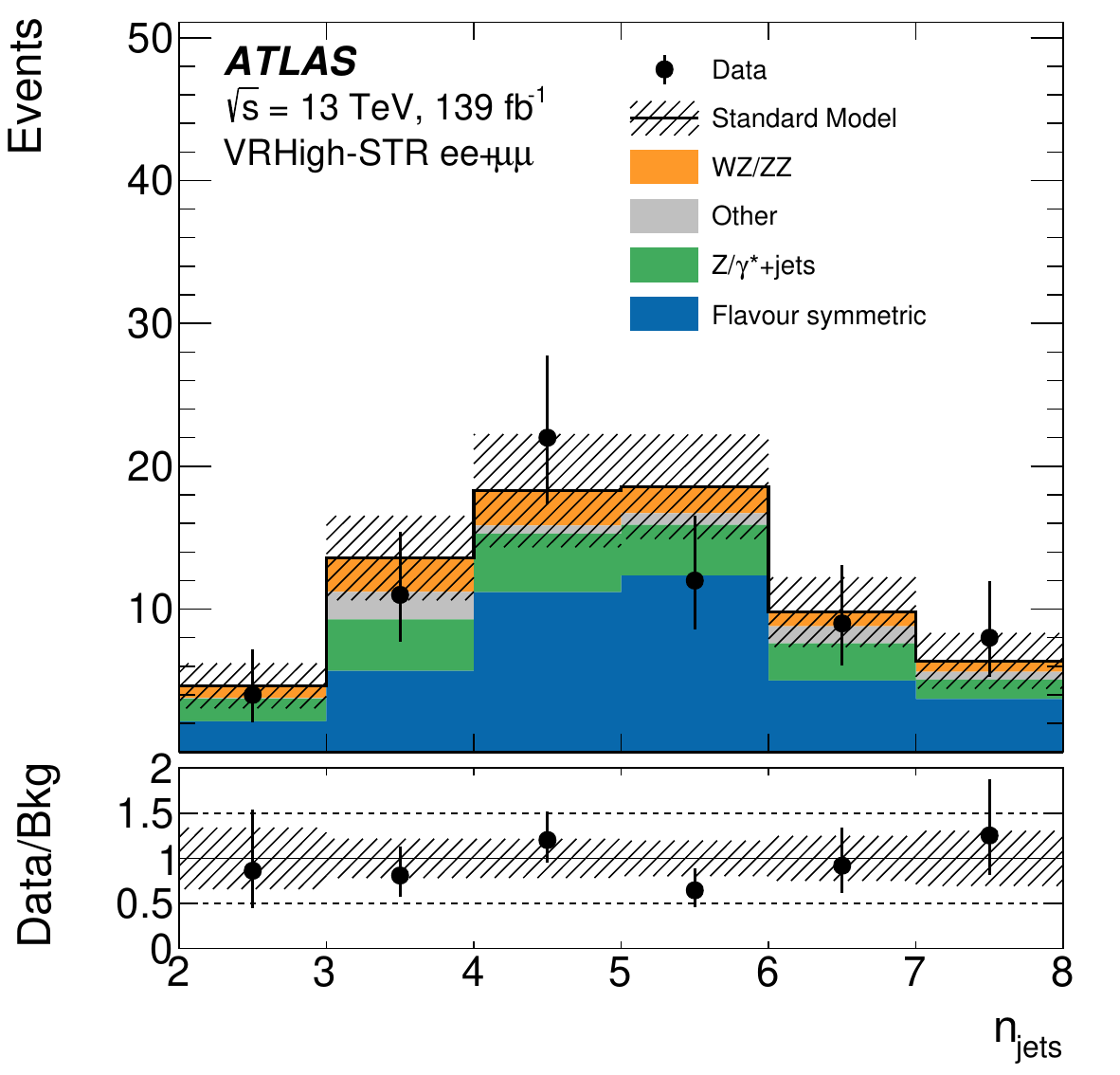}
\caption{Observed and expected jet multiplicity in VRLow-STR (top-left), VRMed-STR (top-right), and VRHigh-STR (bottom) after a fit performed on the \mll\ distribution and corresponding \acl{CR}.
All statistical and systematic uncertainties are included in the hatched band.
The last bin contains the overflow.
}
\label{fig:strong_VR_njet}
\end{figure}

\begin{figure}[htbp]
\centering
\includegraphics[width=0.49\textwidth]{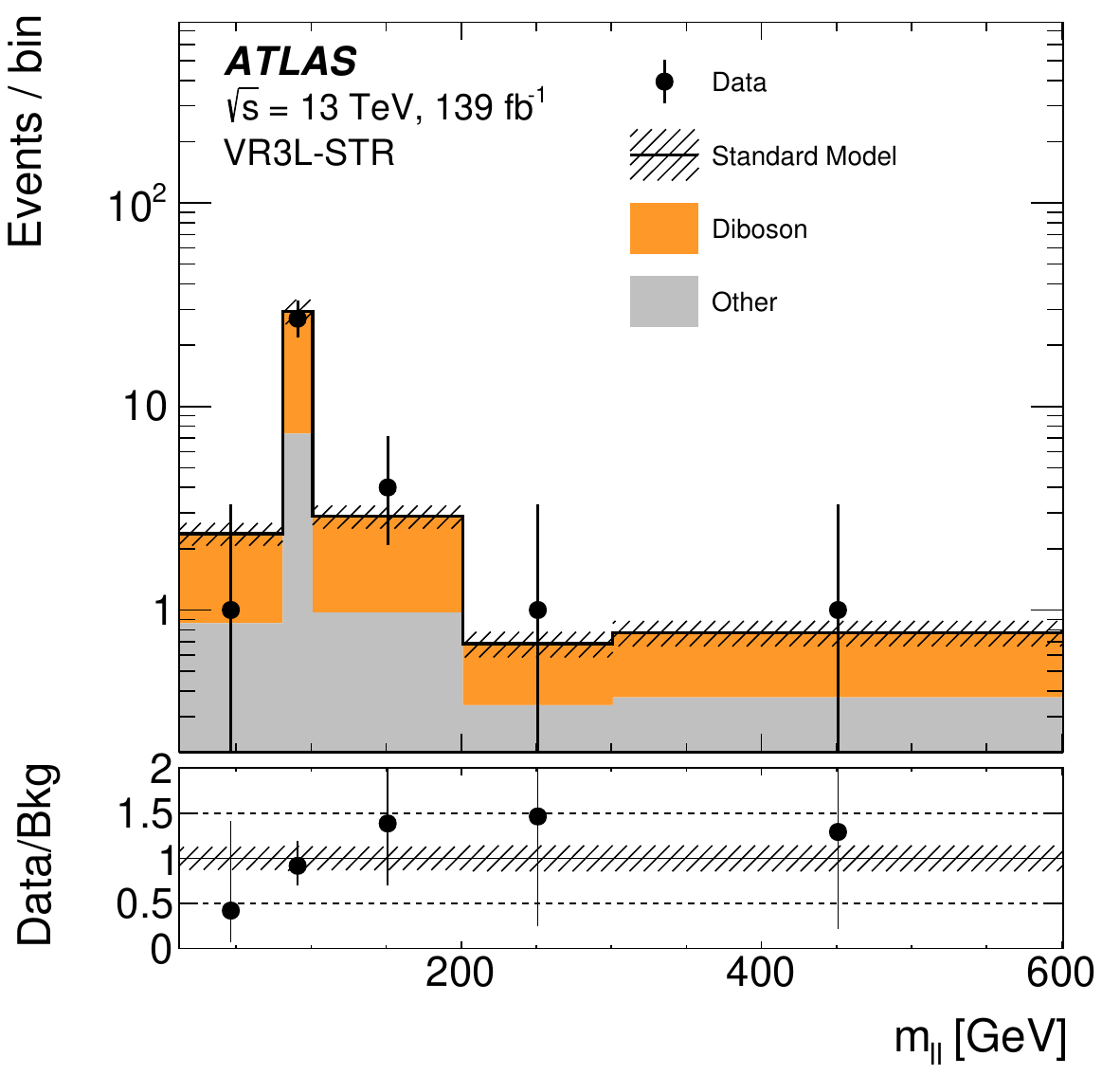}
\caption{Observed and expected dilepton mass distributions in VR3L-STR without a fit to the data.
The `Other' category includes the negligible contributions from \ttbar\ and \zjets\ processes.
The hatched band contains the statistical uncertainty and the theoretical systematic uncertainties of the $WZ$/$ZZ$ prediction, which are the dominant sources of uncertainty in VR3L-STR.
No fit is performed.
The last bin contains the overflow.}
\label{fig:strong_VR3L}
\end{figure}



\FloatBarrier
\section{Systematic uncertainties}
\label{sec:systs}

Theoretical and experimental uncertainties are taken into account for the signal and background models.
For processes that are not normalized to data, a 1.7\% uncertainty is assigned to the integrated luminosity \cite{ATLAS-CONF-2019-021,LUCID2}.
The estimated efficiencies from \ac{MC} simulation to trigger on, reconstruct, and measure the objects required for this analysis are all subject to uncertainties to account for differences between data and simulation.
For the jet energy scale, uncertainties due to the jet flavour composition, pile-up, and the jet and event kinematics \cite{PERF-2016-04} are considered.
The \ac{JER} is also subject to additional uncertainties to account for differences between data and simulation \cite{JETM-2018-05}.
Uncertainties in the lepton energy scales, resolutions, identification efficiencies, isolation efficiencies, reconstruction efficiencies, and trigger efficiencies are also considered.
The uncertainties associated with the objects used to compute the \ptmiss\ are propagated through the computation, and additional uncertainties in the scale and resolution of the contribution from low-momentum tracks not associated with the primary objects are also included \cite{ATLAS-CONF-2018-023}.  
These experimental uncertainties are correlated between regions and processes, including signal models.
The largest experimental uncertainties are related to the measurement of jets and typically result in an uncertainty in the total background of a few percent, but reach 10\%--20\% in a couple of search regions.
All of the uncertainties quoted below are for the expected yield of the process in analysis regions.

Uncertainties in the predicted cross-sections and modelling of background processes predicted via \ac{MC} simulation are considered.
If the background process is normalized to data, the total cross-section uncertainty is not applied, but instead an uncertainty for the extrapolation from the \ac{CR} to the \ac{SR} is evaluated.
For background processes making large contributions to regions in the searches, acceptance uncertainties due to various choices, such as the \ac{PDF} set and QCD scales, are applied.
For small-background processes, these are ignored and only the total cross-section uncertainty is applied.
In a few cases, negligibly small uncertainties are ignored in order to simplify the statistical evaluation of the level of data-to-background agreement.
As an example, uncertainties due to the assumed value of the strong coupling, $\alphas$, are ignored as they are found to be negligible.
These uncertainties from theoretical predictions are correlated between regions for a given process, but uncorrelated between processes.
 
For the RJR and EWK searches, uncertainties in the \ttbar\ \ac{ME} are evaluated by a comparison of the default \textsc{Powheg} generator and \textsc{MG5\_aMC@NLO}.
The effects of uncertainties in the \ac{PS} are evaluated by comparing the default \Pythia[8] showering with \Herwig[7] showering \cite{Bellm:2015jjp}.
Uncertainties due to the amount of initial- and final-state radiation are evaluated by varying the relevant parameters in the generator.
The effects of uncertainties in the QCD renormalization and factorization scales are evaluated by representing these scale uncertainties by three nuisance parameters in the fit:
two nuisance parameters to independently vary each scale upwards and downwards by a factor of two, and one nuisance parameters for a correlated variation of both.
Finally, uncertainties due to the choice of PDF are evaluated following the recommendations of Ref.~\cite{pdf4lhc}.
The largest \ttbar\ uncertainties in the RJR search regions SR$2\ell$-Low-RJR and SR$2\ell$-ISR-RJR are due to the uncertainty in the \ac{PS}, 20\% and 30\% respectively.
The uncertainty due to the amount of initial-state radiation is also large, 20\% in SR$2\ell$-ISR-RJR.
The largest \ttbar\ uncertainties in the EWK \acp{SR} tend to be from the \ac{ME} comparison and range from 20\% to 60\%.
The \ac{PS} and radiation uncertainties are of similar size.

For the diboson processes, uncertainties from the choice of PDF set and QCD scales are evaluated in the same way as for \ttbar.
Additional uncertainties are evaluated by varying the resummation scale upwards and downwards by a factor of two and by varying the merging scale between 15 and 30~\GeV.
Finally, an alternative \ac{PS} recoil scheme in \textsc{Sherpa} is evaluated, and the difference from the nominal sample is taken as the one standard deviation variation.
For the Strong search, where the diboson processes are not normalized to data, a 6\% uncertainty \cite{ATL-PHYS-PUB-2016-002} is applied to the total cross-section.
The uncertainty due to the choice of various scales is generally the largest one for the diboson backgrounds
and varies from 10\% to 30\% depending on the region.
 
For the EWK and Strong searches, the \zjets\ processes are also subject to the same uncertainties from the choice of PDF, QCD scales, resummation scale, and merging scale as the diboson processes.
An additional uncertainty from the \ac{ME} calculation and is evaluated by comparing the nominal \textsc{Sherpa} sample with an alternative \textsc{MG5\_aMC@NLO} sample.
Uncertainties in the \zjets\ background in the EWK search regions range from 10\% to 20\% from the choice of scale and up to 60\% from the \ac{ME} comparison in regions with at least one predicted \zjets\ event.
The largest uncertainty in the \zjets\ background in the Strong search regions is typically from the \ac{ME} comparison, and ranges from 20\% to 60\% in regions with at least one predicted \zjets\ event.
 
In the SR-$\ell\ell bb$ region of the EWK search a large fraction of the background events are from $t\bar{t}$\,+\,$Z$.
For $t\bar{t}$\,+\,$Z$ in the EWK search regions, uncertainties from the \ac{ME} and \ac{PS} are evaluated by comparisons with a \textsc{Sherpa} sample and a \textsc{MG5\_aMC@NLO}\,+\,\Herwig[7] sample respectively.
The same renormalization scale, factorization scale, and PDF uncertainties described above are also evaluated.
The largest uncertainty is from the \ac{ME} comparison and ranges from 10\% to 50\%, depending on the region.
In addition, a 6\% uncertainty on the $Wt$ background is applied.
 
Uncertainties in the cross-sections of smaller backgrounds estimated directly from \ac{MC} simulation are applied as follows.
A conservative uncertainty of 10\% is applied to the entire Higgs boson sample, based on the $t\bar{t} H$ cross-section uncertainty in Ref.~\cite{deFlorian:2016spz}.
The Other Top sample is predominantly $t\bar{t}$\,+\,$V$, so a cross-section uncertainty of 13\%, also from Ref.~\cite{deFlorian:2016spz}, is applied to this background category.
An uncertainty of 32\% is applied to the $VVV$ background, based on comparisons between \textsc{Sherpa} and \textsc{VBFNLO} \cite{ATL-PHYS-PUB-2016-002}.
 
For data-driven background estimates, uncertainties in the prediction are derived from the limited number of events used in the prediction and by varying the assumptions used.
In cases where \acp{CR} are used to normalize the yield of a process, theoretical uncertainties are evaluated, as described above, for the acceptance of the \ac{CR}  relative to the \ac{SR}.
 
Uncertainties in the matrix-method estimate of the FNP leptons arise from the size of the samples used and the composition of the FNP-lepton sources, and the total uncertainty
of the estimate is obtained by combining uncertainties from several sources.
The statistical uncertainties arise from the limited number of events in regions used to derive the various efficiencies and are propagated to the final estimate.
The variations of the lepton efficiency corrections are propagated to the light-flavour efficiencies because \ac{MC} simulation is used in the derivation.
For heavy-flavour efficiencies, an uncertainty in the prompt-lepton contamination from \ac{MC} simulation is taken into account.
Finally, an uncertainty in the estimated proportions of light-flavour jets, heavy-flavour jets, and photon conversions contributing to the processes producing FNP leptons is included.
This is computed by varying the composition, using \ac{MC} simulation, of the regions with selections similar to various analysis regions used to construct the weights.
In cases where the number of events in a region is not sufficient to predict the FNP lepton contribution with the matrix method, the central value and uncertainty are taken from the larger of two computations.
The first computation is the mean of the weights in the matrix method with the root mean square of these weights as the uncertainty.
This would be the average prediction if there were one event for the matrix method.
The second computation is the FNP-lepton yield predicted by MC simulation with a 100\% uncertainty.
These extra computations of the FNP-lepton background are only used in the EWK search, where the search regions have small yields.

For the flavour-symmetric background estimate in the Strong search regions, the uncertainties due to a limited number of events in the \ac{DF} regions and the regions used to derive the correction factors are propagated to the final estimate.
An additional uncertainty of 10\% is applied to the estimate, based on closure tests.
The closure tests were performed on both data and MC events in the region used to derive the weights, with \met\ restricted to be between 100 and 200~\GeV, and on MC events in the \acp{SR}.
Finally, for the \onz\ regions where the \mll\ window used to collect \ac{DF} events is expanded, an additional uncertainty is applied to account for the limited number of events in the \ac{MC} simulation used to evaluate this extrapolation.
 
Uncertainties are applied to the estimate of \zjets\ with the ABCD method in the RJR search to account for the limited number of events in the regions used to extract the estimate and the theoretical and experimental uncertainties, discussed above, in the non-\zjets\ \ac{MC} simulation backgrounds.
The correlations between the ABCD variables are found to be very small and the related systematic uncertainties are treated as negligible.
 
For signal models with the production of squarks or gluinos, the nominal cross-section and its uncertainty due to scale variations and the choice of PDF are determined as described in Ref.~\cite{Beenakker:2016lwe}.
For signal models containing the direct production of electroweakinos, the cross-section and its uncertainty are taken from Ref.~\cite{Fiaschi:2018hgm}.
The uncertainties in the acceptance of signal model events, from scale variations and the \ac{PS} tuning and radiation, are estimated and summed in quadrature.

The most important uncertainties in the total background estimate across the various searches are related to the theoretical prediction of the largest backgrounds (e.g.\ diboson and \ttbar\ production), experimental uncertainties related to jets, and the number of simulated \ac{MC} events for the estimation of \ac{SM} processes.
The most important uncertainties in the RJR search \acp{SR} (SR$2\ell$-Low-RJR / SR$2\ell$-ISR-RJR), relative to the total expected background, are from the normalization of \zjets\ via the ABCD background estimate (19\%\,/\,28\%), the size of the simulated samples (7\%), and the matching scale for the diboson processes (5\%\,/\,6\%).
The largest jet-related experimental uncertainty is a component of the \ac{JER} (1\%) in SR$2\ell$-Low-RJR, and the $b$-jet tagging efficiency (2\%) in SR$2\ell$-ISR-RJR.
The sizes of the uncertainties in the total background, grouped into categories after fits to the \acp{SR}, are summarized in Figure~\ref{subfig:rjr_sys}.
 
The most important uncertainties in the EWK search \acp{SR} depend on the region.
The largest uncertainties in the total background in the SR-High-EWK regions arise from the size of the simulated samples (4\%--23\%), the prediction of the fake and non-prompt leptons (3\%--17\%), and in SR-1J-High-EWK the uncertainty in the jet mass measurement (37\%).
The largest jet-related experimental uncertainties in regions that do not use the jet mass are 2\%--3\%.
In SR-Int\_a-EWK, the dominant uncertainty is from the amount of final-state radiation in the simulation of \ttbar\ production (16\%).
In SR-Int\_b-EWK, the dominant uncertainty is related to the \ac{JER}, with several components in the 4\%--6\% range.
In the SR-Low-EWK regions, the largest uncertainties are from the size of the simulated samples (6\%--35\%), the \ac{JER} (6\%--18\%), and the choice of generator for the \zjets\ \ac{ME} (13\%--23\%).
In the SR-OffShell-EWK and SR-$\ell\ell bb$-EWK regions, the largest uncertainties arise from the size of the simulated samples (7\%--12\%), the choice of generator for the \zjets\ \ac{ME} (4\%--10\%), and the \ac{JER} (4\%--11\%).
Additionally, in SR-$\ell\ell bb$-EWK, the uncertainties from the \ttbar\ \ac{PS} and \ac{ME} calculation are 12\% for each.
The sizes of the uncertainties in the total background, grouped into categories after a fit to the regions, are summarized in Figure~\ref{subfig:ewk_sys}.
 
The most important uncertainties in the Strong search \acp{SR} are those from the various scales used in the simulation of diboson production and the \ac{PS} recoil scheme,
typically 3\%--6\% in the edge \acp{SR} and up to 18\% in the on-$Z$ \acp{SR}; the non-closure of the flavour-symmetric background estimate (2\%--6\%) in the edge \acp{SR}; the \ac{ME} generator for \zjets, up to 7\% in SRZ-High-STR; and the size of the simulated samples (2\%--11\%).
The largest jet-related experimental uncertainties are 1\%--5\% from uncertainties in the parton origin, or flavour, of the jet, depending on the region.
The sizes of the uncertainties in the total background, grouped into categories after fits to the \acp{SR}, are summarized in Figure~\ref{subfig:strong_sys}.

\begin{figure}[htbp]
\centering
\subfloat[]{\includegraphics[width=0.49\textwidth]{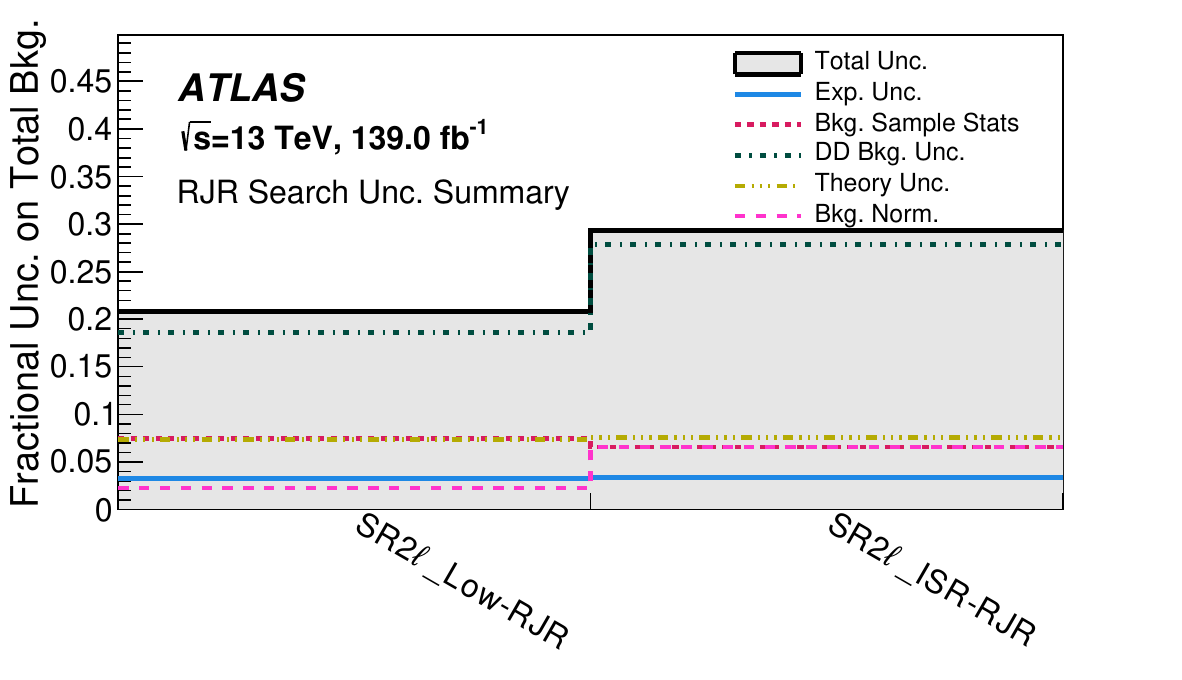}\label{subfig:rjr_sys}}
\subfloat[]{\includegraphics[width=0.49\textwidth]{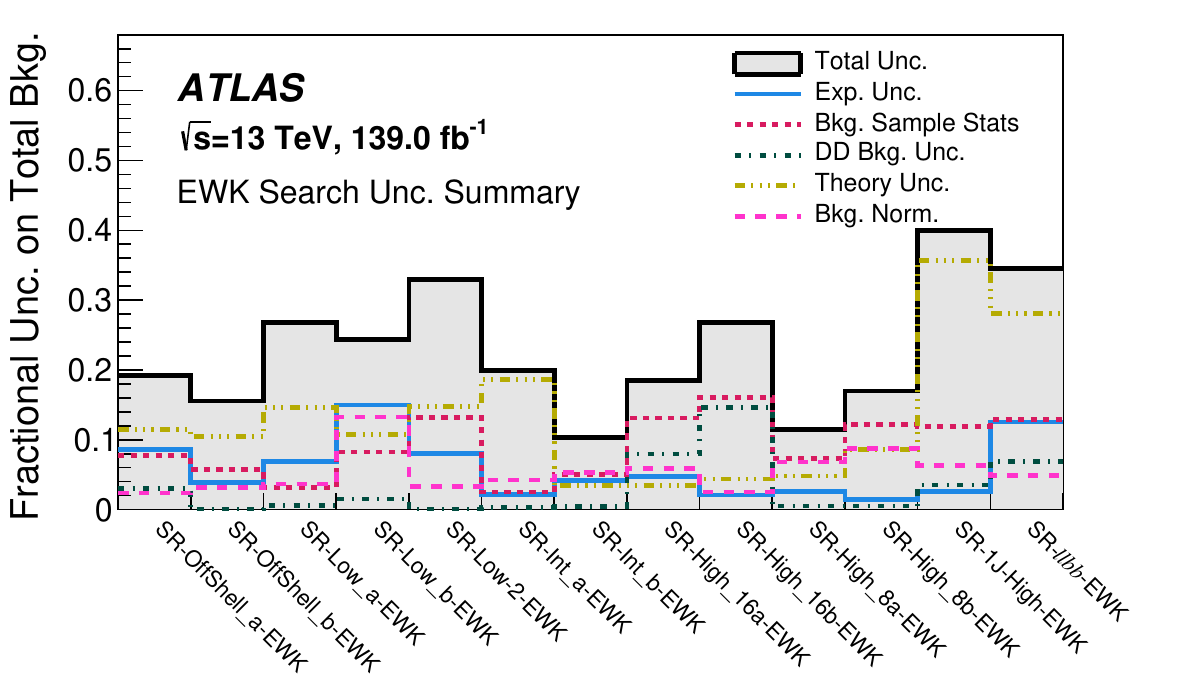}\label{subfig:ewk_sys}}\\
\subfloat[]{\includegraphics[width=0.49\textwidth]{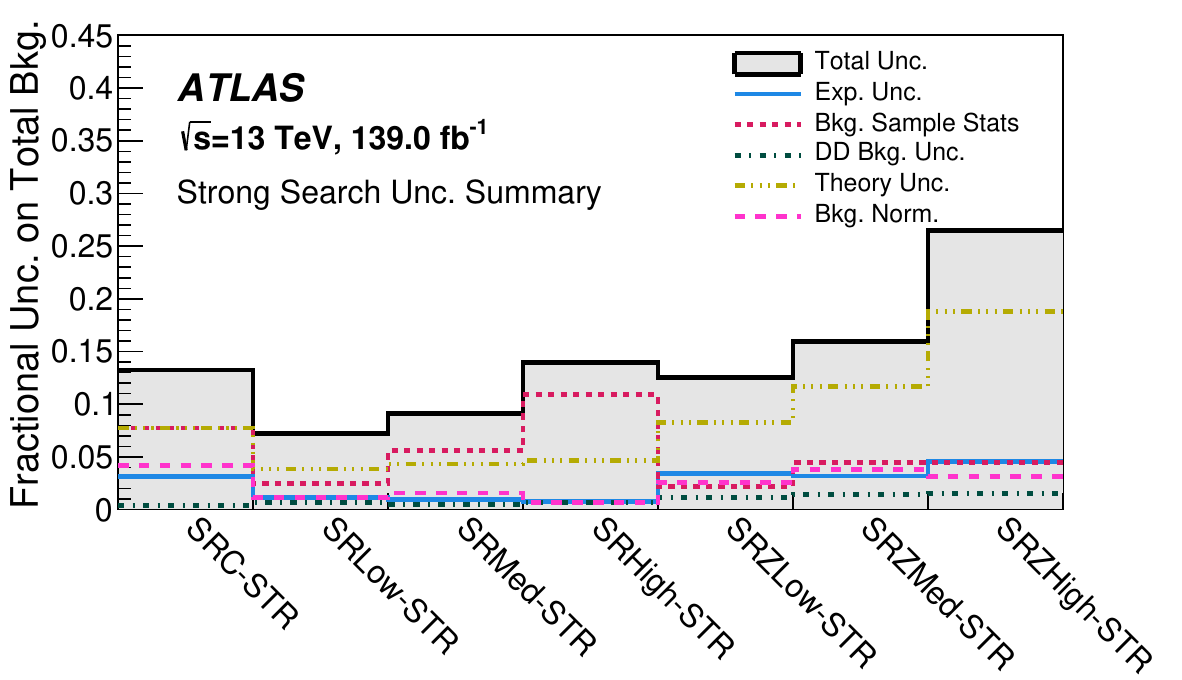}\label{subfig:strong_sys}}
\caption{Breakdown of the uncertainties in each of the (a) RJR, (b) EWK, and (c) Strong searches into categories.
The total fractional uncertainty of the background is shown as a shaded histogram.
The experimental uncertainties include those from object reconstruction and identification efficiencies, object resolutions and scales, pile-up distribution, and trigger efficiencies.
The background sample statistical uncertainty includes the uncertainty in the number of \acl{MC} and data-driven background events.
The data-driven background uncertainty includes additional uncertainties, e.g.\ the uncertainty in the \ac{FNP}-lepton estimate.
The theory uncertainty includes any theory uncertainties in background estimates.
The background normalization category includes the uncertainty associated with the normalization factors derived from \acp{CR}.
}
\label{fig:sys_summtary}
 
\end{figure}


\FloatBarrier
\section{Results}
\label{sec:result}

The data are compared with \ac{SM} predictions using the profile likelihood method \cite{Cowan:2010js}.
The HistFitter framework \cite{Baak:2014wma}, built on RooFit, RooStats, HistFactory, and ROOT \cite{RooFit, Moneta:2010pm, Cranmer:1456844, Brun:1997}, is used to produce the presented statistical results.
Model-independent upper limits on the number of events $(S^{95})$ at 95\% \ac{CL} that could be attributed to \ac{BSM} processes are evaluated using the CL$_\text{s}$ prescription \cite{Read:2002hq}.
Yields are reported after a fit to the background-only model, i.e.\ the zero signal-strength model.
Compatibility tests are reported for the observed data and the background-only hypothesis ($p(s=0)$) and for the observed data and the hypothesis with the signal strength at the 95\% \ac{CL} limit (CL$_\text{b}$).
 
\subsection{Recursive-jigsaw reconstruction search results}

 
A breakdown of the observed and expected yields in the RJR search SRs is shown in Table~\ref{tab:RJR_yields}.
The background yields result from a simultaneous profile likelihood fit of the \acp{CR}, excluding the SRs themselves.
The fits for SR$2\ell$-Low-RJR and SR$2\ell$-ISR-RJR are performed separately and are not combined.
The background estimate agrees with the observed data in both cases.
Figure~\ref{fig:rjr_summary} summarizes the event yields in the \acp{CR}, \acp{VR}, and \acp{SR} of the RJR search.
The asymptotic approximation \cite{Cowan:2010js} is used to set model-independent upper limits on the number of events, or cross-section, which could arise from \ac{BSM} processes in the two SRs.
These are presented in Table~\ref{tab:rjr_discovery}.
The asymptotic approximation was validated against pseudo-experiments.

\begin{table}[htbp]
\centering
\caption{Breakdown of expected and observed yields in the two \acl{RJR} signal regions after a simultaneous fit of the CRs.
The two sets of regions are fit separately.
The uncertainties include both the statistical and systematic sources.}
\label{tab:RJR_yields}
\begin{tabular*}{\textwidth}{@{\extracolsep{\fill}}lP{-1}P{-1}}
\noalign{\smallskip}\hline\noalign{\smallskip}
& \multicolumn{1}{c}{SR$2\ell$-Low-RJR}    &   \multicolumn{1}{c}{SR$2\ell$-ISR-RJR} \\
\noalign{\smallskip}\hline\noalign{\smallskip}
Observed events          & 39     & 30  \\
\noalign{\smallskip}\hline\noalign{\smallskip}
Total expected background events         & 42,9     & 31,9   \\
\noalign{\smallskip}\hline\noalign{\smallskip}
Diboson events      & 10.6,3.4  & 8.9,2.5  \\
Top events          & 3.5,1.7   & 8.2,2.3  \\
\zjets\ events      & 27,8      & 12,9 \\
Other events        & 0.3\rlap{$_{-0.3}^{+0.5}$} & 0.11,0.04 \\
\noalign{\smallskip}\hline\noalign{\smallskip}
\end{tabular*}
\end{table}

\begin{figure}[htbp]
\centering
\includegraphics[width=\textwidth]{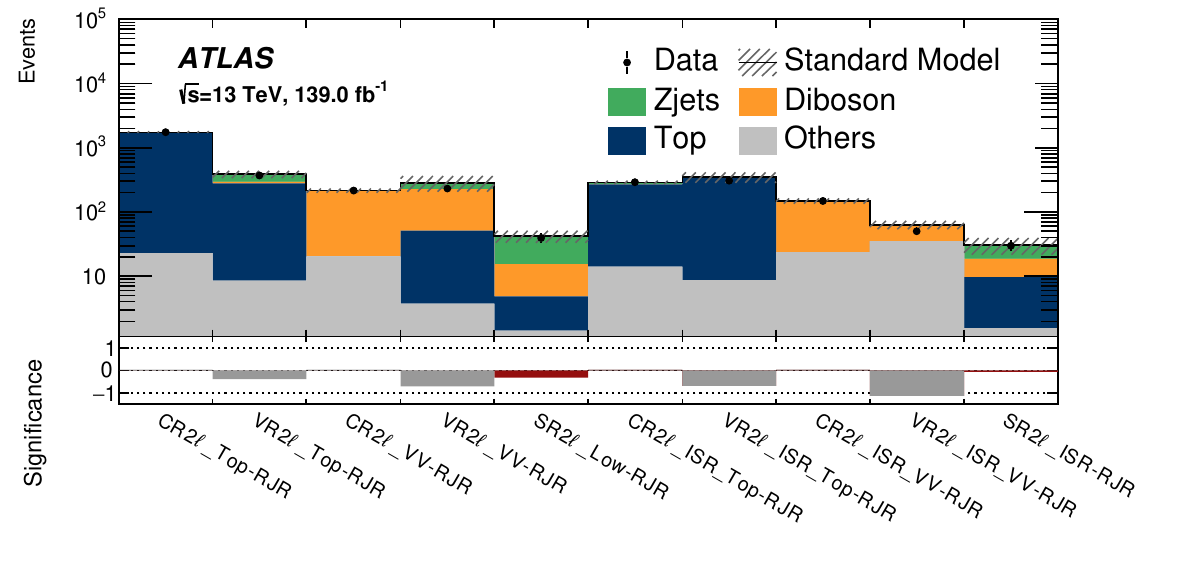}
\caption{The observed and expected yields in the control regions, validation regions, and signal regions of the RJR search after a simultaneous fit of the background-only model to \aclp{CR}.
The standard regions are shown to the left, and the ISR regions are shown to the right.
These two sets of regions are fit separately.
The hatched band includes the statistical and systematic uncertainties of the background prediction in each region.
The significance of the difference between the observed data and the expected yield in each region is shown in the lower panel using the profile likelihood method of Ref.~\cite{Cousins:2007bmb};
the colours black, grey, and red separate the CRs, VRs, and SRs, respectively.
For the cases where the expected yield is larger than the data, a negative significance is shown.
The relevant backgrounds are normalized to data in the CR by the fit and the significances are therefore zero.
}
\label{fig:rjr_summary}
\end{figure}

\begin{table}[htbp]
\centering
\caption{Model-independent upper limits on the observed visible cross-section in the two \acl{RJR} signal regions, derived using the asymptotic approximation.
Left to right: background-only model post-fit total expected background, with the combined statistical and systematic uncertainties; observed data;
95\% CL upper limits on the visible cross-section ($\langle A\epsilon\sigma\rangle_{\text{obs}}^{95}$) and on the number of signal events ($S_{\text{obs}}^{95}$ ).
The sixth column ($S_{\text{exp}}^{95}$) shows the expected 95\% CL upper limit on the number of signal events, given the expected number (and $\pm 1\sigma$ excursions of the expectation) of background events.
The last two columns indicate the confidence level of the background-only hypothesis (CL$_\text{b}$) and discovery $p$-value with the corresponding Gaussian significance ($Z(s=0)$).
CL$_\text{b}$ provides a measure of compatibility of the observed data with the signal strength hypothesis at the 95\% CL limit relative to fluctuations of the background, and $p(s=0)$ provides a measure of compatibility of the observed data with the background-only hypothesis relative to fluctuations of the background.
The $p$-value is capped at 0.5.}
\setlength{\tabcolsep}{0.0pc}
\begin{tabular*}{\textwidth}{@{\extracolsep{\fill}}lP{-1}cccccc l}
\noalign{\smallskip}\hline\noalign{\smallskip}{Signal Region} & \multicolumn{1}{c}{Total Bkg.} & Data & $\langle A\epsilon{ \sigma}\rangle_{\text{obs}}^{95}$ [fb]  &  $S_{\text{obs}}^{95}$  & $S_{\text{exp}}^{95}$ & CL$_\text{b}$& $p(s=0)~(Z)$ \\
\noalign{\smallskip}\hline\noalign{\smallskip}
SR$2\ell$-Low-RJR  & 42,9  & 39 & 0.13 &  17.9 & $ { 20.0 }^{ +7.0 }_{ -5.4 }$ & $0.40$ & 0.50~(0.00)\hphantom{3}\\[.5ex]
SR$2\ell$-ISR-RJR  & 31,9  & 30 & 0.13 &  18.2 & $ { 18.2 }^{ +7.8 }_{ -2.6 }$ & $0.50$ & 0.50~(0.00)\hphantom{3}\\
\noalign{\smallskip}\hline\noalign{\smallskip}
\end{tabular*}
\label{tab:rjr_discovery}
\end{table}


\FloatBarrier
 
\subsection{Electroweak search results}

The yields in the High, $\ell\ell bb$, Int, Low, and OffShell \acp{SR} of the EWK search are listed in Tables~\ref{tab:ewkyieldsSRHigh-STR} and \ref{tab:ewkyieldsSRIntLowOffShell}.
Each SR in the tables is split by the binning described in Section~\ref{sec:ewk_selection}.
For SR-High-EWK the 8 or 16 suffix denotes the $\Delta R_{jj}$ ranges of 0--0.8 and 0.8--1.6 respectively.
Similarly, the a and b suffixes denote the \metsig\ ranges of 18--21 and $>21$ respectively.
For SR-Int-EWK and SR-Low-EWK, the a and b suffixes denote the lower and upper ranges of \metsig\ in the binning $(12,15,18)$ and $(6,9,12)$ respectively.
For SR-OffShell-EWK, the a and b suffixes denote the lower and upper \mll\ ranges of 12--40 and 40--70~\GeV.
The results are extracted after a simultaneous profile likelihood fit to the four \acp{CR} and all of the \acp{SR}.
SR-High-EWK regions have small deficits in the observed data compared to the expected background yield.
No systematic trends are observed in the \acp{VR} around these regions, VR-High-Sideband-EWK and VR-High-R-EWK.
Figure~\ref{fig:ewk_sr} shows distributions in five regions, plotting the variable used for the binning denoted by a and b above.
Example signal distributions are overlaid from models near the edge of sensitivity for this search.
Figure~\ref{fig:ewk_summary} summarizes all of the \acp{CR}, \acp{VR}, and \acp{SR} of the EWK search.
 
In order to perform model-independent tests, the five \ac{SR} categories are merged into single \ac{DR} bins.
DR-High-EWK is SR-High\_8-EWK merged over its \metsig\ bins, and DR-Low-EWK is SR-Low-EWK merged over its \metsig\ bins.
DR-High-EWK has the largest deficit of data compared to the expected background, resulting in a local significance of $-2.8\sigma$.
The resulting upper limits, based on pseudo-experiments, on the number of possible \ac{BSM} events in each region are summarized in Table~\ref{tab:ewk_discovery}.

\begin{table}[htbp]
\centering
\caption{Breakdown of expected and observed yields in the electroweak search High and $\ell\ell bb$ signal regions after a simultaneous fit to the \aclp{SR} and \aclp{CR}.
All statistical and systematic uncertainties are included. }
\label{tab:ewkyieldsSRHigh-STR}
\resizebox{\textwidth}{!}{
\small
\begin{tabular}{l P{-1}P{-1}P{-1}P{-1}}
\noalign{\smallskip}\hline\noalign{\smallskip}
& \multicolumn{1}{c}{SR-High\_16a-EWK} & \multicolumn{1}{c}{SR-High\_8a-EWK}  & \multicolumn{1}{c}{SR-1J-High-EWK} & \multicolumn{1}{c}{SR-$\ell\ell bb$-EWK} \\
\noalign{\smallskip}\hline\noalign{\smallskip}
Observed events           & 4             & 0                & 1         & 0                    \\
\noalign{\smallskip}\hline\noalign{\smallskip}
Total exp.\ bkg.\ events    & 3.9,0.7  & 2.00,0.23  & 0.85,0.34 & 0.58,0.20 \\
\noalign{\smallskip}\hline\noalign{\smallskip}
Diboson events       & 3.2,0.6  & 1.86,0.22    & 0.80,0.31 & 0.13,0.03 \\
Top events           & 0.00\rlap{$^{+0.01}_{-0.00}$} & 0.0,0.0  & 0.03\rlap{$^{+0.04}_{-0.03}$} & 0.05\rlap{$^{+0.08}_{-0.05}$} \\
\zjets events        & 0.0,0.0  & 0.0,0.0      & 0.0,0.0 & 0.0,0.0\\
Other events         & 0.7,0.4  & 0.15,0.07 & 0.02\rlap{$^{+0.04}_{-0.02}$} & 0.39,0.16 \\
\noalign{\smallskip}\hline\noalign{\smallskip}
\noalign{\smallskip}\hline\noalign{\smallskip}
&  \multicolumn{1}{c}{SR-High\_16b-EWK} &  \multicolumn{1}{c}{SR-High\_8b-EWK} &  &  \\
\noalign{\smallskip}\hline\noalign{\smallskip}
Observed events                & 3       & 0         &          &                        \\
\noalign{\smallskip}\hline\noalign{\smallskip}
Total exp.\ bkg.\ events    & 3.4,0.9 & 2.00,0.33 &  & \\
\noalign{\smallskip}\hline\noalign{\smallskip}
Diboson events       & 2.5,0.6  & 1.94,0.33 &  &  \\
Top events           & 0.0,0.0  & 0.0,0.0 &  &  \\
\zjets events        & 0.0,0.0  & 0.0,0.0   &    &  \\
Other events         & 0.9,0.7  & 0.06,0.04 &  &  \\
\noalign{\smallskip}\hline\noalign{\smallskip}
\end{tabular}
}
\end{table}

\begin{table}[htbp]
\centering
\caption{Breakdown of expected and observed yields in the electroweak search Int, Low, and OffShell signal regions after a simultaneous fit to the \aclp{SR} and \aclp{CR}.
All statistical and systematic uncertainties are included.}
\label{tab:ewkyieldsSRIntLowOffShell}
\small
\resizebox{\textwidth}{!}{
\begin{tabular}{l P{-1}P{-1}P{-1} P{-1}}
\noalign{\smallskip}\hline\noalign{\smallskip}
& \multicolumn{1}{c}{SR-Int\_a-EWK}       & \multicolumn{1}{c}{SR-Low\_a-EWK}    & \multicolumn{1}{c}{SR-Low-2-EWK}            & \multicolumn{1}{c}{SR-OffShell\_a-EWK}     \\
\noalign{\smallskip}\hline\noalign{\smallskip}
Observed events           & 24             & 10            & 8   & 6              \\
\noalign{\smallskip}\hline\noalign{\smallskip}
Total exp.\ bkg.\ events    & 22.8,3.5  & 12.8,3.4 & 9,4 & 9.2,1.7\\
\noalign{\smallskip}\hline\noalign{\smallskip}
Diboson events       & 16.5,1.7      & 7.3,1.3  & 4.0,2.1 & 4.9,1.3 \\
Top events           & 4,4           & 0.06\rlap{$^{+0.14}_{-0.06}$}  & 1.0\rlap{$^{+1.2}_{-1.0}$} & 1.4,0.7  \\
\zjets events        & 2.1,0.7       & 3.7,3.3  & 4,4 & 1.2,1.2  \\
Other events         & 0.44,0.13     & 1.7,0.4  & 0.58,0.3 & 1.6,0.4  \\
\noalign{\smallskip}\hline\noalign{\smallskip}
\noalign{\smallskip}\hline\noalign{\smallskip}
& \multicolumn{1}{c}{SR-Int\_b-EWK}             & \multicolumn{1}{c}{SR-Low\_b-EWK}  &        & \multicolumn{1}{c}{SR-OffShell\_b-EWK} \\
\noalign{\smallskip}\hline\noalign{\smallskip}
Observed events                 & 14              & 8        &           & 15         \\
\noalign{\smallskip}\hline\noalign{\smallskip}
Total exp.\ bkg.\ events     & 10.1,1.0 & 10.5,2.5 & & 12.5,1.9\\
\noalign{\smallskip}\hline\noalign{\smallskip}
Diboson events       & 9.2,1.0   & 8.6,1.2 &   & 6.1,1.5\\
Top events           & 0.22,0.13 & 0.0,0.0 &   & 2.8,1.4 \\
\zjets events        & 0.51,0.31 & 1.3\rlap{$^{+2.2}_{-1.3}$} &   & 3.1,1.4 \\
Other events         & 0.19,0.08 & 0.70,0.11 &  & 0.54,0.24 \\
\noalign{\smallskip}\hline\noalign{\smallskip}
\end{tabular}
}
\end{table}

\begin{figure}[htbp]
\centering
\includegraphics[width=0.4\textwidth]{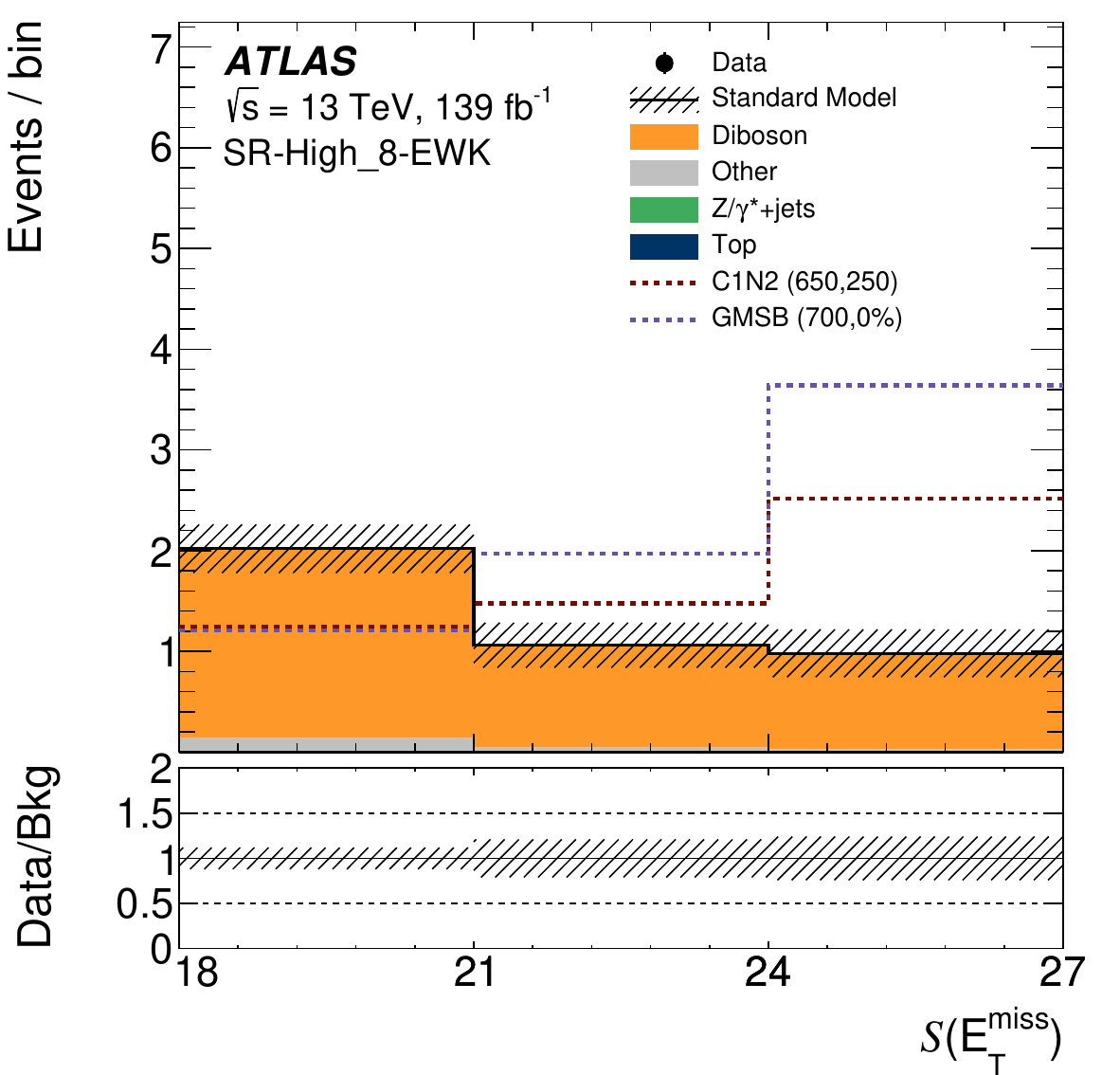}
\includegraphics[width=0.4\textwidth]{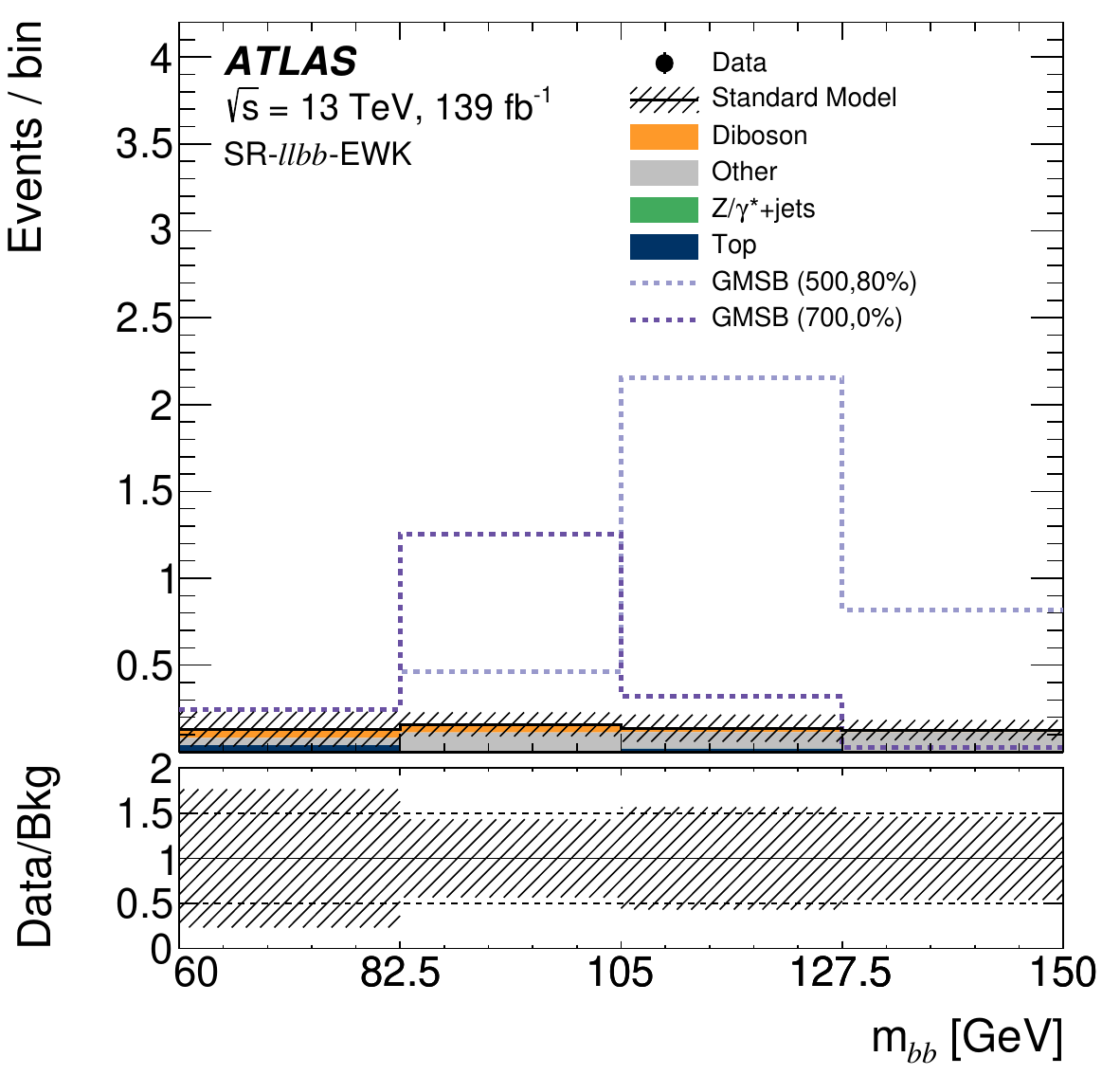}
\includegraphics[width=0.4\textwidth]{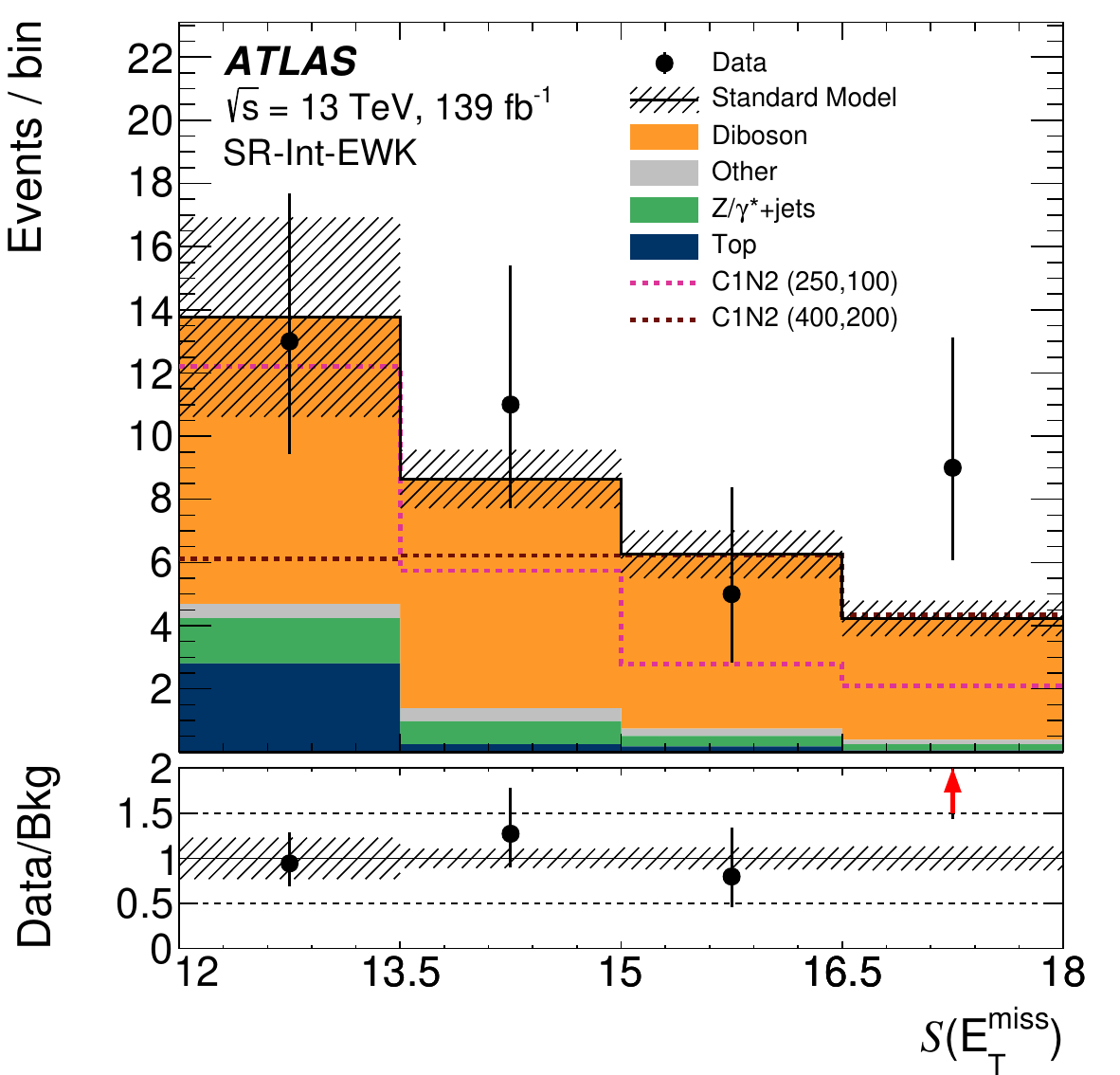}
\includegraphics[width=0.4\textwidth]{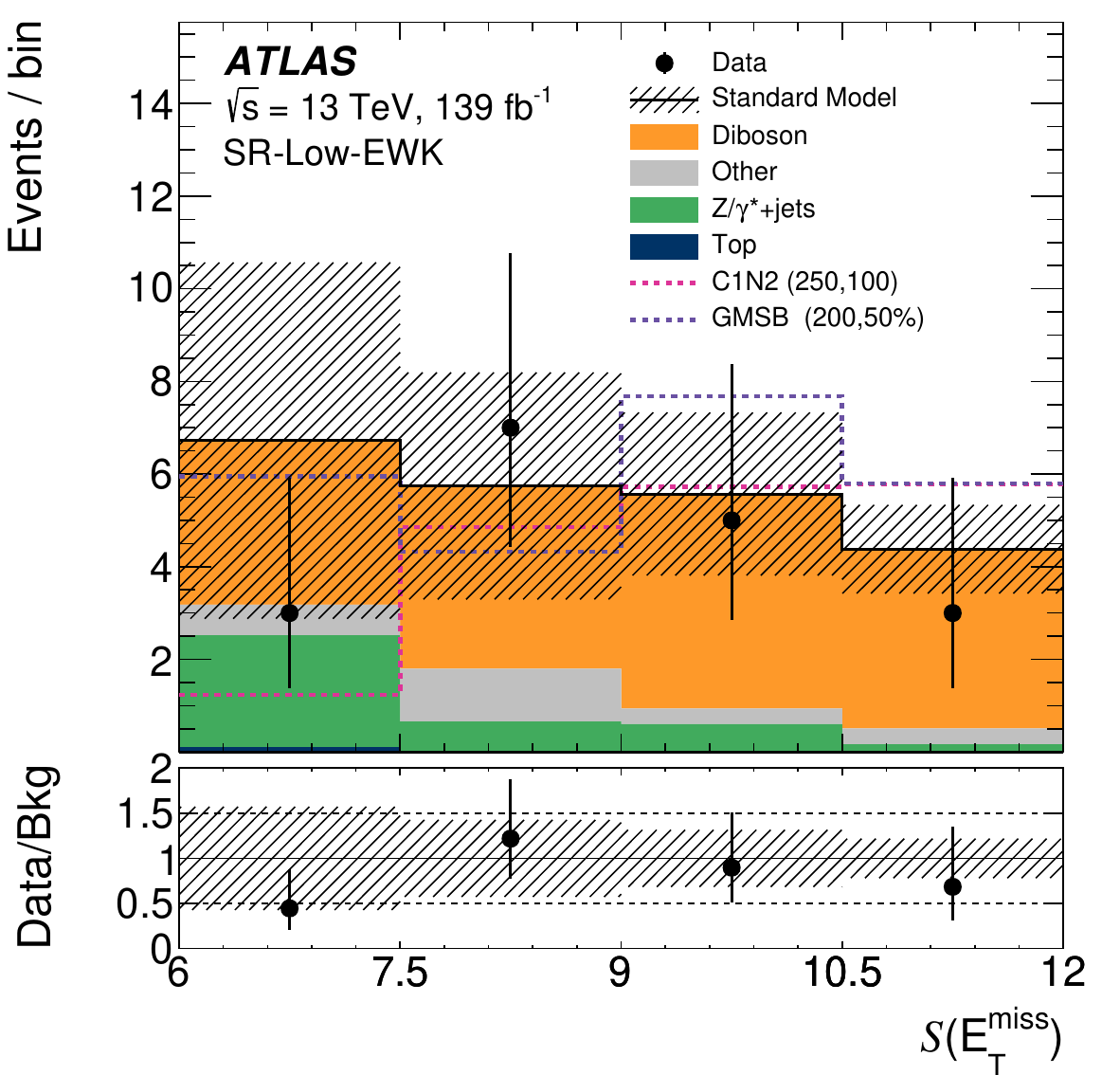}
\includegraphics[width=0.4\textwidth]{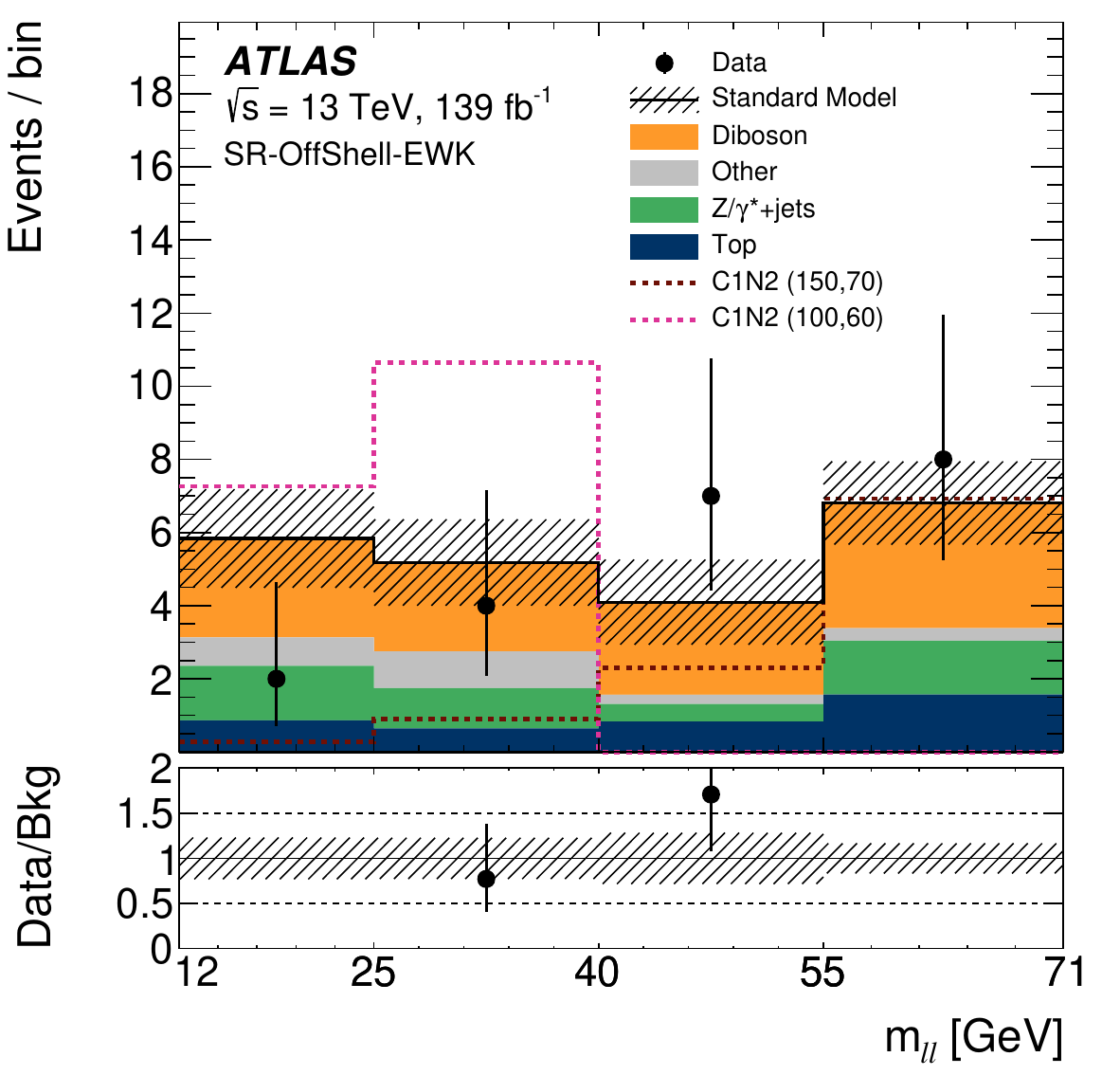}
\caption{Observed and expected distributions in five EWK search regions after a simultaneous fit to the \aclp{SR} and \aclp{CR}.
In the top row, left-to-right, are $\metsig$ in SR-High\_8-EWK and $\mbb$ in SR-$\ell\ell bb$-EWK.
In the middle row, left-to-right, are $\metsig$ in SR-Int-EWK and $\metsig$ in SR-Low-EWK.
In the bottom row is $\mll$ in SR-OffShell-EWK.
Overlaid are example C1N2 and GMSB signal models, where the numbers in the brackets indicate the masses, in \GeV, of the \chionepm\ and \chitwozero\ or the mass of the \chionezero\ and branching ratio to the Higgs boson respectively.
All statistical and systematic uncertainties are included in the hatched bands.
The last bin includes the overflow.}
\label{fig:ewk_sr}
\end{figure}

\begin{figure}[htbp]
\centering
\includegraphics[width=\textwidth]{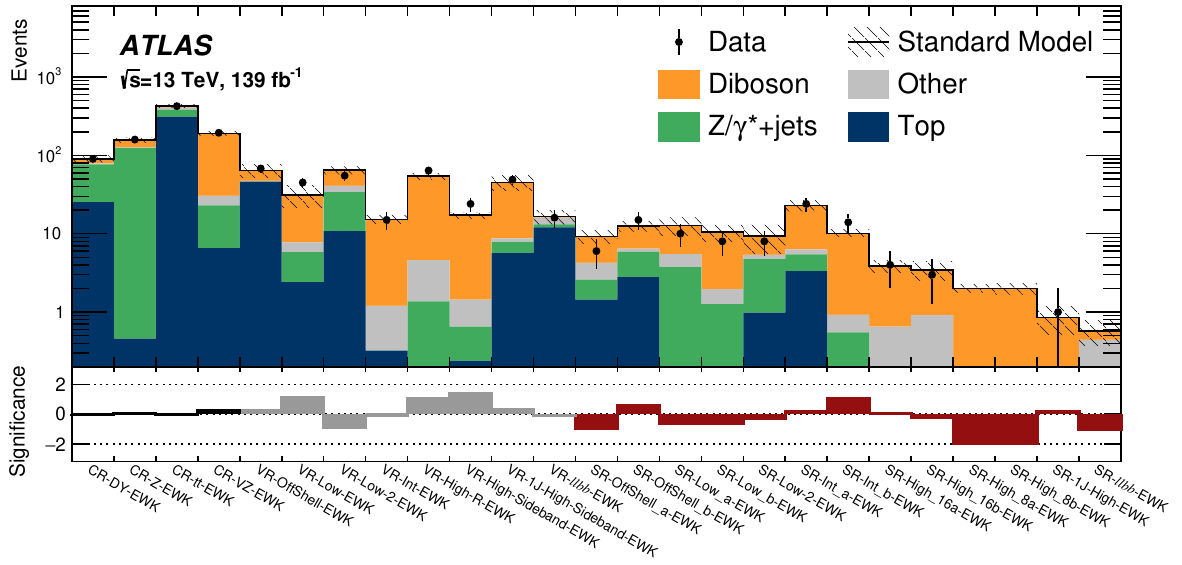}
\caption{The observed and expected yields in the control regions (left), validation regions (middle), and signal regions (right) of the EWK search  after a simultaneous fit to the \aclp{SR} and \aclp{CR}.
The hatched band includes the statistical and systematic uncertainties of the background prediction in each region.
The significance of the difference between the observed data and the expected yield in each region is shown in the lower panel using the profile likelihood method of Ref.~\cite{Cousins:2007bmb};
the colours black, grey, and red separate the CRs, VRs, and SRs, respectively.
For the cases where the expected yield is larger than the data, a negative significance is shown.
}
\label{fig:ewk_summary}
\end{figure}

\begin{table}[htbp]
\centering
\caption{
Model-independent upper limits on the observed visible cross-section in the five electroweak search discovery regions, derived using pseudo-experiments.
Left to right: background-only model post-fit total expected background, with the combined statistical and systematic uncertainties; observed data;
95\% CL upper limits on the visible cross-section ($\langle A\epsilon\sigma\rangle_{\text{obs}}^{95}$) and on the number of signal events ($S_{\text{obs}}^{95}$ ).
The sixth column ($S_{\text{exp}}^{95}$) shows the expected 95\% CL upper limit on the number of signal events, given the expected number (and $\pm 1\sigma$ excursions of the expectation) of background events.
The last two columns indicate the confidence level of the background-only hypothesis (CL$_\text{b}$) and discovery $p$-value with the corresponding Gaussian significance ($Z(s=0)$).
CL$_\text{b}$ provides a measure of compatibility of the observed data with the signal strength hypothesis at the 95\% CL limit relative to fluctuations of the background, and $p(s=0)$ measures compatibility of the observed data with the background-only hypothesis relative to fluctuations of the background.
The $p$-value is capped at 0.5.
}
\label{tab:ewk_discovery}
\setlength{\tabcolsep}{0.0pc}
\begin{tabular*}{\textwidth}{@{\extracolsep{\fill}}lP{-1}P{-1}cP{-1}P{-1}ccc}
\noalign{\smallskip}\hline\noalign{\smallskip}{Signal Region} & \multicolumn{1}{c}{Total Bkg.} & \multicolumn{1}{l}{Data} & $\langle A\epsilon{ \sigma}\rangle_{\text{obs}}^{95}$ [fb]  &  \multicolumn{1}{c}{$S_{\text{obs}}^{95}$}  & \multicolumn{1}{c}{$S_{\text{exp}}^{95}$} & CL$_\text{b}$& $p(s=0)~(Z)$  \\
\noalign{\smallskip}\hline\noalign{\smallskip}
DR-OffShell-EWK      & 22.1,2.7 & 21 & $0.10$ & 14.3 & 12.3\rlap{$^{+4.7}_{-3.1}$} & $0.68$ & $0.50~(0.0)$ \\[.5ex]
DR-Low-EWK           & 22,4 & 18 & $0.08$ & 10.8 & 15.3\rlap{$^{+5.7}_{-4.0}$} & $0.09$ & $0.50~(0.0)$ \\[.5ex]
DR-Int-EWK           & 35,4 & 38 & $0.15$ & 20.9 & 17.5\rlap{$^{+5.9}_{-3.9}$} & $0.73$ & $0.23~(0.8)$ \\[.5ex]
DR-High-EWK          & 3.9,0.5  & 0  & $0.02$ & 3.0 & 5.6\rlap{$^{+2.2}_{-1.5}$} & $0.00$ & $0.50~(0.0)$ \\[.5ex]
DR-$\ell\ell bb$-EWK & 0.51,0.20  & 0  & $0.02$ & 3.0 & 3.0\rlap{$^{+1.3}_{-0.0}$} & $0.19$ & $0.50~(0.0)$ \\[.5ex]
\noalign{\smallskip}\hline\noalign{\smallskip}
\end{tabular*}
\end{table}


\FloatBarrier
 
\subsection{Strong search results}

The integrated data yields in the edge and \onz\ regions are compared to the expected background yields in Tables~\ref{tab:strong_edge_yields} and \ref{tab:strong_onz_yields} respectively.
The \mll\ distributions in the four edge regions are shown in Figure~\ref{fig:strong_SR_mll}.
The yields and distributions are shown after a profile likelihood fit to the SR and its respective CR.
The \ac{SM} expectations agree well with the observed data.
Figure~\ref{fig:strong_summary} summarizes the observed and expected yields in all of the \acp{CR}, \acp{VR}, and \acp{SR} in the Strong search.
 
\begin{table}[htbp]
\centering
\caption{Breakdown of expected and observed yields in the four edge signal regions, integrated over the \mll\ distribution after a separate simultaneous fit to each \acl{SR} and \acl{CR} pair.
The uncertainties include both the statistical and systematic sources.}
\label{tab:strong_edge_yields}
\begin{tabular*}{\textwidth}{@{\extracolsep{\fill}}lP{-1}P{-1}P{-1}P{-1}}
\noalign{\smallskip}\hline\noalign{\smallskip}
& \multicolumn{1}{c}{SRC-STR}    &   \multicolumn{1}{c}{SRLow-STR} & \multicolumn{1}{c}{SRMed-STR} & \multicolumn{1}{c}{SRHigh-STR}            \\
\noalign{\smallskip}\hline\noalign{\smallskip}
Observed events            & 38              &    118  & 65    &  19              \\
\noalign{\smallskip}\hline\noalign{\smallskip}
Total expected background events  & 35,4 &  116,8   & 68,6  & 24.1,3.4   \\
\noalign{\smallskip}\hline\noalign{\smallskip}
$WZ$/$ZZ$ events            & 15.8,3.4        & 76,8     & 25,4     & 6.1,1.6       \\
Flavour-symmetric events & 15.3,2.4        & 23.9,2.8 & 35,4     & 14.5,2.3        \\
\zjets\ events          & 1.8\rlap{$_{-1.8}^{+3.0}$}      & 4.8,2.7  & 2.9,1.7 & 1.7,1.3 \\
Other events            & 1.6,0.4         & 11.2,2.4 & 4.3,1.1  & 1.8,0.5 \\
\noalign{\smallskip}\hline\noalign{\smallskip}
\end{tabular*}

\end{table}
 
\begin{table}[htbp]
\centering
\caption{Breakdown of expected and observed yields in the three \onz\ signal regions after a separate simultaneous fit to each \acl{SR} and \acl{CR} pair.
The uncertainties include both the statistical and systematic sources.}
\label{tab:strong_onz_yields}
\begin{tabular*}{\textwidth}{@{\extracolsep{\fill}}lP{-1}P{-1}P{-1}}
\noalign{\smallskip}\hline\noalign{\smallskip}
&   \multicolumn{1}{c}{SRZLow-STR} & \multicolumn{1}{c}{SRZMed-STR} & \multicolumn{1}{c}{SRZHigh-STR}            \\
\noalign{\smallskip}\hline\noalign{\smallskip}
Observed events            & 35              &  15    &    3               \\
\noalign{\smallskip}\hline\noalign{\smallskip}
Total expected background events  & 33,4  &  15.2,2.4    &  4.5,1.2    \\
\noalign{\smallskip}\hline\noalign{\smallskip}
$WZ$/$ZZ$ events              & 24,4         & 10.0,2.5         & 3.2,1.2              \\
Flavour-symmetric events   & 1.5,0.7      & 2.3,0.4          & 0.32,0.07                \\
\zjets\ events            & 1.7,1.3      & 0.9,0.5          & 0.3\rlap{$_{-0.3}^{+0.4}$} \\
Other events              & 5.1,1.1      & 2.0,0.5          & 0.72,0.14   \\
\noalign{\smallskip}\hline\noalign{\smallskip}
\end{tabular*}
 
\end{table}

\begin{figure}[htbp]
\centering
\includegraphics[width=0.49\textwidth]{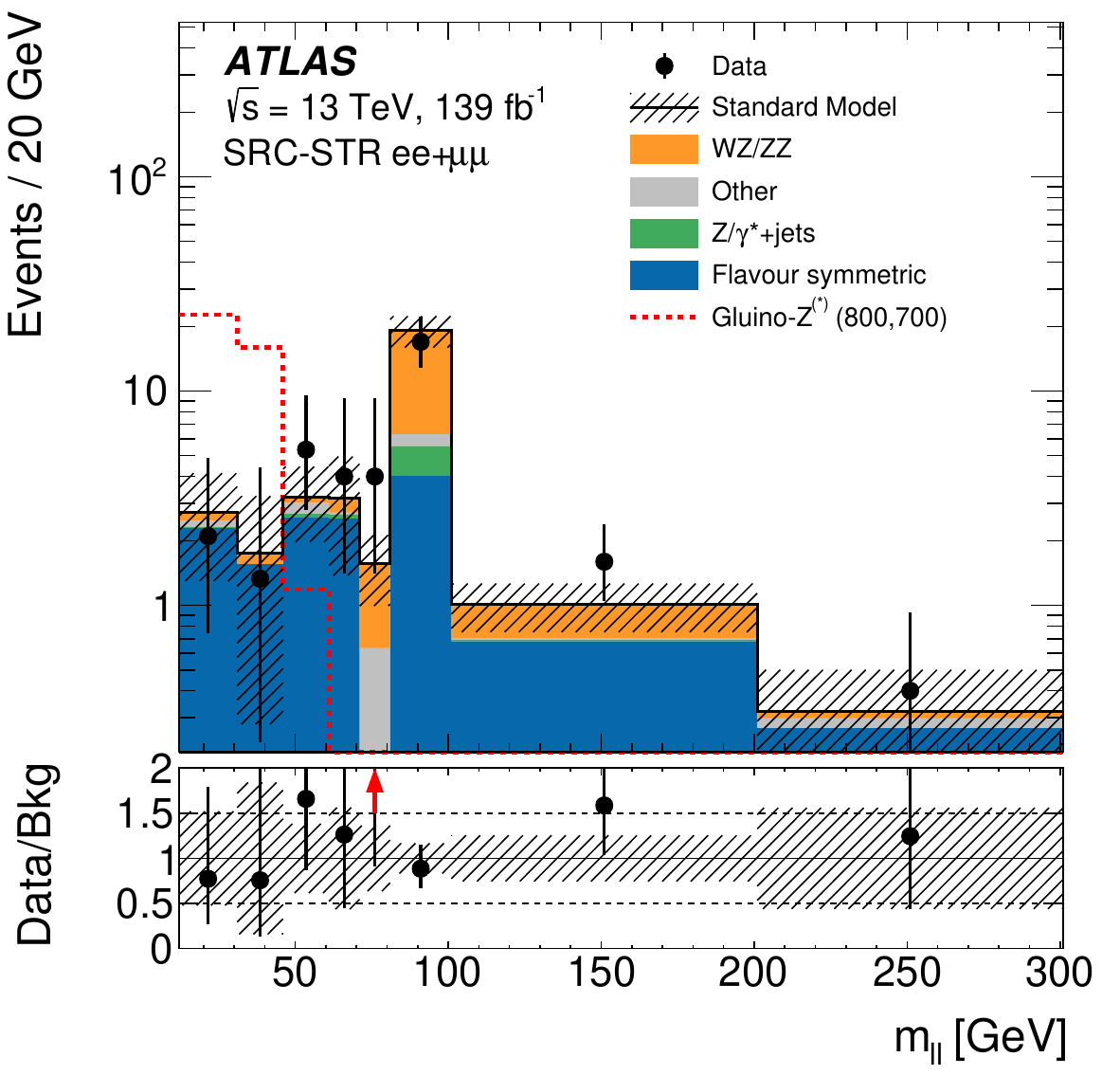}
\includegraphics[width=0.49\textwidth]{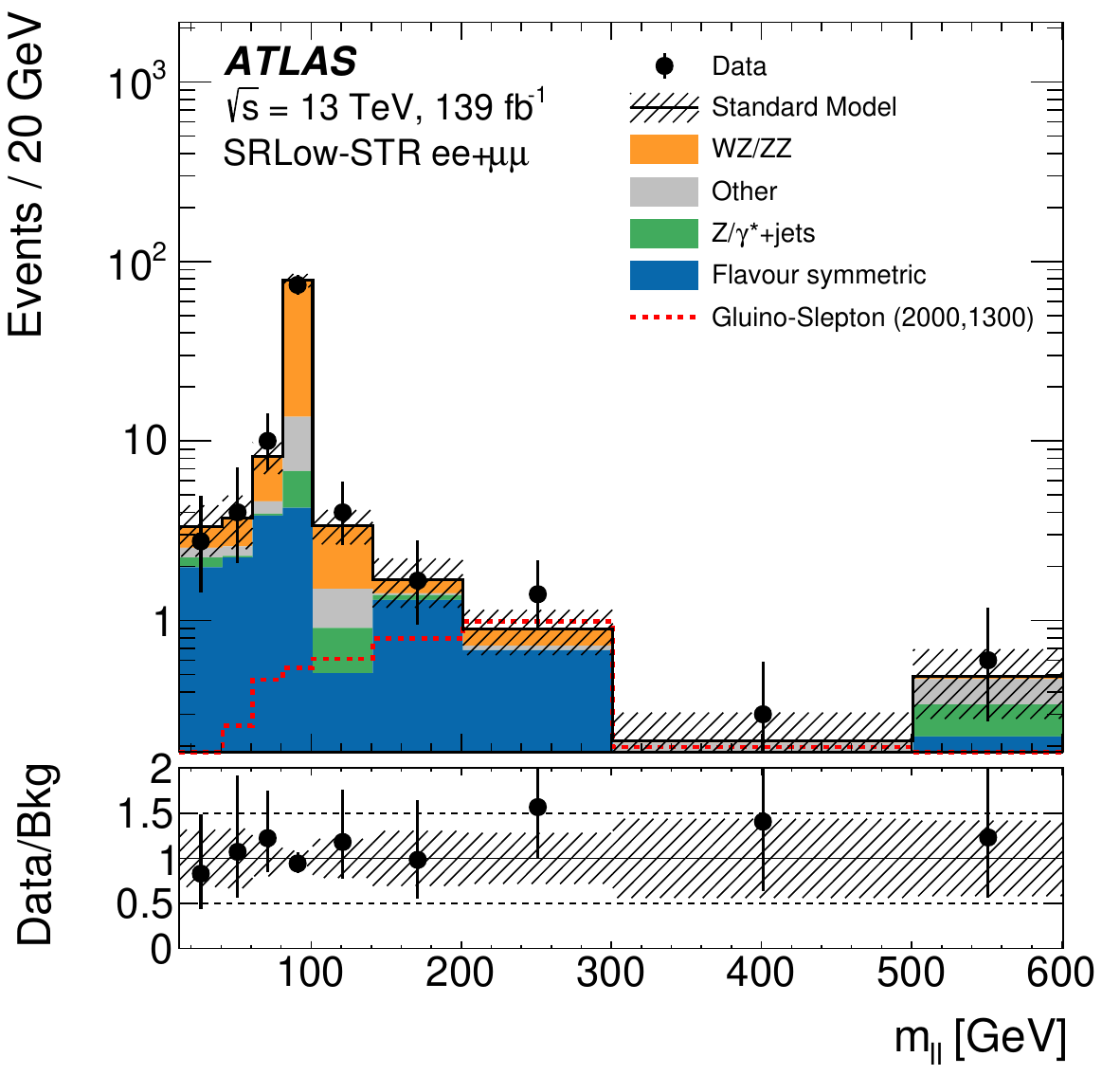}
\includegraphics[width=0.49\textwidth]{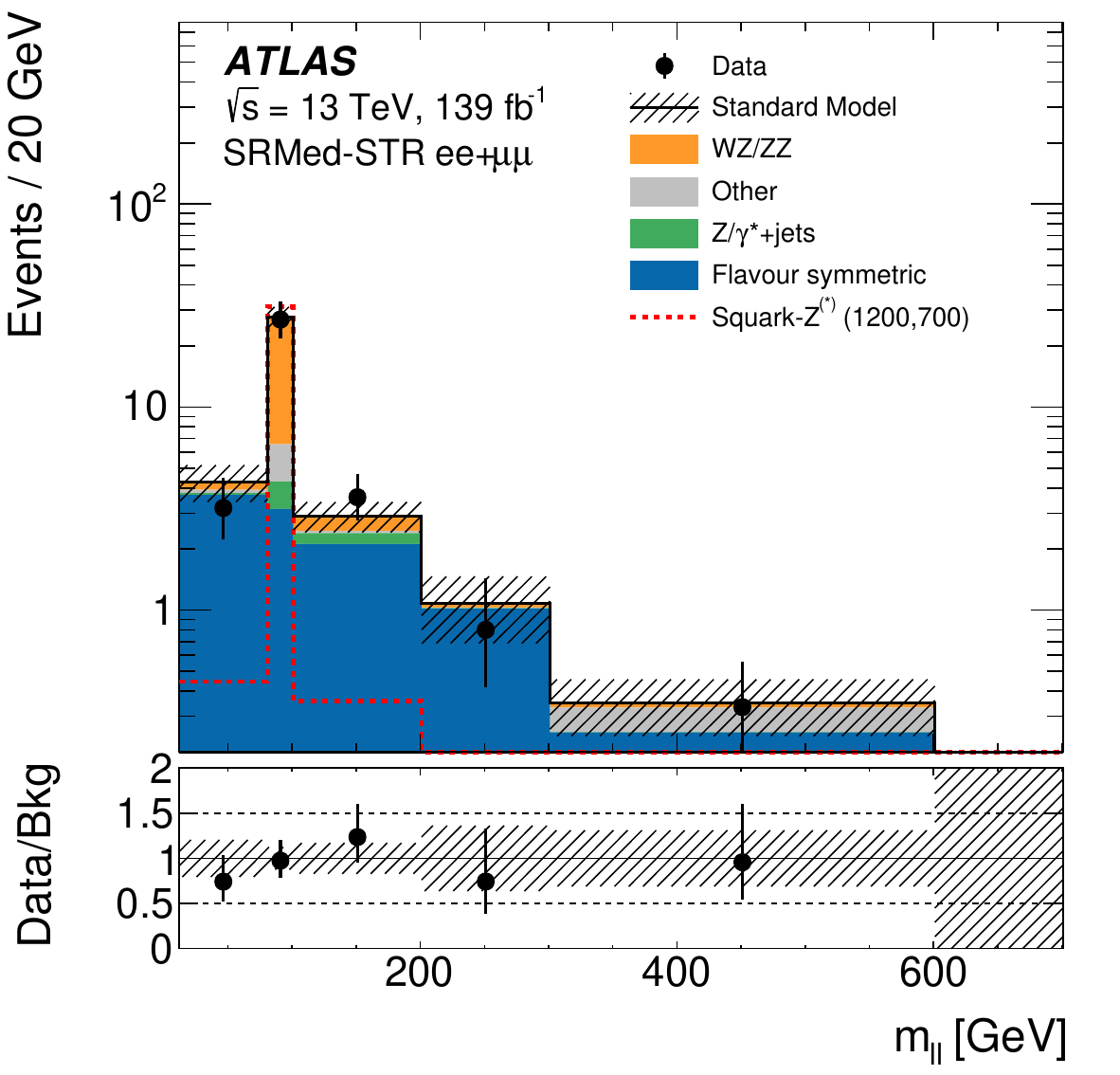}
\includegraphics[width=0.49\textwidth]{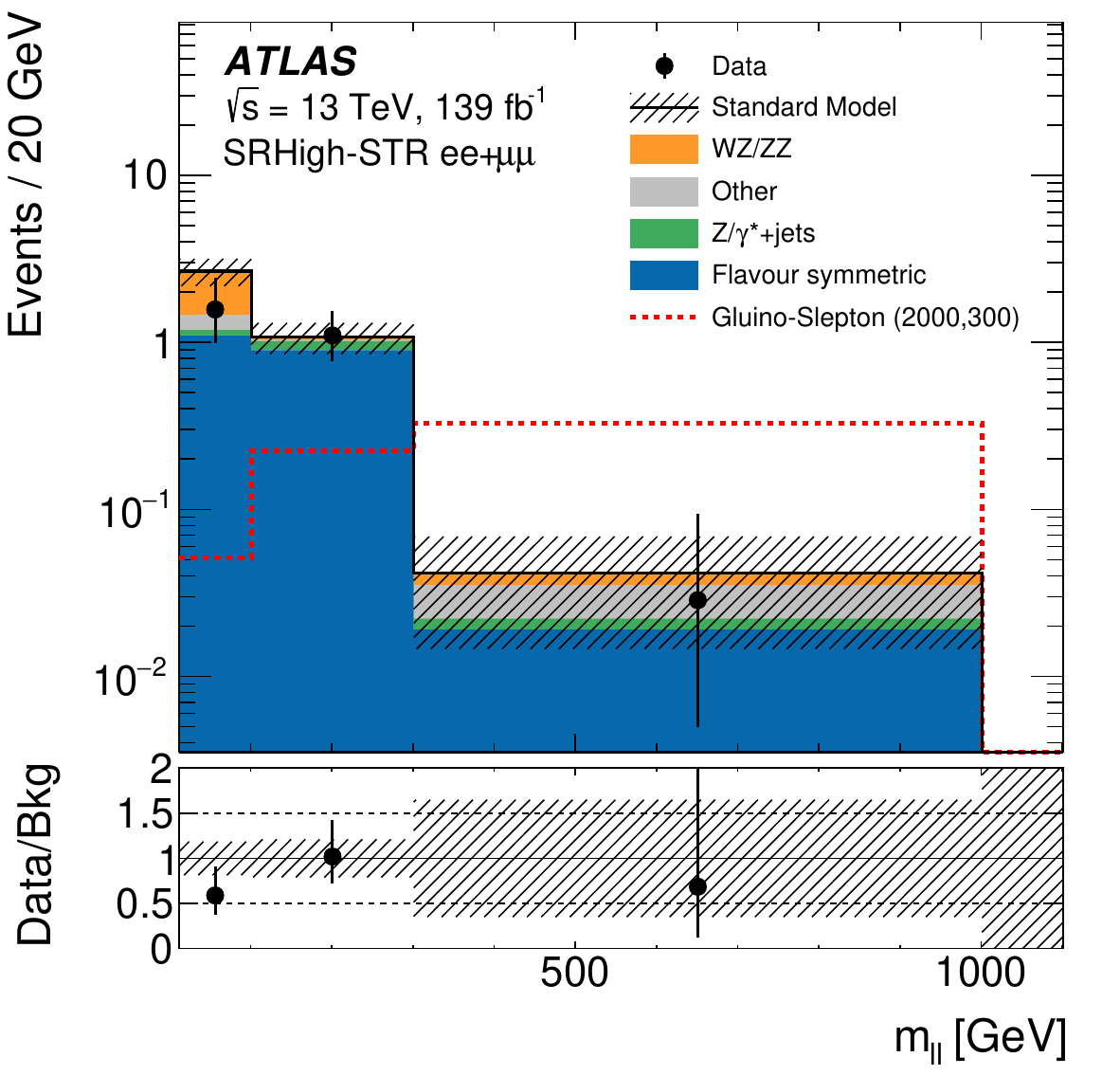}
\caption{Observed and expected dilepton mass distributions in SRC-STR (top-left), SRLow-STR (top-right), SRMed-STR (bottom-left), and SRHigh-STR (bottom-right), with the binning used for interpretations after a separate simultaneous fit to each \acl{SR} and \acl{CR} pair.
The red dashed lines are example signal models overlaid on the figure.
All statistical and systematic uncertainties are included in the hatched bands.
The last bins are the overflow.
}
\label{fig:strong_SR_mll}
\end{figure}

\begin{figure}[htbp]
\centering
\includegraphics[width=\textwidth]{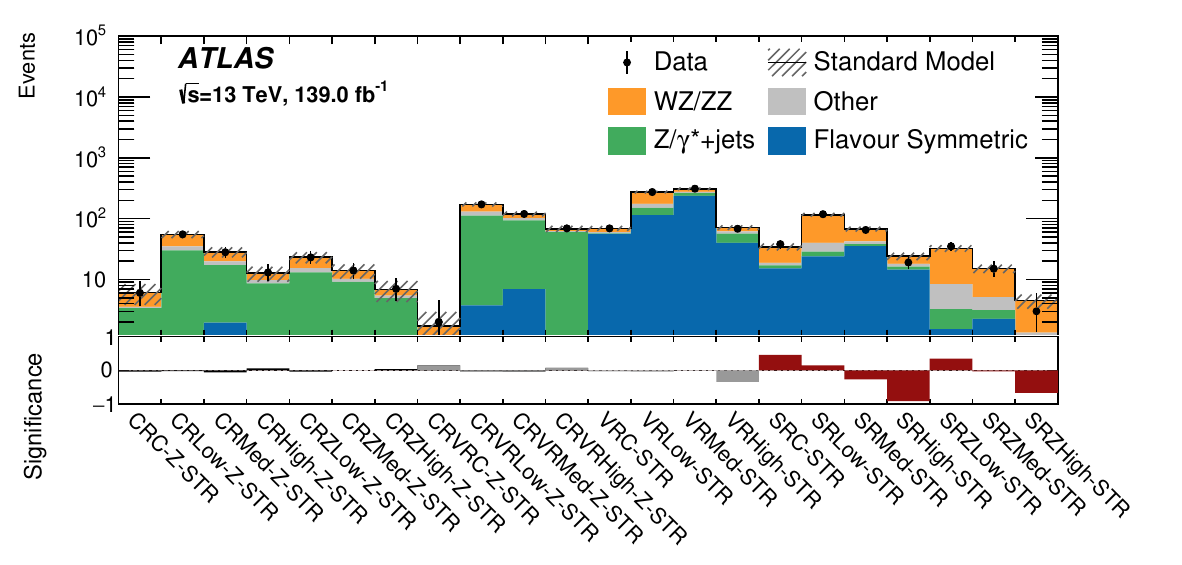}
\caption{The observed and expected yields in the control regions (left), validation regions (middle), and signal regions (right) of the Strong search after a separate simultaneous fit to each \acl{SR} and \acl{CR} pair.
The hatched band includes the statistical and systematic uncertainties of the background prediction in each region.
The significance of the difference between the observed data and the expected yield in each region is shown in the lower panel using the profile likelihood method of Ref.~\cite{Cousins:2007bmb};
the colours black, grey, and red separate the CRs, VRs, and SRs, respectively.
For the cases where the expected yield is larger than the data, a negative significance is shown.
The absence of significant yield differences in the control regions is by construction.
}
\label{fig:strong_summary}
\end{figure}
 
Since the signal models may produce a kinematic edge anywhere along the \mll\ distribution, below the $Z$ boson mass for those with a $Z$ boson in the decay, several \mll\ windows are used to search for an excess of data events above the expectation.
Twelve windows covering various \mll\ ranges in the edge \acp{SR} are considered, along with the three single-bin \onz\ \acp{SR}.
Results in each of these windows are summarized in Table~\ref{tab:strong:UL}.
The data are generally consistent with the background to one standard deviation.
The high \mll\ bins of SRLow-STR have a slightly higher local deviation above the expectation.
The table also includes model-independent limits on the number of possible \ac{BSM} events, as described at the beginning of this section, evaluated with pseudo-experiments.

\begin{table}[htbp]
\centering
\caption{Expected and observed yields in 12 \mll\ windows of the edge signal regions and model-independent upper limits derived using pseudo-experiments.
The \mll\ range is indicated in the leftmost column of the table, with units of \GeV.
Left to right: background-only model post-fit total expected background, with the combined statistical and systematic uncertainties; observed data;
95\% CL upper limits on the visible cross-section ($\langle A\epsilon\sigma\rangle_{\text{obs}}^{95}$) and on the number of signal events ($S_{\text{obs}}^{95}$ ).
The sixth column ($S_{\text{exp}}^{95}$) shows the expected 95\% CL upper limit on the number of signal events, given the expected number (and $\pm 1\sigma$ excursions of the expectation) of background events.
The last two columns indicate the confidence level of the background-only hypothesis (CL$_\text{b}$) and discovery $p$-value with the corresponding Gaussian significance ($Z(s=0)$).
CL$_\text{b}$ provides a measure of compatibility of the observed data with the signal strength hypothesis at the 95\% CL limit relative to fluctuations of the background, and $p(s=0)$ measures compatibility of the observed data with the background-only hypothesis relative to fluctuations of the background.
The $p$-value is capped at 0.5.}
\setlength{\tabcolsep}{0.0pc}
\begin{tabular*}{\textwidth}{@{\extracolsep{\fill}}lP{-1}P{-1}cP{-1}P{-1}cc}
\noalign{\smallskip}\hline\noalign{\smallskip}{ Signal Region}  & \multicolumn{1}{c}{Total Bkg.} & \multicolumn{1}{l}{Data}  & $\langle A\epsilon{ \sigma}\rangle_\text{obs}^{95}$ [fb]  &  \multicolumn{1}{c}{$S_\text{obs}^{95}$}  & \multicolumn{1}{c}{$S_\text{exp}^{95}$} & CL$_\text{b}$ & $p(s=0)$ ($Z$)  \\
\noalign{\smallskip}\hline\noalign{\smallskip}
\noalign{\smallskip}\hline\noalign{\smallskip}
\multicolumn{8}{l}{SRC-STR}\\
\noalign{\smallskip}\hline\noalign{\smallskip}
12--31   & 2.6,1.3 & 2 & $0.03$ &  4.7 & 5.7\rlap{$^{ +2.1 }_{ -1.5 }$} & $0.33$ & $ 0.50$~$(0.0)$ \\[.5ex]
12--61   & 6.3,1.9 & 7 & $0.06$ &  8.4 & 8.3\rlap{$^{ +3.1 }_{ -2.3 }$} & $0.51$ & $ 0.50$~$(0.0)$ \\[.5ex]
31--81   & 6.8,2.8 & 9 & $0.08$ &  11.0 & 8.0\rlap{$^{ +3.2 }_{ -1.7 }$} & $0.83$ & $ 0.18$~$(0.9)$ \\[.5ex]
81--     & 25.4,3.5 & 27 & $0.12$ &  16.4 & 15.3\rlap{$^{ +5.0 }_{ -4.6 }$} & $0.61$ & $ 0.34$~$(0.4)$ \\[.5ex]
\noalign{\smallskip}\hline\noalign{\smallskip}
\multicolumn{8}{l}{SRLow-STR}\\
\noalign{\smallskip}\hline\noalign{\smallskip}
12--81   & 16.8,2.6  & 18 & $0.09$ &  12.6 & 11.6\rlap{$^{ +4.5 }_{ -3.6 }$} & $0.61$ & $ 0.39$~$(0.3)$ \\[.5ex]
101--201 & 12.0,2.3 & 13 & $0.08$ &  11.2 & 9.6\rlap{$^{ +3.8 }_{ -2.5 }$} & $0.67$ & $ 0.30$~$(0.5)$ \\[.5ex]
101--301 & 16.5,2.6 & 20 & $0.11$ &  15.4 & 11.4\rlap{$^{ +4.1 }_{ -3.6 }$} & $0.84$ & $ 0.17$~$(1.0)$ \\[.5ex]
301--    & 4.6,1.4  & 6  & $0.07$ &  9.5  & 6.6\rlap{$^{ +2.8 }_{ -1.8 }$} & $0.85$ & $ 0.09$~$(1.3)$ \\[.5ex]
\noalign{\smallskip}\hline\noalign{\smallskip}
\multicolumn{8}{l}{SRMed-STR}\\
\noalign{\smallskip}\hline\noalign{\smallskip}
12--101  & 42,5  & 38 & $0.10$ &  13.6 & 18.8\rlap{$^{ +7.2 }_{ -5.0 }$} & $0.15$ & $ 0.50$~$(0.0)$ \\[.5ex]
101--    & 26,4 & 27 & $0.12$ &  16.7 & 15.8\rlap{$^{ +6.2 }_{ -4.2 }$} & $0.56$ & $ 0.41$~$(0.2)$ \\[.5ex]
\noalign{\smallskip}\hline\noalign{\smallskip}
\multicolumn{8}{l}{SRHigh-STR}\\
\noalign{\smallskip}\hline\noalign{\smallskip}
12--301  & 22.7,3.2 & 18 & $0.06$ &  8.9 & 13.0\rlap{$^{ +5.8 }_{ -3.0 }$} & $0.11$ & $ 0.50$~$(0.0)$ \\[.5ex]
301--    & 1.5,1.0 & 1 & $0.03$ &  3.7 & 4.4\rlap{$^{ +1.8 }_{ -1.0 }$} & $0.32$ & $ 0.50$~$(0.0)$ \\[.5ex]
\noalign{\smallskip}\hline\noalign{\smallskip}
\noalign{\smallskip}\hline\noalign{\smallskip}
\multicolumn{8}{l}{On-$Z$}\\
\noalign{\smallskip}\hline\noalign{\smallskip}
SRZLow-STR  & 33,4 & 35 & $0.15$ &  20.5 & 17.0\rlap{$^{ +5.4 }_{ -4.5 }$} & $0.76$ & $ 0.25$~$(0.7)$ \\[.5ex]
SRZMed-STR  & 15.2,2.4 & 15 & $0.08$ &  10.7 & 10.3\rlap{$^{ +4.4 }_{ -2.8 }$} & $0.53$ & $ 0.50$~$(0.0)$ \\[.5ex]
SRZHigh-STR & 4.5,1.2 & 3  & $0.04$ &  5.0 & 6.4\rlap{$^{ +2.5 }_{ -1.9 }$} & $0.24$ & $ 0.50$~$(0.0)$ \\[.5ex]
\noalign{\smallskip}\hline\noalign{\smallskip}
\end{tabular*}
\label{tab:strong:UL}
\end{table}



\FloatBarrier
\section{Interpretation}
\label{sec:interpretation}

 
This section contains the model-dependent interpretations of the EWK and Strong searches.
The RJR search does not interpret any signal models as it is a model-independent follow-up to a previously observed excess.
The exclusion contours are presented on a two-dimensional plane of sparticle masses, or in the case of the GMSB model the branching ratio to the \chionezero\ to a  Higgs boson and gravitino and the mass of the \chionezero.
They are evaluated at 95\% \ac{CL} using the CL$_\text{s}$ method.
The dashed lines represent the expected exclusion contour given the background estimates and is surrounded by a yellow-shaded region identifying the $\pm1\sigma$ variation of the median expected limit.
The red solid line denotes the observed exclusion contours.
Uncertainties in the signal cross-sections are not directly included in this contour, but instead shown as red dashed lines denoting $\pm1\sigma$ variations of the signal cross-section.
Grey-shaded areas cover regions excluded by previous similar analyses.
 
\subsection{Electroweak search interpretation}

This section presents mass and branching fraction exclusion limits for the C1N2 and GMSB models described in Section~\ref{sec:susy}.
All of the \acp{SR} and \acp{CR} are fitted simultaneously in order to test each model.
 
The left panel of Figure~\ref{fig:ewk_limits} shows the exclusion contours in the $m_{\chionepm,\chitwozero}$--$m_{\chionezero}$ plane for the C1N2 model.
Masses of the \chitwozero\ and \chionepm\ up to 820~\GeV\ (740~\GeV\ expected) are excluded for \chionezero\ masses up to 375~\GeV.
In the region where the $Z$ boson decay is off-shell at $m_{\chitwozero}=m_{\chionepm}=100~\GeV$, masses of the \chionezero\ up to 100~\GeV\ are excluded.
The observed limit at high \chitwozero\ and \chionepm\ masses is stronger than expected due to the data event deficits observed in SR-High-EWK.
 
The right panel of Figure~\ref{fig:ewk_limits} shows the exclusion contours in the $m_{\chionezero}$--$B(\chionezero\rightarrow h \tilde G)$ plane.
For a branching fraction entirely to $Z$ bosons, masses of the \chionezero\ up to 900~\GeV\ (800~\GeV\ expected) are excluded.
Branching fractions up to 95\% (85\% expected) of the \chionezero\ to Higgs bosons are excluded for \chionezero\ masses around 450~\GeV.
Sensitivity to models with decays entirely through Higgs bosons is limited since the majority of Higgs boson decays do not produce a dilepton system with a mass around the $Z$ boson mass.
The observed limit at moderate to high \chionezero\ mass is stronger than expected due to the deficits observed in SR-$\ell\ell bb$-EWK and SR-High-EWK.

\begin{figure}[htbp]
\centering
\includegraphics[width=0.49\textwidth]{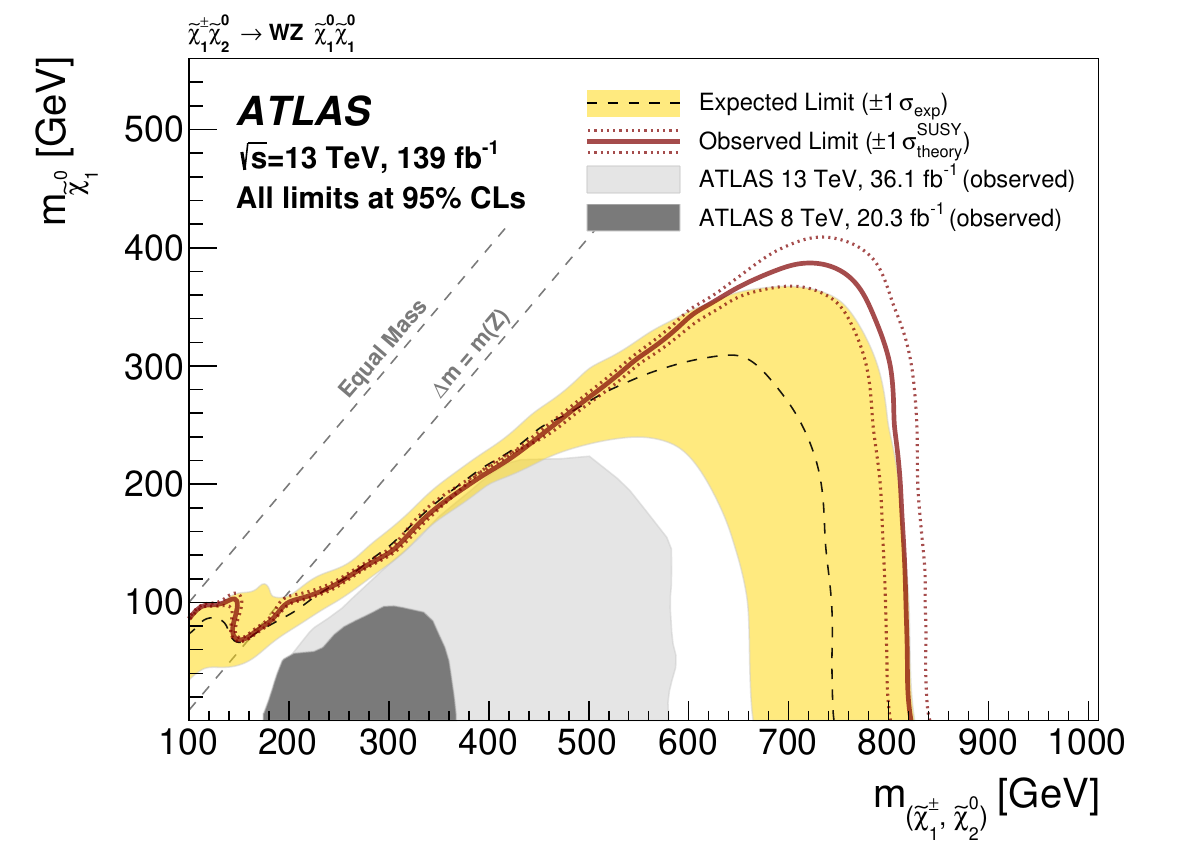}
\includegraphics[width=0.49\textwidth]{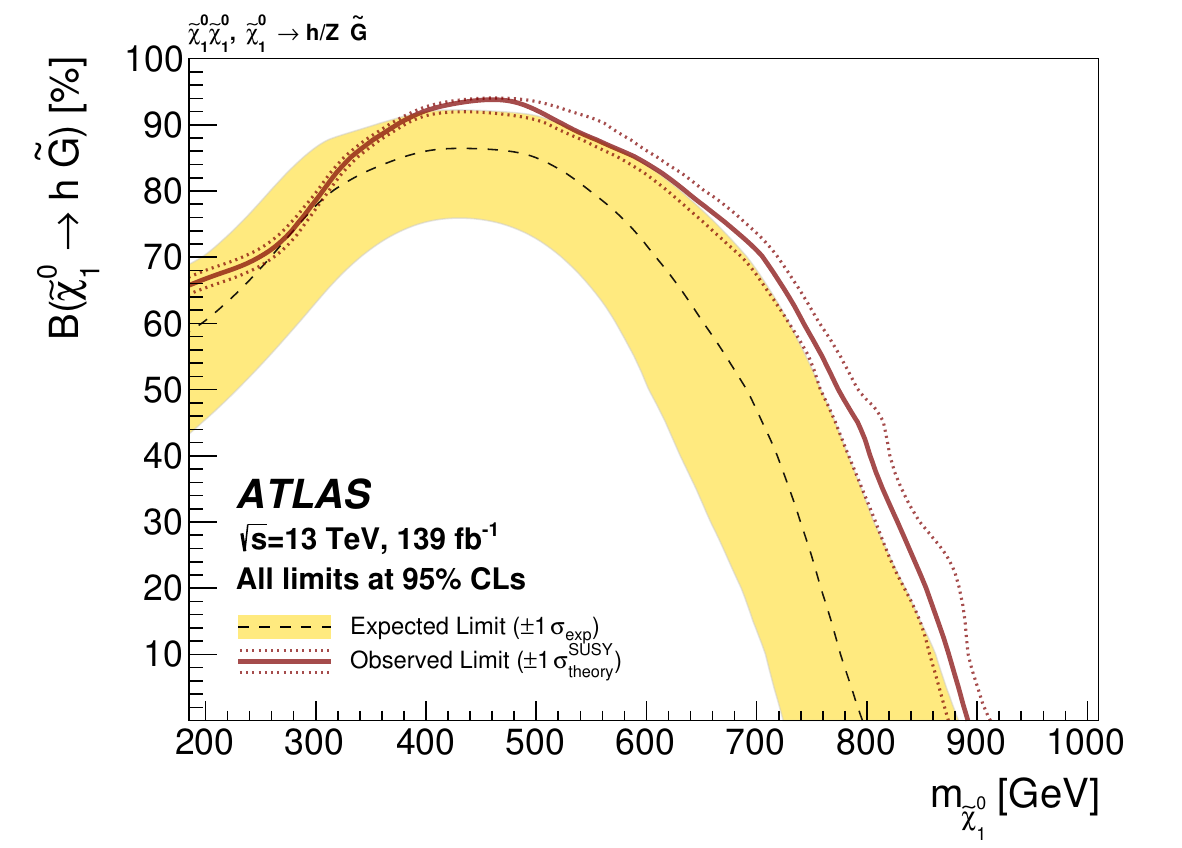}
\caption{Expected and observed exclusion contours from the EWK analysis for the C1N2 model (left) and GMSB model (right).
The dashed line indicates the expected limits at 95\% CL and the surrounding band shows the $1\sigma$ variation of the expected limit as a consequence of the uncertainties in the background prediction and experimental uncertainties of the signal ($\pm1\sigma_\text{exp}$).
The red dotted lines surrounding the observed limit contours indicate the variation resulting from changing the signal cross-section within its uncertainty ($\pm1\sigma_\text{theory}^\text{SUSY}$).
The grey-shaded areas indicate observed limits on these models from the two-lepton channels in Ref.~\cite{SUSY-2016-24} and Ref.~\cite{SUSY-2013-11}.
}
\label{fig:ewk_limits}
\end{figure}


\FloatBarrier

\subsection{Strong search interpretation}

This section presents mass exclusion limits for the squark- and gluino-initiated SUSY models described in Section~\ref{sec:susy}.
These model-dependent exclusion limits are computed separately for each SR, and then combined according to which SR has the best expected sensitivity for a particular set of the model parameters.
This can result in sharp contour features where the best expected sensitivity switches between SRs with different data-to-background ratios.
A shape fit of the binned \mll\ distribution and the \zjets\ CR is performed for the four edge SRs, shown in Figure~\ref{fig:strong_SR_mll}.
A fit to the single-bin SR and its \zjets\ CR is performed for three \onz\ regions in Table~\ref{tab:strong_onz_yields}.
 
The first plot in Figure~\ref{fig:strong_limits} shows the exclusion contours in the $m(\gluino)$--$m(\chionezero)$ plane for a simplified model where the gluino decays via sleptons.
The limit is derived from only the four edge SRs.
SRHigh-STR drives the limit for large splittings between the \gluino\ and \chionezero\ masses.
Moving up in \chionezero\ mass, SRLow-STR takes over around $m(\gluino)=2.0~\TeV$, $m(\chionezero)=0.8~\TeV$.
In SRLow-STR, the \mll\ bins above a value of 200~\GeV\ have a slight excess, as seen in Table~\ref{tab:strong:UL}, hence the weaker observed exclusion limit in this part of the combined contour.
The largest excluded gluino mass is observed (expected) to be 2.25 (2.20)~\TeV.
 
The second plot in Figure~\ref{fig:strong_limits} shows the exclusion contours in the $m(\gluino)$--$m(\chionezero)$ plane for a simplified model where the gluino decays via a $Z$ boson.
The limit is derived using the four edge SRs and three \onz\ SRs.
The three kinks in the observed limit, from left to right, are the transition from SRLow-STR to SRZLow-STR, then to SRZMed-STR, and then to SRZHigh-STR.
The largest excluded gluino mass is observed (expected) to be 1.95 (1.90)~\TeV.

The third plot in Figure~\ref{fig:strong_limits} shows the exclusion contours in the $m(\squark)$--$m(\chionezero)$ plane for a simplified model where the squark decays via a $Z$ boson.
As above, all of the SRs are included.
The three kinks in the observed limit, from left to right, are the same transitions as in the $m(\gluino)$--$m(\chionezero)$ contour.
The largest excluded squark mass is observed (expected) to be 1.55 (1.50)~\TeV.

\begin{figure}[htbp]
\centering
\includegraphics[width=0.49\textwidth]{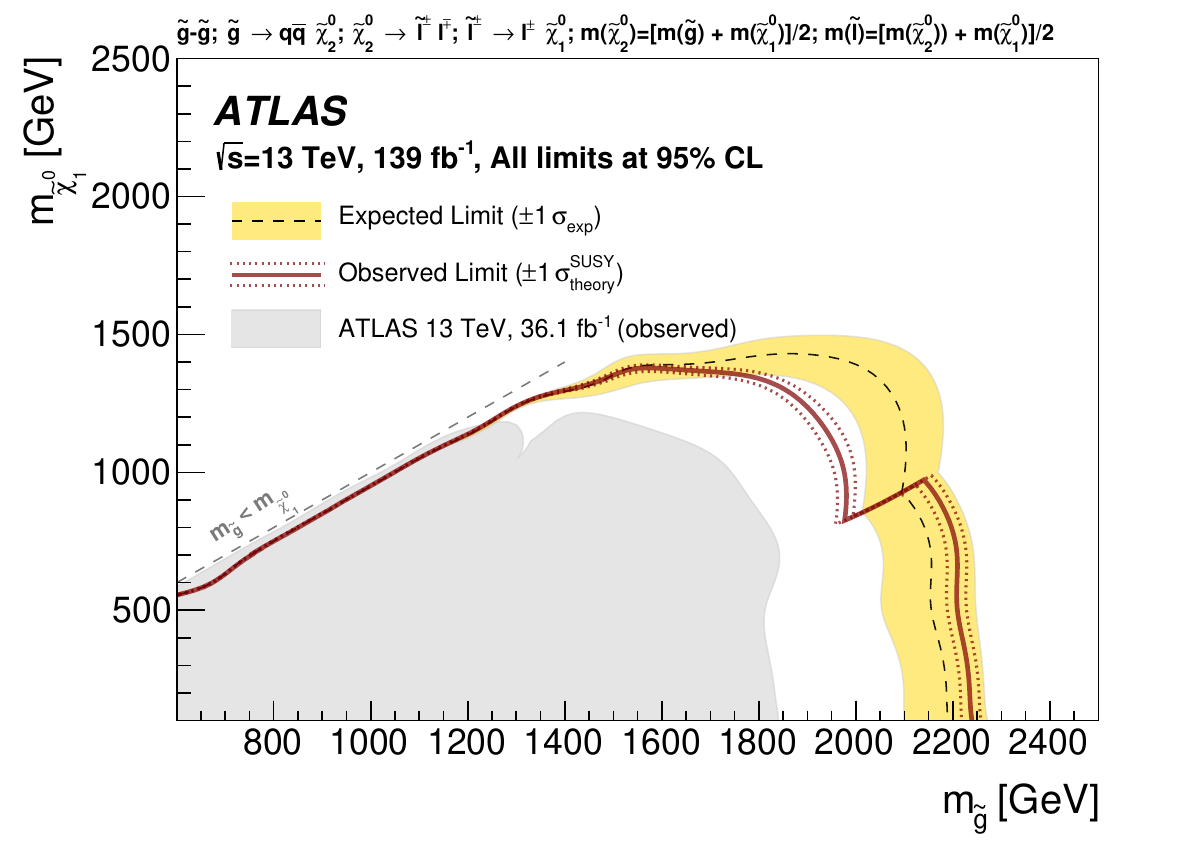}
\includegraphics[width=0.49\textwidth]{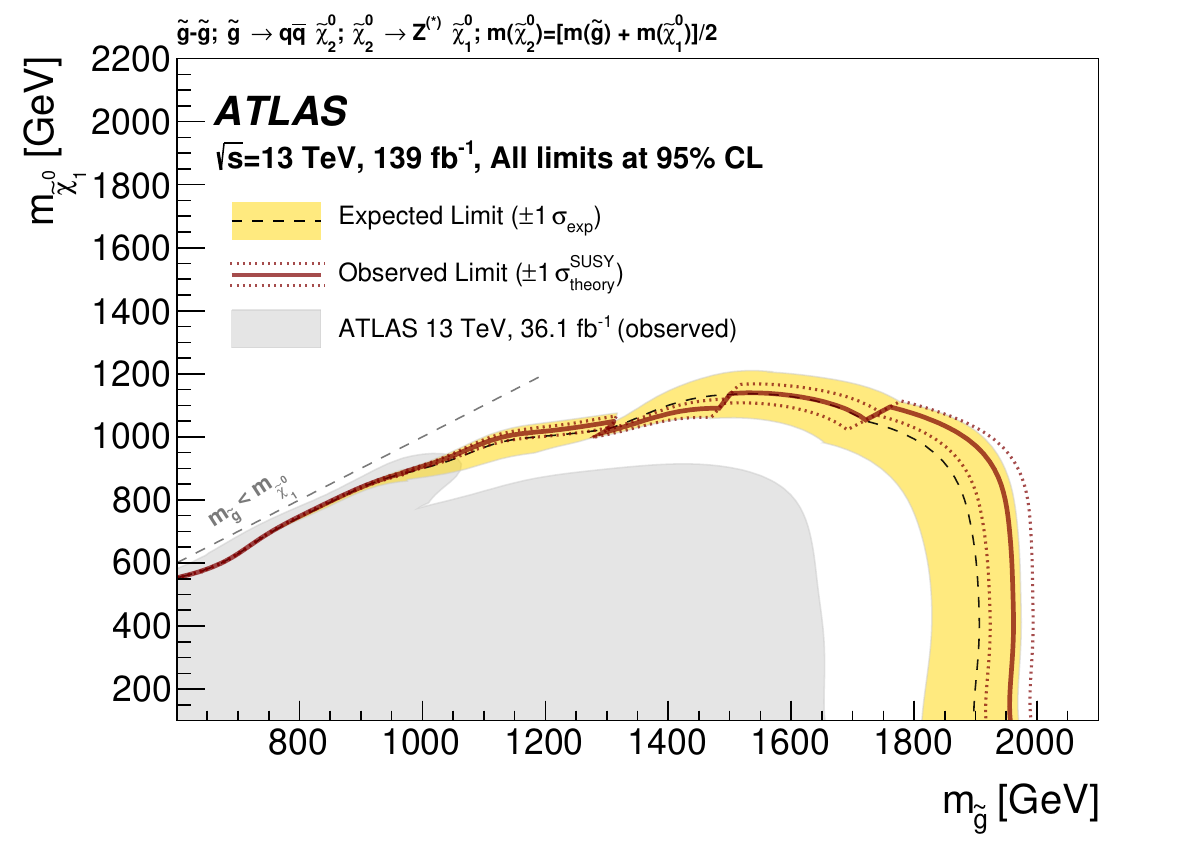}
\includegraphics[width=0.49\textwidth]{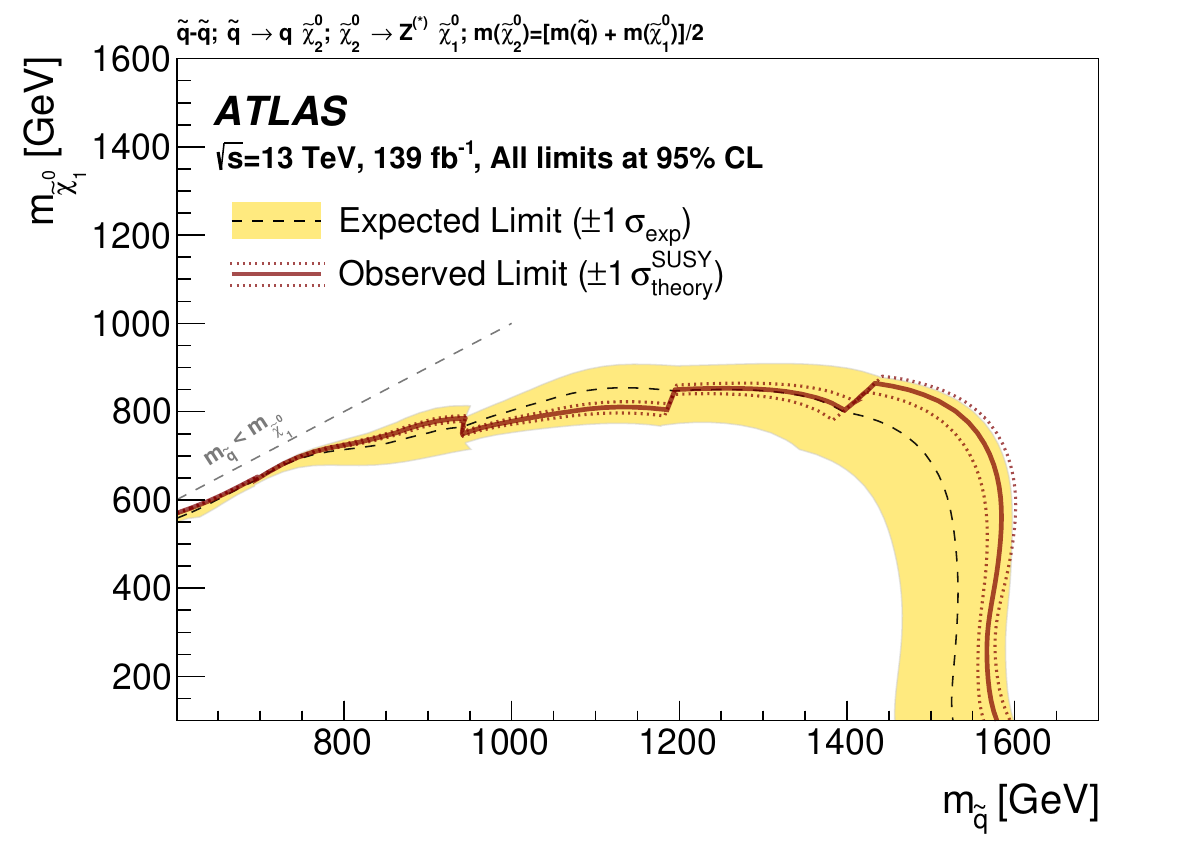}
\caption{Expected and observed exclusion contours derived from the combination of all of the Strong search SRs for the \gluino--\slepton (top-left), \gluino--$Z$ (top-right), and \squark--Z (bottom) models.
The dashed line indicates the expected limits at 95\% CL and the surrounding band shows the $1\sigma$ variation of the expected limit as a consequence of the uncertainties in the background prediction and experimental uncertainties of the signal ($\pm1\sigma_\text{exp}$).
The red dotted lines surrounding the observed limit contours indicate the variation resulting from changing the signal cross-section within its uncertainty ($\pm1\sigma_\text{theory}^\text{SUSY}$).
The grey-shaded area indicates the observed limits on these models from Ref.~\cite{SUSY-2016-33}.
}
\label{fig:strong_limits}
\end{figure}



\FloatBarrier
\section{Conclusion}
\label{sec:conclusion}

This paper presents searches for new phenomena in final states with exactly two oppositely charged same-flavour leptons, jets, and missing transverse momentum using the 139~\ifb\ Run~2 dataset of $\sqrt{s}=13~\TeV$ proton--proton collision data collected by the ATLAS detector at the LHC.
The analysis is split into three searches, probing both strong and electroweak production of sparticles.
The data are found to be consistent with \ac{SM} expectations in each, and strong limits are set on possible \acl{BSM} contributions to the signal regions.
The RJR search does not see significant excesses above the background expectation.
Thus the small excesses in the two-lepton channel of the 36~\ifb\ dataset did not persist with more data.
The EWK search and Strong search interpret the data with simplified models inspired by SUSY and set exclusion limits on these models, expanding the sensitivity reach compared to previous analyses.
The EWK search targets electroweak production of chargino--neutralino pairs and a \ac{GMSB} model with higgsino \acp{NLSP}.
Limits up to 820~\GeV\ and 900~\GeV\ are set on the masses of the mass-degenerate chargino/neutralino and higgsino \ac{NLSP} respectively.
This is an improvement on the previous result by roughly 200~\GeV\ in chargino/neutralino masses.
The EWK search covers masses and branching ratios of the \ac{GMSB} model that are between the limits from the ATLAS four-lepton and all-hadronic searches.
The Strong search targets strong production of gluino- or squark-pairs decaying via a $Z$ boson or the \ac{NLSP} to lepton-pairs and jets.
Limits up to 2250~\GeV\ and 1550~\GeV\ are set on the masses of the gluino and squarks respectively.
Compared to the previous result, the limits on the gluino mass improved by 400~\GeV\ and the limits on squark masses by 300~\GeV.
Improvements to both the Strong and EWK searches result from the increased size of the dataset and subsequent optimizations of analysis requirements.


\section*{Acknowledgements}


We thank CERN for the very successful operation of the LHC, as well as the
support staff from our institutions without whom ATLAS could not be
operated efficiently.
 
We acknowledge the support of
ANPCyT, Argentina;
YerPhI, Armenia;
ARC, Australia;
BMWFW and FWF, Austria;
ANAS, Azerbaijan;
CNPq and FAPESP, Brazil;
NSERC, NRC and CFI, Canada;
CERN;
ANID, Chile;
CAS, MOST and NSFC, China;
Minciencias, Colombia;
MEYS CR, Czech Republic;
DNRF and DNSRC, Denmark;
IN2P3-CNRS and CEA-DRF/IRFU, France;
SRNSFG, Georgia;
BMBF, HGF and MPG, Germany;
GSRI, Greece;
RGC and Hong Kong SAR, China;
ISF and Benoziyo Center, Israel;
INFN, Italy;
MEXT and JSPS, Japan;
CNRST, Morocco;
NWO, Netherlands;
RCN, Norway;
MEiN, Poland;
FCT, Portugal;
MNE/IFA, Romania;
MESTD, Serbia;
MSSR, Slovakia;
ARRS and MIZ\v{S}, Slovenia;
DSI/NRF, South Africa;
MICINN, Spain;
SRC and Wallenberg Foundation, Sweden;
SERI, SNSF and Cantons of Bern and Geneva, Switzerland;
MOST, Taiwan;
TENMAK, T\"urkiye;
STFC, United Kingdom;
DOE and NSF, United States of America.
In addition, individual groups and members have received support from
BCKDF, CANARIE, Compute Canada and CRC, Canada;
PRIMUS 21/SCI/017 and UNCE SCI/013, Czech Republic;
COST, ERC, ERDF, Horizon 2020 and Marie Sk{\l}odowska-Curie Actions, European Union;
Investissements d'Avenir Labex, Investissements d'Avenir Idex and ANR, France;
DFG and AvH Foundation, Germany;
Herakleitos, Thales and Aristeia programmes co-financed by EU-ESF and the Greek NSRF, Greece;
BSF-NSF and MINERVA, Israel;
Norwegian Financial Mechanism 2014-2021, Norway;
NCN and NAWA, Poland;
La Caixa Banking Foundation, CERCA Programme Generalitat de Catalunya and PROMETEO and GenT Programmes Generalitat Valenciana, Spain;
G\"{o}ran Gustafssons Stiftelse, Sweden;
The Royal Society and Leverhulme Trust, United Kingdom.
 
The crucial computing support from all WLCG partners is acknowledged gratefully, in particular from CERN, the ATLAS Tier-1 facilities at TRIUMF (Canada), NDGF (Denmark, Norway, Sweden), CC-IN2P3 (France), KIT/GridKA (Germany), INFN-CNAF (Italy), NL-T1 (Netherlands), PIC (Spain), ASGC (Taiwan), RAL (UK) and BNL (USA), the Tier-2 facilities worldwide and large non-WLCG resource providers. Major contributors of computing resources are listed in Ref.~\cite{ATL-SOFT-PUB-2021-003}.


\clearpage

\printbibliography

\clearpage
\input{atlas_authlist}

\end{document}

%% file: atlas_authlist.tex
 
\begin{flushleft}
\hypersetup{urlcolor=black}
{\Large The ATLAS Collaboration}

\bigskip

\AtlasOrcid[0000-0002-6665-4934]{G.~Aad}$^\textrm{\scriptsize 101}$,
\AtlasOrcid[0000-0002-5888-2734]{B.~Abbott}$^\textrm{\scriptsize 119}$,
\AtlasOrcid[0000-0002-7248-3203]{D.C.~Abbott}$^\textrm{\scriptsize 102}$,
\AtlasOrcid[0000-0002-2788-3822]{A.~Abed~Abud}$^\textrm{\scriptsize 36}$,
\AtlasOrcid[0000-0002-1002-1652]{K.~Abeling}$^\textrm{\scriptsize 55}$,
\AtlasOrcid[0000-0002-2987-4006]{D.K.~Abhayasinghe}$^\textrm{\scriptsize 94}$,
\AtlasOrcid[0000-0002-8496-9294]{S.H.~Abidi}$^\textrm{\scriptsize 29}$,
\AtlasOrcid[0000-0002-9987-2292]{A.~Aboulhorma}$^\textrm{\scriptsize 35e}$,
\AtlasOrcid[0000-0001-5329-6640]{H.~Abramowicz}$^\textrm{\scriptsize 150}$,
\AtlasOrcid[0000-0002-1599-2896]{H.~Abreu}$^\textrm{\scriptsize 149}$
% ...
\bigskip
\\

% ...
$^{29}$Physics Department, Brookhaven National Laboratory, Upton NY; United States of America.\\
% ...
$^{35}$$^{(a)}$Facult\'e des Sciences Ain Chock, R\'eseau Universitaire de Physique des Hautes Energies - Universit\'e Hassan II, Casablanca;$^{(b)}$Facult\'{e} des Sciences, Universit\'{e} Ibn-Tofail, K\'{e}nitra;$^{(c)}$Facult\'e des Sciences Semlalia, Universit\'e Cadi Ayyad, LPHEA-Marrakech;$^{(d)}$LPMR, Facult\'e des Sciences, Universit\'e Mohamed Premier, Oujda;$^{(e)}$Facult\'e des sciences, Universit\'e Mohammed V, Rabat;$^{(f)}$Institute of Applied Physics, Mohammed VI Polytechnic University, Ben Guerir; Morocco.\\
$^{36}$CERN, Geneva; Switzerland.\\
% ...
$^{55}$II. Physikalisches Institut, Georg-August-Universit\"{a}t G\"ottingen, G\"ottingen; Germany.\\
% ...
$^{94}$Department of Physics, Royal Holloway University of London, Egham; United Kingdom.\\
% ...
$^{101}$CPPM, Aix-Marseille Universit\'e, CNRS/IN2P3, Marseille; France.\\
$^{102}$Department of Physics, University of Massachusetts, Amherst MA; United States of America.\\
% ...
$^{119}$Homer L. Dodge Department of Physics and Astronomy, University of Oklahoma, Norman OK; United States of America.\\
% ...
$^{149}$Department of Physics, Technion, Israel Institute of Technology, Haifa; Israel.\\
$^{150}$Raymond and Beverly Sackler School of Physics and Astronomy, Tel Aviv University, Tel Aviv; Israel.\\
% ...
$^{a}$ Also Affiliated with an institute covered by a cooperation agreement with CERN.\\
$^{b}$ Also at Borough of Manhattan Community College, City University of New York, New York NY; United States of America.\\
$^{c}$ Also at Bruno Kessler Foundation, Trento; Italy.\\
$^{d}$ Also at Center for High Energy Physics, Peking University; China.\\
$^{e}$ Also at Centro Studi e Ricerche Enrico Fermi; Italy.\\
$^{f}$ Also at CERN, Geneva; Switzerland.\\
% ...

\end{flushleft}
